\documentclass[aps,floatfix,prb,twocolumn,a4paper,10pt,showpacs,citeautoscript,preprintnumbers,superscriptaddress]{revtex4-1}
\usepackage[english]{babel}%%% LaTeX for English

\usepackage{amsmath}	%%% Mathematics AMS package 
\usepackage{amssymb}	%%% More math symbols
\usepackage{bbm}		%%% Font for ensembes
\usepackage{graphicx}	%%% Include graphics
\usepackage{graphics}	%%% Include graphics
\DeclareGraphicsExtensions{.jpg,.pdf}
\usepackage{grffile}        	%%% For complicated file names
\usepackage{xcolor}		%%% Color options for eps
\usepackage[colorlinks=true, a4paper=true, pdfstartview=FitV,linkcolor=blue, citecolor=blue, urlcolor=blue]{hyperref}

\renewcommand{\theequation}{\arabic{section}.\arabic{equation}}

\makeatletter
\newcommand*{\Equation}{\@ifstar\sEquation\oEquation}
\newcommand{\sEquation}[1]{\begin{equation*}#1\end{equation*}}
\newcommand{\oEquation}[2]{  \begin{equation}\label{#1}#2\end{equation} }
\makeatother

\newcommand{\Align}[2]{\begin{align}\label{#1}#2\end{align}}
\newcommand{\SubAlign}[2]{\begin{subequations}\label{#1}\begin{align}#2\end{align}\end{subequations}}
\newcommand{\Vector}[1]{\left( \begin{array}{c}#1\end{array}\right)}

\newcommand{\mbf}{\mathbf}
\newcommand{\bs}{\boldsymbol}

\newcommand{\Partref}[1]{Sec.~\ref{#1}}
\newcommand{\Appref}[1]{App.~\ref{#1}}
\newcommand{\Figref}[1]{Fig.~\ref{#1}}
\newcommand{\Eqref}[1]{\eqref{#1}}

\newcommand{\eg}{{\it e.g.~}}
\newcommand{\ie}{{\it i.e.~}} 
\newcommand{\Ie}{{\it I.e.~}} 

\newcommand{\Uone}{\mbox{U(1)}}
\newcommand{\Real}{\mathbbm{R}}
\newcommand{\CPtwo}{$\ {\mathbbm{C}}{{P}}^2\ $}
\newcommand{\Ztwo}{{\mathbbm{Z}}_2}

\newcommand{\dd}{\text{d}}

\renewcommand\Re{\mathrm{Re}}
\renewcommand\Im{\mathrm{Im}}

\newcommand{\oz}{{(0)}}
\newcommand{\order}{o}

\newcommand{\Q}{\mathcal{Q}}
\newcommand{\F}{\mathcal{F}}

\newcommand{\FLondon}{\mathcal{F}_{\mbox{\tiny London}}}
\newcommand{\FGS}{F_{\mbox{\tiny GS}}}
\newcommand{\FSG}{F_{\mbox{\tiny SG}}}

\def \d{\mathrm{d}}
\newcommand{\VV}{{\mathbb{V}}} 
\newcommand{\PP}{{\cal P}}
\newcommand{\R}{{\mathbb{R}}}

\newcommand{\C}{{\mathbb{C}}} 
 
\newcommand{\I}{{\mathbb{I}}} 
 
\newcommand{\CP}{{\mathbb{C}}{{P}}}

\newcommand{\beq}{\begin{equation}} 
\newcommand{\eeq}{\end{equation}} 
\newcommand{\bea}{\begin{eqnarray}} 
\newcommand{\eea}{\end{eqnarray}} 
\newcommand{\ben}{\begin{eqnarray*}}
\newcommand{\een}{\end{eqnarray*}}
\newcommand{\ra}{\rightarrow}

\newcommand{\cd}{\partial} 
\newcommand{\wt}{\widetilde}

\newcommand{\eps}{{\varepsilon}}

\newcommand{\hess}{{\sf Hess}}

\renewcommand{\ol}{\overline} 
\newcommand{\less}{\backslash}

\newcommand{\nvec}{{\bf n}}

\newcommand{\tauvec}{\mbox{\boldmath $\tau$}}

\newcommand{\ip}[1]{ \langle #1 \rangle }
\renewcommand{\varrho}{\rho}
\graphicspath{{.}}
\begin{document}

%%%%%%%%%%%%%%%%%%%%%%%%%%%%%%%%%%%%%%%%%%%%%%%%%%%%%%%%%%%%%%%%%%%%%%
%%%%%%%%%%%%%%%%%%%%%%%%%%%%%%%%%%%%%%%%%%%%%%%%%%%%%%%%%%%%%%%%%%%%%%
%%%% Title informations and authors
%\preprint{Phys. Rev. B. {\bf XX}, XXXXXX (2012)}
\title{\texorpdfstring{ Chiral \CPtwo skyrmions in three-band superconductors}{Chiral CP2 skyrmions in three-band superconductors}}
\author{Julien~Garaud}
%\email{garaud.phys@gmail.com}
\affiliation{Department of Physics, University of Massachusetts Amherst, MA 01003 USA }
\affiliation{Department of Theoretical Physics, The Royal Institute of Technology, Stockholm, SE-10691 Sweden}
\author{Johan~Carlstr\"om}
\affiliation{Department of Theoretical Physics, The Royal Institute of Technology, Stockholm, SE-10691 Sweden}
\author{Egor~Babaev}
\affiliation{Department of Physics, University of Massachusetts Amherst, MA 01003 USA }
\affiliation{Department of Theoretical Physics, The Royal Institute of Technology, Stockholm, SE-10691 Sweden}
\author{Martin~Speight}
\affiliation{School of Mathematics, University of Leeds, Leeds LS2 9JT, UK}
\date{\today}

%%%%%%%%%%%%%%%%%%%%%%%%%%%%%%%%%%%%%%%%%%%%%%%%%%%%%%%%%%%%%%%%%%%%%%
%%%% The abstract
\begin{abstract}

It is shown that under certain conditions, three-component 
superconductors (and in particular three-band systems) allow 
stable topological defects different from vortices. We 
demonstrate the existence of these excitations, characterized by a 
\CPtwo topological invariant, in models for three-component 
superconductors with broken time reversal symmetry. 
We term these topological defects 
``chiral $GL^{(3)}$ skyrmions'', where ``chiral'' 
refers to the fact that due to broken time reversal 
symmetry, these defects come in inequivalent 
left- and right-handed versions. In certain cases these objects 
are energetically cheaper than vortices and should be induced 
by an applied magnetic field. In other  situations these skyrmions 
are metastable states, which can be produced by a quench. 
Observation of these defects can signal broken time reversal 
symmetry in three-band superconductors or in Josephson-coupled
bilayers of $s_\pm$ and $s$-wave superconductors.

\end{abstract}


\maketitle

%%%%%%%%%%%%%%%%%%%%%%%%%%%%%%%%%%%%%%%%%%%%%%%%%%%%%%%%%%%%%%%%%%%%%%
%%%%%%%%%%%%%%%%%%%%%%%%%%%%%%%%%%%%%%%%%%%%%%%%%%%%%%%%%%%%%%%%%%%%%%
\section{Introduction}

Experiments on the recently discovered  iron pnictide 
superconductors suggest the existence of positive coefficient 
of Josephson coupling between superconducting components 
in two bands ($s_\pm$ state) and possibly more than two 
superconducting bands \cite{iron2}. 
Under these circumstances,  new physics can appear. That is, 
frustration of competing interband Josephson couplings in 
three-component superconductors, can lead to spontaneously 
Broken Time Reversal Symmetry (BTRS) \cite{nagaosa,stanev} 
(another scenario for BTRS states in pnictides was discussed in 
Refs.~\onlinecite{zhang,honerkamp}). 
There, the ground state explicitly breaks the discrete 
$\Uone\times\Ztwo$ symmetry  
\cite{Garaud.Carlstrom.ea:11,Carlstrom.Garaud.ea:11a}. 
Related multicomponent states were also recently discussed, 
in connection with other materials \cite{agterberg2011}.
If superconductivity in iron pnictides is described by just a 
two-band $s_\pm$ models, BTRS states can nonetheless be 
obtained in a Josephson-coupled bilayer of $s_\pm$ 
superconductor and ordinary $s$-wave material \cite{nagaosa}. 
Such bilayer systems can be effectively described by a 
three-component model where the third component is coupled 
through a ``real-space" inter-layered Josephson coupling.

Due to a number of unconventional phenomena, which are 
not possible in two-band superconductors, the possible 
experimental realization of three component superconductors 
(either with or without BTRS) recently started to attract substantial 
interest  \cite{stanev,hu,Garaud.Carlstrom.ea:11,
Carlstrom.Garaud.ea:11a,stanev2,machida3,levchenko,nitta,Lin:12}.
These phenomena include: exotic collective modes which are 
different from the Leggett's mode 
\cite{massless,Carlstrom.Garaud.ea:11a,stanev2}; the existence 
of a large disparity in coherence lengths even when intercomponent 
Josephson coupling is very strong, leading to type-1.5 regimes 
\cite{Carlstrom.Garaud.ea:11a} (where some coherence lengths 
are smaller and some are larger than the magnetic field penetration 
length \cite{bs1}); the possibility of flux-carrying topological solitons 
different from Abrikosov vortices \cite{Garaud.Carlstrom.ea:11}. 

\begin{figure*}[!htbp]
\hbox to \linewidth{ \hss
  \includegraphics[width=0.1\linewidth]{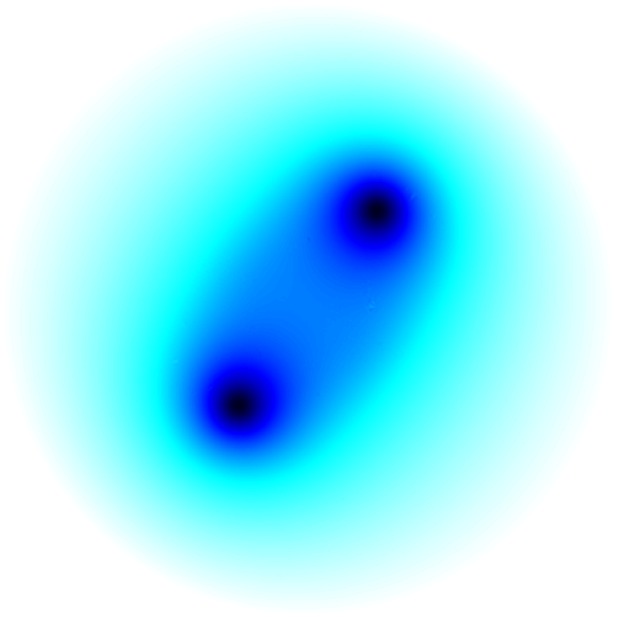} 
  \includegraphics[width=0.1\linewidth]{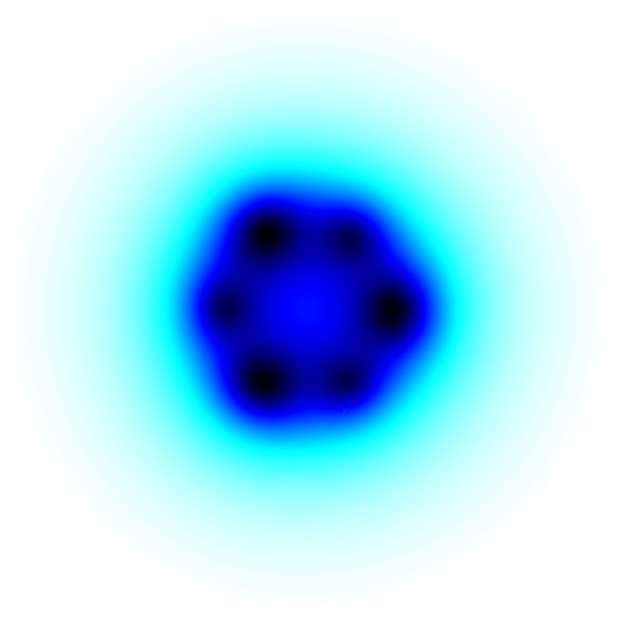}
  \includegraphics[width=0.1\linewidth]{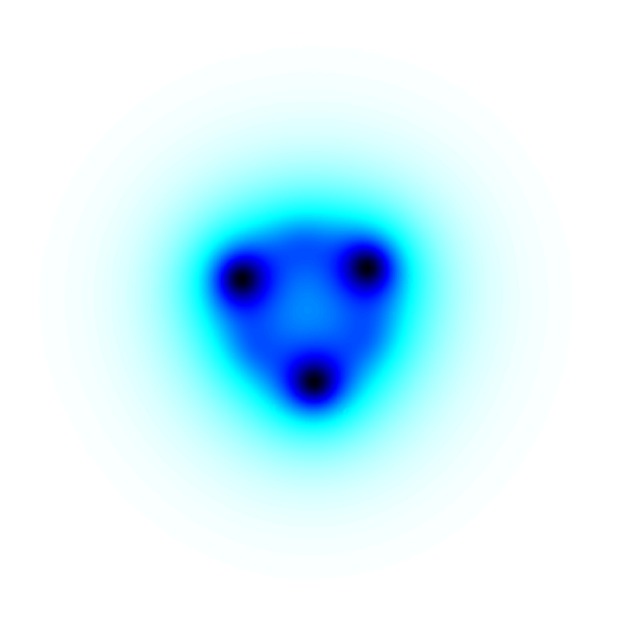}
  \includegraphics[width=0.1\linewidth]{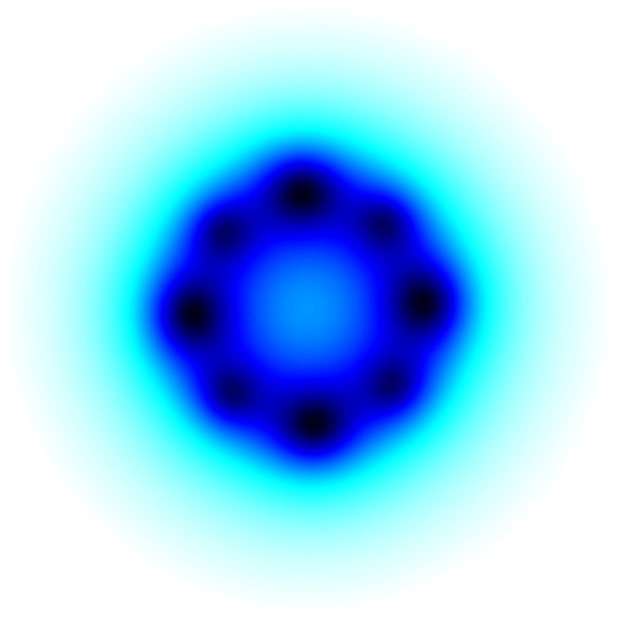}
  \includegraphics[width=0.1\linewidth]{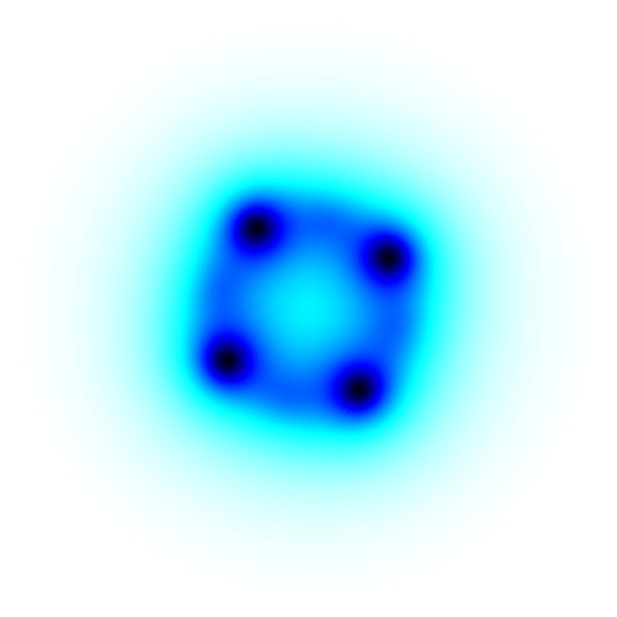}
  \includegraphics[width=0.1\linewidth]{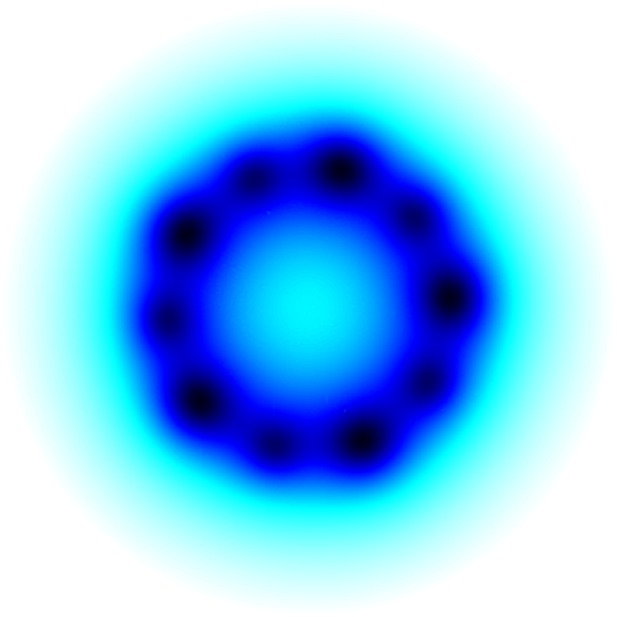}
  \includegraphics[width=0.1\linewidth]{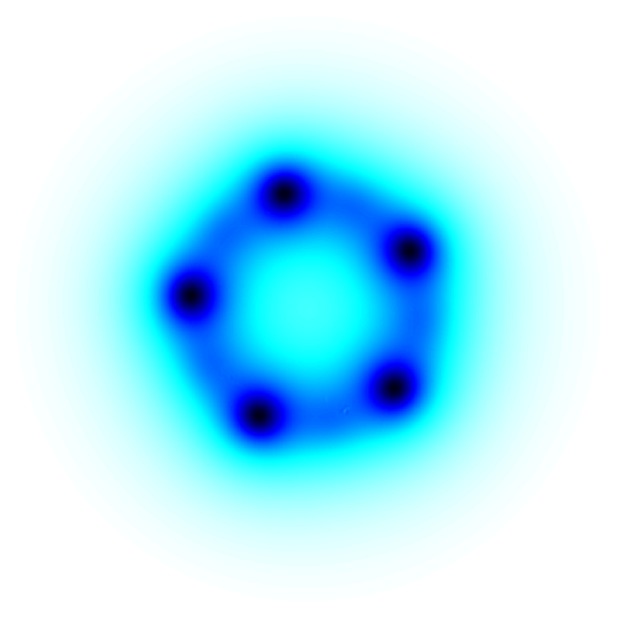}
  \includegraphics[width=0.1\linewidth]{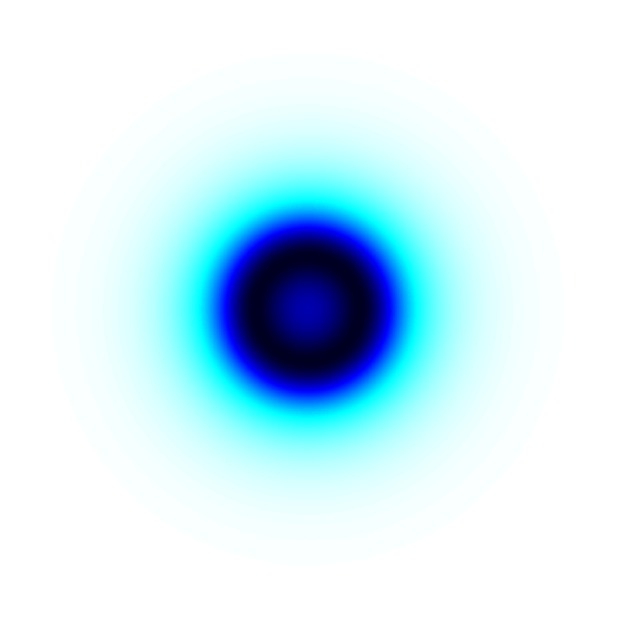}
  \includegraphics[width=0.1\linewidth]{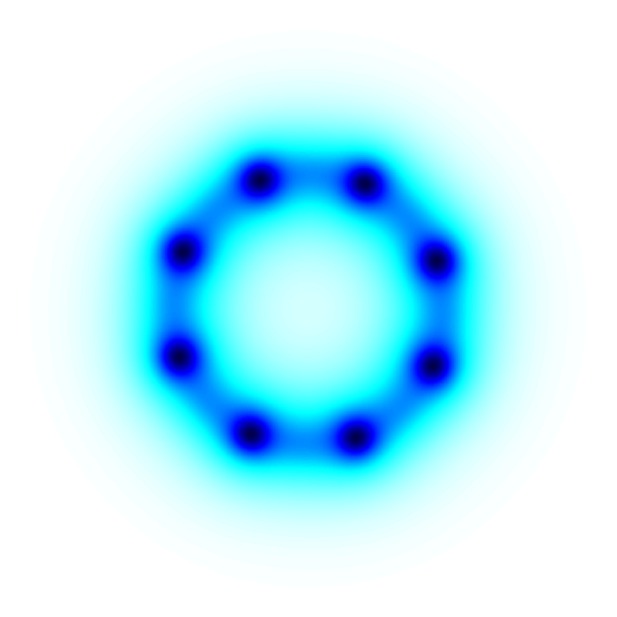}
  \includegraphics[width=0.1\linewidth]{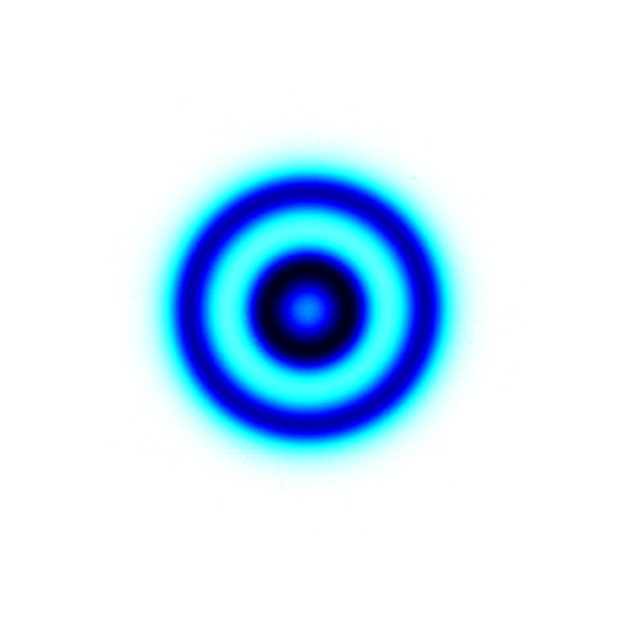}
\hss}
\caption{
(Color online) -- Example of unusual observable magnetic field 
configuration of chiral skyrmions.
}
\label{Signature}
\end{figure*}

This paper is a follow-up to Ref.~\onlinecite{Garaud.Carlstrom.ea:11} 
where we introduced new flux-carrying topological solitons. Here we 
study in detail, these topological solitons which we term chiral 
$GL^{(3)}$ skyrmions (chiral skyrmions for short). 
They are magnetic flux-carrying excitations characterized by a \CPtwo 
topological invariant, (by contrast this invariant  is trivial for ordinary 
vortices). The topological properties, motivating the denomination 
skyrmion are rigorously discussed. As the terminology suggests, the 
soliton itself has a given \emph{chiral} state of the Broken Time 
Reversal Symmetry. More precisely, different arrangements of the 
fractional vortices constituting a skyrmion carrying integer flux define 
different chirality of the skyrmion.
Finally $GL^{(3)}$ refers to the physical context of the three-component 
Ginzburg--Landau theory. The thermodynamic and energetic 
(meta)stability of chiral skyrmions are discussed, as well as their 
perturbative stability.
In scanning SQUID, scanning Hall or magnetic force microscopy 
experiments, chiral $GL^{(3)}$ skyrmions can (under certain conditions) 
be distinguished from vortices by their very exotic magnetic field 
profile. \Figref{Signature} shows examples of such exotic magnetic field 
signatures of chiral skyrmions in three band superconductors with 
various parameters of the model.

%%%%%%%%%%%%%%%%%%%%%%%%%%%%%%%%%%%%%%%%%
%%% Plan
The paper is organized as follows. In \Partref{section:theoretical_framework} 
we introduce a Ginzburg-Landau model for three-component superconductors 
where phase frustration due to competing Josephson interactions leads to 
Broken Time Reversal Symmetry states. 
The structure of the domain walls which are possible due to this new 
spontaneously broken $\Ztwo$ symmetry is discussed in \Partref{TRSB}. 
The essential concepts of the topological excitations in multi-band 
superconductors are discussed in \Partref{Fractional-vortices}. After that, 
the new kind of topological excitations, chiral $GL^{(3)}$ skyrmions, 
are discussed \Partref{Skyrmions}.
The physical properties: (i) energy of formation of a skyrmion versus 
vortex lattice, (ii) thermodynamical stability of the chiral skyrmions 
and (iii) their perturbative stability are investigated \Partref{Properties}. 
In the next part,  \Partref{Interactions}, the very rich interactions between 
the chiral skyrmions and between skyrmions and vortices are investigated. 
The model has many interesting mathematical aspects as well. 
\Partref{section:math} is devoted to the most formal aspects and rigorous 
justifications of the physics and mathematical properties of the 
three-component Ginzburg--Landau model and the skyrmionic excitations 
therein. This section aims at a more mathematical audience. Thus, readers 
less interested in formal justification of the physics can skip these discussions, 
and go straight after \Partref{Interactions} to our conclusions in 
\Partref{Conclusion}. There we conclude this paper by addressing, 
in more detail, the possible experimental signatures of our chiral 
$GL^{(3)}$ skyrmions.

%%%%%%%%%%%%%%%%%%%%%%%%%%%%%%%%%%%%%%%%%%%%%%%%%%%%%%%%%%%%%%%%%%%%%%
%%%%%%%%%%%%%%%%%%%%%%%%%%%%%%%%%%%%%%%%%%%%%%%%%%%%%%%%%%%%%%%%%%%%%%
\section{The model} \label{section:theoretical_framework}

In this paper we consider various realizations of three-component 
superconductivity described by the following three-component 
Ginzburg--Landau (GL) model:
\Align{freeEnergy}{
 \F&= \frac{1}{2}(\nabla \times \bs A)^2 
+\sum_{a}\frac{1}{2}|\bs D\psi_a|^2 
+\alpha_a|\psi_a|^2+\frac{1}{2}\beta_a|\psi_a|^4 \nonumber \\
+&\sum_{a,b>a}\gamma_{ab}|\psi_a|^2|\psi_b|^2
-\eta_{ab}|\psi_a||\psi_b|\cos(\varphi_b-\varphi_a)\,.
}
Here $\bs D=\nabla+ie{\bs A}$ and $\psi_a=|\psi_a| e^{i\varphi_a}$ 
are complex fields representing the superconducting components. 
The component indices $a,b$ take the values $1,2,3$. In the particular 
case of a three-band superconductor, different superconducting 
components arise due to Cooper pairing in three different bands. 
The bands are coupled by their interaction with the vector potential 
$\bs A$ and also through potential interactions. The coefficients 
$\eta_{ab}$ are the intercomponent Josephson couplings. We also consider
 the more general case which includes  bi-quadratic density interactions 
with the couplings $\gamma_{ab}$. Here, the London magnetic field 
penetration length is parametrized by the gauge coupling constant $e$.
Functional variation of the free energy \Eqref{freeEnergy} 
with respect to the fields gives Ginzburg-Landau equations  
\Equation{EOM}{
\bs D\bs D \psi_a =2\frac{\partial V}{\partial\psi_a^*}\,,~~~~ 
\partial_i\left(\partial_iA_j -\partial_jA_i\right)=  J_i\,.
}
where $V$ is the collection of all non-gradient terms and 
the supercurrent is defined as
\Equation{Currents}{
  \bs J\equiv\sum_{a=1,2,3}\bs J^{(a)}= 
   \sum_{a=1,2,3}e\Im\left(\psi_a^*\bs D\psi_a  \right)\,.
}
In multiband superconductors,  a Ginzburg--Landau expansion of this 
kind can in certain cases be formally justified microscopically (see \eg 
corresponding discussion in two-band case \cite{silaev}). In what follows, 
different physical realizations of the model \Eqref{freeEnergy} with 
different broken symmetries are considered. Note that in some of the 
physical realizations of multicomponent GL models,  some of the 
couplings are forbidden (for example on symmetry grounds). This can 
occur for intercomponent Josephson couplings, in some realizations
\cite{frac,*smiseth}. 
More terms, consistent with symmetries, can be included to extend 
the GL functional. Alternatively a microscopic approach can provide 
a more quantitatively accurate picture at lower temperatures. However, 
the properties of the topological objects which are discussed, should 
then differ only quantitatively and not qualitatively  in the framework 
of \eg microscopic approach for a system with a given symmetry 
(some examples how phenomenological multiband GL models give 
good results even at low temperature can be found in 
Ref.~\onlinecite{silaev}).  

The field configurations considered in the following are 
two-dimensional, as well as  three dimensional systems with 
translation invariance along the third axis.

\subsection{\texorpdfstring{
Broken Time Reversal Symmetry, the $U(1)\times \Ztwo$ states}
{Broken Time Reversal Symmetry, the U(1)xZ2 states}} 
\label{TRSB}

For a given parameter set $(\alpha_a,\beta_a,\eta_{ab},\gamma_{ab})$, 
the ground state is the field configuration which minimizes the 
potential energy. The corresponding values of $|\psi_a|$'s and 
$\varphi_a$'s, together with the gauge coupling $e$ determine 
the physical length scales of the theory. The particularly interesting 
property of the model \Eqref{freeEnergy}, is that the ground 
state can be qualitatively different from its two band counterparts. 
While in two bands systems with Josephson interactions the 
phase-locking is trivial (either $0$ or $\pi$), the phase-locking 
in three bands can be much more involved. Indeed, competition 
between different phase-locking terms possibly leads to phase 
frustration. 
When $\eta_{ab}>0$, the corresponding Josephson term is 
minimal for zero phase difference, while if $\eta_{ab}<0$ it is 
minimal for $\varphi_{ab}\equiv\varphi_b-\varphi_a=\pi$. Now
 if the signs of $\eta_{ab}$'s are all positive (we denote it as 
$[+++]$), the ground state has $\varphi_1=\varphi_2=\varphi_3$. 
Similarly for $[+--]$ couplings, the phase locking pattern 
$\varphi_1=\varphi_2=\varphi_3+\pi$. However for $[++-]$ or 
$[---]$, the phase locking terms are \emph{frustrated}. That is: 
all three Josephson terms cannot simultaneously attain their 
minimal values. As a result ground state phase differences are 
neither $0$ nor $\pi$.
For example, consider the case $\alpha_a=-1,\;\beta_a=1$ and 
$\eta_{ab}=-1$. Symmetry under global $\Uone$ phase rotations 
allows to set $\varphi_1=0$ without loss of generality (for the 
below considerations). There, two ground states are possible 
$\varphi_2=2\pi/3,\;\varphi_3=-2\pi/3$ or 
$\varphi_2=-2\pi/3,\;\varphi_3=2\pi/3$. The two ground states are 
each other's complex conjugate. The actual values of the ground 
state phases depend on the potential parameters.

Note that the free energy is invariant under complex conjugation, 
$(\psi_1,\psi_2,\psi_3)\mapsto(\psi_1^*,\psi_2^*,\psi_3^*)$, which 
takes it to a state with different phase locking. Thus the theory has 
a spontaneously broken discrete ($\Ztwo$) symmetry, called Time 
Reversal Symmetry. 
That is, the free energy is still invariant under complex conjugation, 
but the ground state is \emph{not}. By `picking' one of the two 
inequivalent phase-locking patterns, the ground state explicitly 
breaks the discrete $\Ztwo$ symmetry. Such states are	
termed Broken Time Reversal Symmetry (BTRS) states.

\subsection{Domain walls in BTRS states}

\begin{figure}[!htb]
\vspace{0.25cm} 
\hbox to \linewidth{ \hss
 \hspace{1cm} 
%  \resizebox{250pt}{168pt}{ \input{figure2a}   }
 % GNUPLOT: LaTeX picture with Postscript
\begingroup
  \makeatletter
  \providecommand\color[2][]{%
    \GenericError{(gnuplot) \space\space\space\@spaces}{%
      Package color not loaded in conjunction with
      terminal option `colourtext'%
    }{See the gnuplot documentation for explanation.%
    }{Either use 'blacktext' in gnuplot or load the package
      color.sty in LaTeX.}%
    \renewcommand\color[2][]{}%
  }%
  \providecommand\includegraphics[2][]{%
    \GenericError{(gnuplot) \space\space\space\@spaces}{%
      Package graphicx or graphics not loaded%
    }{See the gnuplot documentation for explanation.%
    }{The gnuplot epslatex terminal needs graphicx.sty or graphics.sty.}%
    \renewcommand\includegraphics[2][]{}%
  }%
  \providecommand\rotatebox[2]{#2}%
  \@ifundefined{ifGPcolor}{%
    \newif\ifGPcolor
    \GPcolortrue
  }{}%
  \@ifundefined{ifGPblacktext}{%
    \newif\ifGPblacktext
    \GPblacktexttrue
  }{}%
  % define a \g@addto@macro without @ in the name:
  \let\gplgaddtomacro\g@addto@macro
  % define empty templates for all commands taking text:
  \gdef\gplbacktext{}%
  \gdef\gplfronttext{}%
  \makeatother
  \ifGPblacktext
    % no textcolor at all
    \def\colorrgb#1{}%
    \def\colorgray#1{}%
  \else
    % gray or color?
    \ifGPcolor
      \def\colorrgb#1{\color[rgb]{#1}}%
      \def\colorgray#1{\color[gray]{#1}}%
      \expandafter\def\csname LTw\endcsname{\color{white}}%
      \expandafter\def\csname LTb\endcsname{\color{black}}%
      \expandafter\def\csname LTa\endcsname{\color{black}}%
      \expandafter\def\csname LT0\endcsname{\color[rgb]{1,0,0}}%
      \expandafter\def\csname LT1\endcsname{\color[rgb]{0,1,0}}%
      \expandafter\def\csname LT2\endcsname{\color[rgb]{0,0,1}}%
      \expandafter\def\csname LT3\endcsname{\color[rgb]{1,0,1}}%
      \expandafter\def\csname LT4\endcsname{\color[rgb]{0,1,1}}%
      \expandafter\def\csname LT5\endcsname{\color[rgb]{1,1,0}}%
      \expandafter\def\csname LT6\endcsname{\color[rgb]{0,0,0}}%
      \expandafter\def\csname LT7\endcsname{\color[rgb]{1,0.3,0}}%
      \expandafter\def\csname LT8\endcsname{\color[rgb]{0.5,0.5,0.5}}%
    \else
      % gray
      \def\colorrgb#1{\color{black}}%
      \def\colorgray#1{\color[gray]{#1}}%
      \expandafter\def\csname LTw\endcsname{\color{white}}%
      \expandafter\def\csname LTb\endcsname{\color{black}}%
      \expandafter\def\csname LTa\endcsname{\color{black}}%
      \expandafter\def\csname LT0\endcsname{\color{black}}%
      \expandafter\def\csname LT1\endcsname{\color{black}}%
      \expandafter\def\csname LT2\endcsname{\color{black}}%
      \expandafter\def\csname LT3\endcsname{\color{black}}%
      \expandafter\def\csname LT4\endcsname{\color{black}}%
      \expandafter\def\csname LT5\endcsname{\color{black}}%
      \expandafter\def\csname LT6\endcsname{\color{black}}%
      \expandafter\def\csname LT7\endcsname{\color{black}}%
      \expandafter\def\csname LT8\endcsname{\color{black}}%
    \fi
  \fi
  \setlength{\unitlength}{0.0500bp}%
 \resizebox{250pt}{168pt}{  \begin{picture}(7200.00,5040.00)%
    \gplgaddtomacro\gplbacktext{%
      \csname LTb\endcsname%
      \put(327,0){\makebox(0,0)[r]{\strut{} \Large  -2$\pi$}}%
      \put(327,1260){\makebox(0,0)[r]{\strut{} \Large $-\pi$}}%
      \put(327,2520){\makebox(0,0)[r]{\strut{} \Large 0}}%
      \put(327,3779){\makebox(0,0)[r]{\strut{} \Large $\pi$}}%
      \put(327,5039){\makebox(0,0)[r]{\strut{} \Large 2$\pi$}}%
      \put(522,5250){\makebox(0,0){\strut{} \Large -2$\pi$}}%
      \put(1782,5250){\makebox(0,0){\strut{} \Large $-\pi$}}%
      \put(3042,5250){\makebox(0,0){\strut{} \Large 0}}%
      \put(4301,5250){\makebox(0,0){\strut{} \Large $\pi$}}%
      \put(5561,5250){\makebox(0,0){\strut{} \Large 2$\pi$}}%
      \put(-311,2519){\rotatebox{-270}{\makebox(0,0){\strut{} \Large $\varphi_3$}}}%
      \put(6600,2519){\makebox(0,0){\strut{} \Large $\F_{\mbox{ pot}}$}}%
      \put(3041,-283){\makebox(0,0){\strut{} \Large $\varphi_2$}}%
    }%
    \gplgaddtomacro\gplfronttext{%
      \csname LTb\endcsname%
      \put(6200,0){\makebox(0,0)[l]{\strut{} \Large -10}}%
      \put(6200,1007){\makebox(0,0)[l]{\strut{} \Large -8}}%
      \put(6200,2015){\makebox(0,0)[l]{\strut{} \Large -6}}%
      \put(6200,3023){\makebox(0,0)[l]{\strut{} \Large -4}}%
      \put(6200,4031){\makebox(0,0)[l]{\strut{} \Large -2}}%
      \put(6200,5039){\makebox(0,0)[l]{\strut{} \Large  0}}%
    }%
    \gplbacktext%
    \put(0,0){\includegraphics{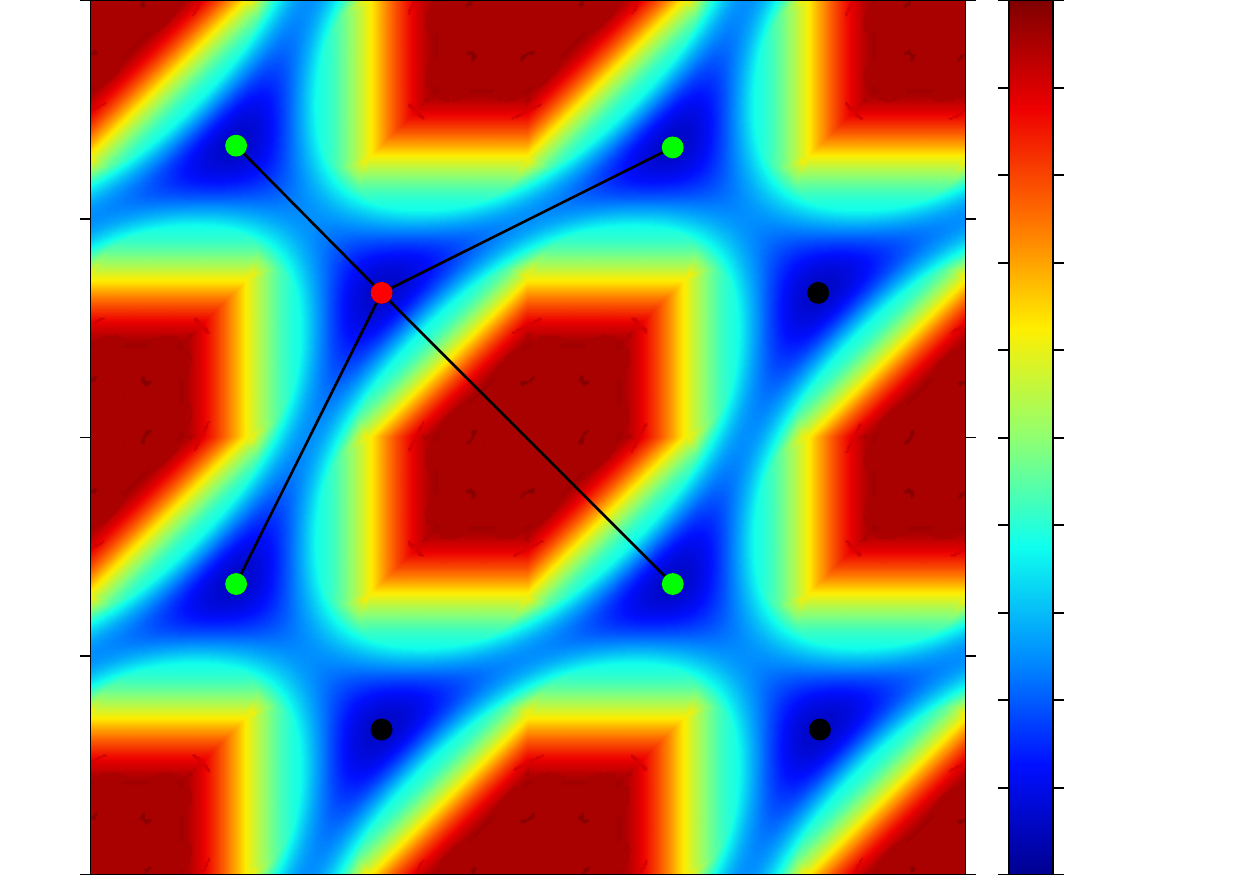}}%
    \gplfronttext
  \end{picture}%
}
\endgroup
%%%%%%%%%%%%%%%%%%%%%%%%%%%%%%%%%%%%%%%%%%%%%%%%%%%%%%%%%%%%%%%%%%%%%%
\hss}
\vspace{0.75cm} 
\hbox to \linewidth{ \hss
    \includegraphics[width=0.25\linewidth]{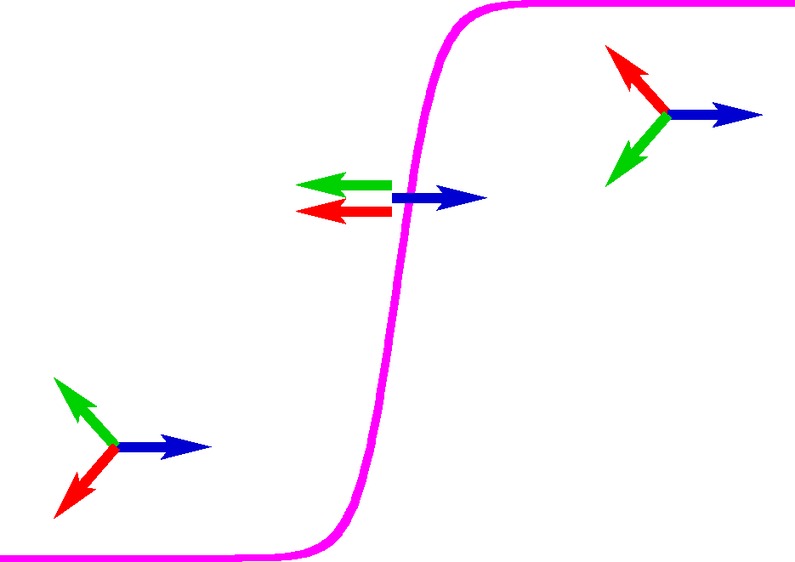}
\hspace{0.35cm} 
   \includegraphics[width=0.25\linewidth]{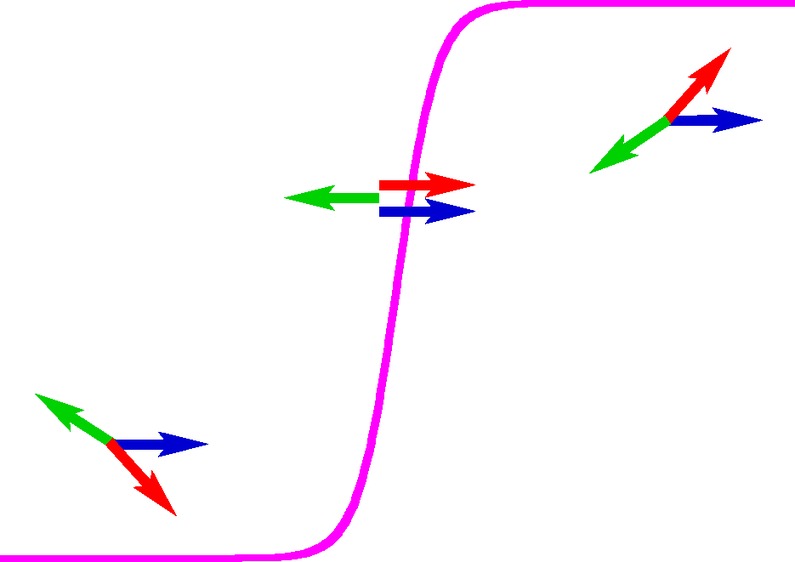}
\hspace{0.35cm} 
    \includegraphics[width=0.25\linewidth]{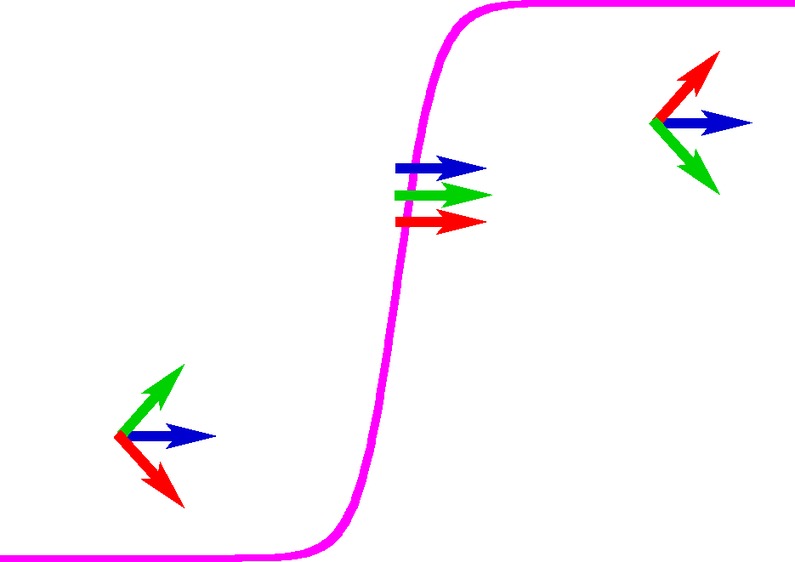}
 \hss}
 \caption{
(Color online) -- Representation of the vacuum submanifold (Top), 
for $(\alpha_a,\beta_a)=(-1,1)$ and $\eta_{ab}=-3$. The image 
shows the potential energy as a function of the phase differences: 
$\varphi_2$ and $\varphi_3$, minimized with respect to all 
moduli degrees of freedom (while $\varphi_1$ is set to zero by 
$\Uone$ invariance associated with simultaneous change of all 
phases). Red and green dots show inequivalent $\Ztwo$ ground 
states. And the lines connecting them represent \emph{four} 
different kinds of domain wall trajectory over the field manifold.  
Black dots are ground states located  farther than $2\pi$ in the 
phase differences.
The second line, gives a schematic representation of various 
$\Ztwo$ domain walls in three-band superconductors with different 
frustrations of phase angles, shown by arrows of different colors. 
The pink line schematically shows the phase difference between 
red and green arrow, interpolating between the two inequivalent 
ground states.
}
\label{Fig:Vacuum-manifold}
\end{figure}

BTRS systems have topological excitations related to the broken 
discrete symmetry in the form of domain walls. The domain walls 
interpolate between  domains of inequivalent ground states. In other 
words they are walls separating regions of different phase locking. 
It is instructive to display more quantitatively the structure of  the 
ground state (or ``vacuum") manifold, 
see \Figref{Fig:Vacuum-manifold}. There, the potential energy is 
minimized with respect to the densities $|\psi_a|$, for uniform fixed 
phase difference configurations. This provides a map of the ground 
state manifold. It appears clearly that there are disconnected 
inequivalent ground states (the red and green dots). Interestingly, 
there is not a unique path to connect inequivalent ground states with 
inequivalent phase locking, but four. The four corresponding 
domain walls will have different line tension (energy per unit length).
Note that, investigating the vacuum manifold with fixed ground 
state densities $|\psi_a|$ (at their true ground state value) provides 
a qualitatively similar picture. Namely, this approximation preserves 
the  positions of the minima. However, the actual values of 
$\F_{\mbox{\tiny pot}}$ are obviously different if $|\psi_a|$'s are 
held constant to the ground state, so this approximation does not 
allow one to calculate the energy of the domain walls. In particular 
the sharp angles appear there for strong Josephson couplings, when 
the ground state densities are not fixed. This property is absent when 
densities are held to their actual ground state values. 

\subsection{Flux-carrying topological defects in three component Ginzburg--Landau model} 
\label{Fractional-vortices}

As previously stated, three component Ginzburg--Landau model can 
exhibit BTRS and domain wall excitations associated with the broken 
$\Ztwo$ symmetry. There are also different topological defects, 
associated with the other broken symmetries.

Our main interest, here, is \emph{three-component} skyrmionic solutions 
of the Ginzburg--Landau model.  Here skyrmions are topological defects 
characterized by a topological invariant which classifies the maps 
$\R^2\to\CP^2$. In contrast to the topological invariant characterizing 
vortices (\ie the winding number which is defined as a line integral over a 
closed path), the topological index associated with skyrmionic excitations 
is given as an integral over $xy$-plane :
\Equation{Topological_Charge}{
   \Q(\Psi)=\int_{\Real^2}\frac{i\epsilon_{ji}}{2\pi|\Psi|^4} \left[
   |\Psi|^2\partial_i\Psi^\dagger\partial_j\Psi
   +\Psi^\dagger\partial_i\Psi\partial_j\Psi^\dagger\Psi
   \right]\dd^2x\,,
}
with $\Psi^\dagger=(\psi_1^*,\psi_2^*,\psi_3^*)$. A detailed
 derivation of this formula is given in \Partref{section:math}. 
If we have an axially symmetry vortex with a core where all 
superconducting condensates simultaneously vanish, then $\Q=0$. 
On the other hand, if singularities happen at different locations, then 
$\Q\neq0$ and the quantization condition $\Q=\bs B/\Phi_0=N$ 
holds ( $\Phi_0$ being the flux quantum and $N$ the number of 
flux quanta). This is rigorously discussed in 
\Partref{section:quantization}.

\subsubsection*{Fractional vortices}

In order to understand the physical properties of the later introduced  
\emph{chiral} skyrmions, it is good to remind oneself of the basic features 
of multi-component superconductors and their topological excitations.  
The elementary vortex excitations in this system are fractional vortices. 
They are defined as field configurations with a $2\pi$ phase winding only 
in one phase (\eg $\varphi_1$ has $\Delta\varphi_1\equiv\oint\nabla\varphi_1=2\pi$ 
winding while $\Delta\varphi_2=\Delta\varphi_3=0$). To better illustrate their physical 
properties, the Ginzburg--Landau free energy \Eqref{freeEnergy} can be
rewritten as 
\SubAlign{GLRewritten}{
   \F&= \frac{1}{2}(\nabla \times \bs A)^2 
	    + \frac{\bs J^2}{2e^2\varrho^2}  \label{JPart} \\
 &+\sum_{a}\frac{1}{2}(\nabla|\psi_a|)^2
	 +\alpha_a|\psi_a|^2+\frac{\beta_a}{2}|\psi_a|^4  
      \label{ModulusPart} \\
 &+\sum_{a,b>a}\frac{|\psi_a|^2|\psi_b|^2}{\varrho^2}\left(
\frac{(\nabla \varphi_{ab})^2}{2}
      -\frac{\eta_{ab}\varrho^2\cos\varphi_{ab}}{|\psi_a||\psi_b|}\right)
      \label{SineGordonPart} \\
 &+\sum_{a,b>a}\gamma_{ab}|\psi_a|^2|\psi_b|^2			
	 \label{Biquadractic}	\,,
}
where  $\varphi_{ab}\equiv\varphi_b-\varphi_a$ are the phase differences 
and $\varrho^2=\sum_a|\psi_a|^2$. The 
indices $a,b$ again denote the different superconducting condensates and 
take value $1,2,3$. The identity 
\Align{}{
   &\sum_{a=1}^n\sum_{b=1}^n |\psi_a|^2|\psi_b|^2\nabla \varphi_a
   \left(\nabla\varphi_a-\nabla\varphi_b  \right) 
\nonumber \\
 &=\sum_{a=1}^n\sum_{b=a+1}^n |\psi_a|^2|\psi_b|^2
   \left(\nabla\varphi_a-\nabla\varphi_b  \right)^2\,,
}
is used to derive this expression. Here, the supercurrent \Eqref{Currents} 
reads, more explicitly
\Equation{Currents2}{
  \bs J/e= e\varrho^2\bs A+\sum_{a}|\psi_a|^2\nabla\varphi_a \,.
}
Consider now a vortex for which the phase of only one component changes by 
$2\pi$: $\oint \nabla \varphi_a = 2\pi$. Such a configuration carries a fraction of 
flux quantum \cite{frac,*smiseth}
\Equation{FractionalFlux}{
   \Phi_a=\oint_\sigma \bs A \dd\bs\ell = \frac{|\psi_a|^2}{\varrho^2}
      \frac{1}{e}\oint_\sigma\nabla\varphi_a = \frac{|\psi_a|^2}{\varrho^2}\Phi_0\,,
}
where $|\psi_a|$ denotes the ground state density of $\psi_a$,
$\sigma$ is a closed curve around the vortex core, and $\Phi_0=2\pi/e$ 
is the flux quantum. For vanishing Josephson interactions, the symmetry 
is $[U(1)]^3$ and each fractional vortex has logarithmically diverging 
energy \cite{frac,*smiseth}. This can be seen easily in  the London limit 
by setting $\psi_a=\mathrm{const}$ everywhere except a sharp cutoff 
in the vortex core. There the terms \Eqref{Biquadractic} and 
\Eqref{ModulusPart} give trivial contribution to the free energy, so that
 the relevant parts now reads
\Align{London}{
   &\FLondon= \frac{1}{2}(\nabla \times \bs A)^2 
      + \frac{\bs J^2}{2e^2\varrho^2} \nonumber \\
 &+\sum_{a,b>a}\frac{|\psi_a|^2|\psi_b|^2}{2\varrho^2}\left((\nabla\varphi_{ab})^2
      -\frac{2\eta_{ab}\varrho^2}{|\psi_a||\psi_b|}\cos\varphi_{ab}\right).
}
In a $[U(1)]^3$ symmetric model, one fractional vortex gives 
logarithmically divergent contribution to the energy through the term 
\Equation{}{
\int_{r_c}^r r^\prime\dd r^\prime\int_0^{2\pi}\dd\theta
\frac{|\psi_a|^2|\psi_b|^2}{2\varrho^2}(\nabla\varphi_{ab})^2
=\pi\frac{|\psi_a|^2|\psi_b|^2}{\varrho^2}\ln{\frac{r}{r_c}}\,,
}
$r_c$ being a sharp cut-off corresponding to the core size of a vortex.
However a bound state of three such vortices (where each phase 
$a=1,2,3$ had $2\pi$ phase winding) has finite energy. Indeed such 
a bound state has no winding in the phase differences. This 
finite-energy bound state is a ``composite" vortex having one core 
singularity where $|\psi_1|+|\psi_2|+|\psi_3|=0$. Around this core all 
three phases have similar winding $\Delta \varphi_a=2\pi$. A vortex 
carrying one quantum $\Phi_0$ of flux is thus a logarithmically bound 
state of fractional vortices. For non-zero Josephson coupling, fractional 
vortices interact linearly, so they are bound much more strongly 
\cite{frac,*smiseth}. It can be seen that, for non-zero Josephson 
coupling, the phase difference sector \Eqref{SineGordonPart} or the 
second line in \Eqref{London} is a sine-Gordon model. There, a given 
fractional vortex excites two Josephson strings (one per phase 
difference sector). Crossections of a string, at a large distance from a 
vortex are sine-Gordon kinks. Such a Josephson string, has an energy 
proportional to its length. Thus for non-zero Josephson coupling one 
fractional vortex has  {linearly} diverging energy (see \Appref{Kink} 
for a detailed derivation). Note that the Josephson strings are different 
topological excitations than the domain walls previously discussed. 
Having linearly diverging energy, fractional vortices interact {linearly}. 
As a result an (composite) integer flux vortex can be seen as a strongly 
bound state of three co-centered fractional vortices. This binding is 
thus much stronger for non-zero Josephson couplings. Because of their 
diverging energies, the fractional vortices are not thermodynamically 
stable in bulk samples \cite{frac,*smiseth}: A group of three different 
fractional vortices is energetically unstable with respect to collapse into 
an integer flux composite vortex. Note however that under certain 
conditions, in a finite sample, they can be thermodynamically stable 
near boundaries \cite{silaev2} with strings terminating on a boundary.
 
Note that in a London limit, magnetic field of fractional vortices is 
exponentially localized. However in  a $[U(1)]^3$ Ginzburg-Landau 
model, the magnetic field of a fractional vortices is in a general
localized only according to  a power law and moreover can 
invert direction \cite{juha}.

\begin{figure*}[!htb]
 \hbox to \linewidth{ \hss
 \includegraphics[width=0.75\linewidth]{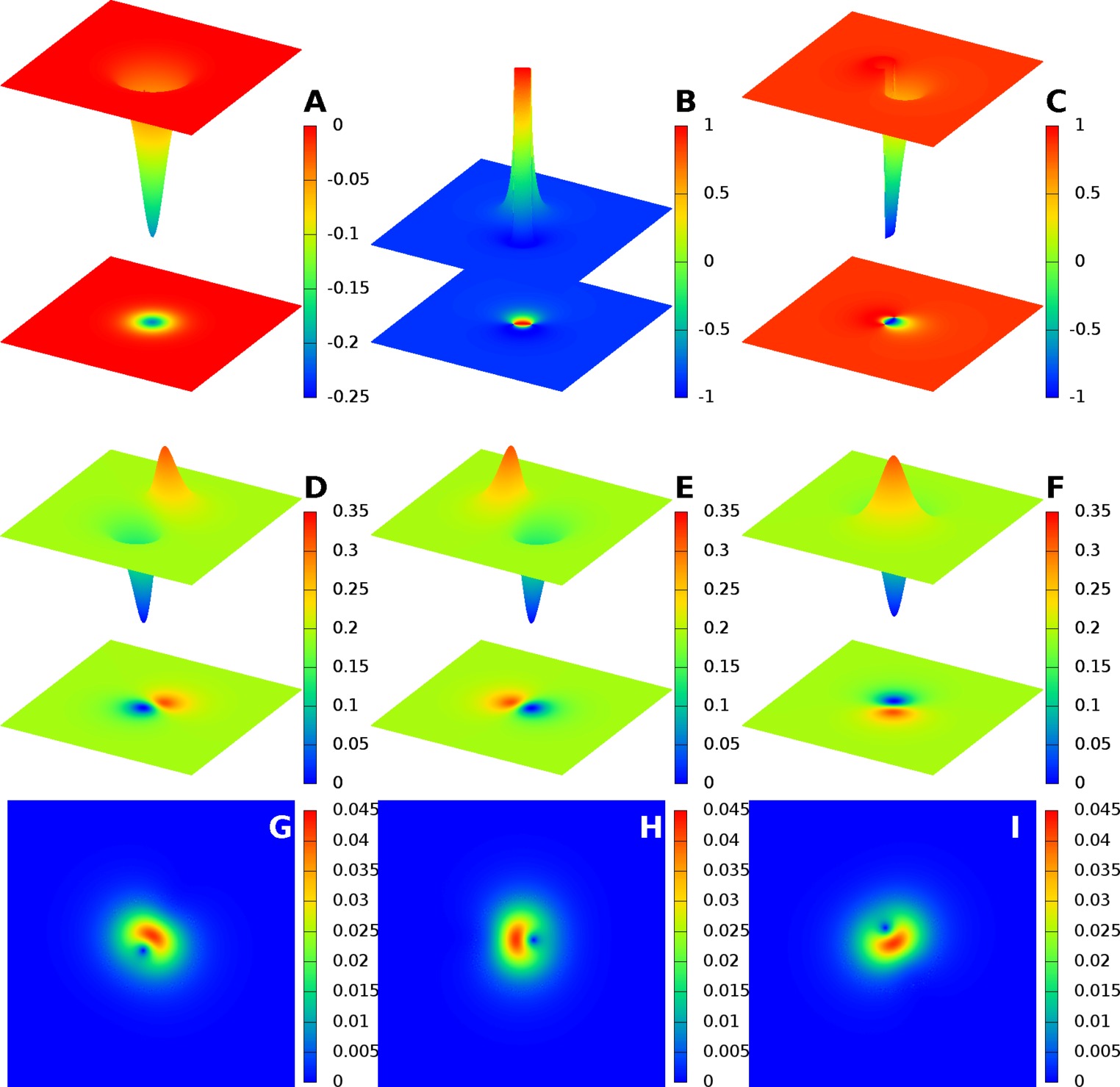}
 \hss}
\caption{
(Color online) -- 
A single charge chiral skyrmion, for 3 mirror passive bands 
$(\alpha_a,\beta_a)=(1,1)$ and Josephson coupling constants 
$\eta_{ab}=-3$. Here $\gamma_{ab}=0.8$ and the gauge coupling 
constant is $e=0.6$.
Displayed quantities are the magnetic flux $(\bf{A})$ and the 
sine of phase differences $\sin(\varphi_{12})$  $(\bf{B})$,  
$\sin(\varphi_{13})$  $(\bf{C})$. Condensate densities 
$|\psi^2_1|$, $(\bf{D})$,  $|\psi^2_2|$, $(\bf{E})$ and 
$|\psi^2_3|$, $(\bf{F})$ are represented on the second line. 
The corresponding supercurrent densities $|J_1|$, $(\bf{G})$, 
$|J_2|$, $(\bf{H})$ and $|J_3|$, $(\bf{I})$ are displayed on the 
third line. 
To avoid redundant informations, the total energy density is 
not displayed. It   qualitatively  follows  the 
magnetic flux shown in panel $(\bf{A})$. 
}
\label{Fig:Single1}
\end{figure*}

\subsection{\texorpdfstring{
Chiral three component Ginzburg--Landau skyrmions
}{
Chiral three component Ginzburg--Landau skyrmions
}} 
\label{Skyrmions}

Domain-walls such as those discussed in \Partref{TRSB} can form 
dynamically in physical systems by a quench. Because of its line 
tension, a closed domain wall collapses to zero size. From the term 
\Eqref{SineGordonPart}, in the rewritten Ginzburg--Landau 
functional, it is clear that in order to decrease the energy cost 
associated with a gradient in the relative phase $\varphi_{ab}$, the 
densities of the components $|\psi_a|$, $|\psi_b|$ should be 
suppressed on the domain wall. Furthermore, on a domain wall, 
the cosines of phase differences $\cos(\varphi_b-\varphi_a)$ are 
energetically unfavorable. Indeed, by definition, it is where they 
are the farthest from their ground state values. As a result, if an 
integer composite vortex is placed on the domain wall, the 
Josephson terms should tend to split it into fractional flux vortices, 
allowing it to attain more favorable phase difference values in 
between the split fractional vortices. As a consequence of these 
circumstances, the domain wall can trap vortices. Recall that 
away from domain walls, fractional vortices are linearly confined 
by Josephson terms. 

\begin{figure*}[!htb]
 \hbox to \linewidth{ \hss
 \includegraphics[width=0.75\linewidth]{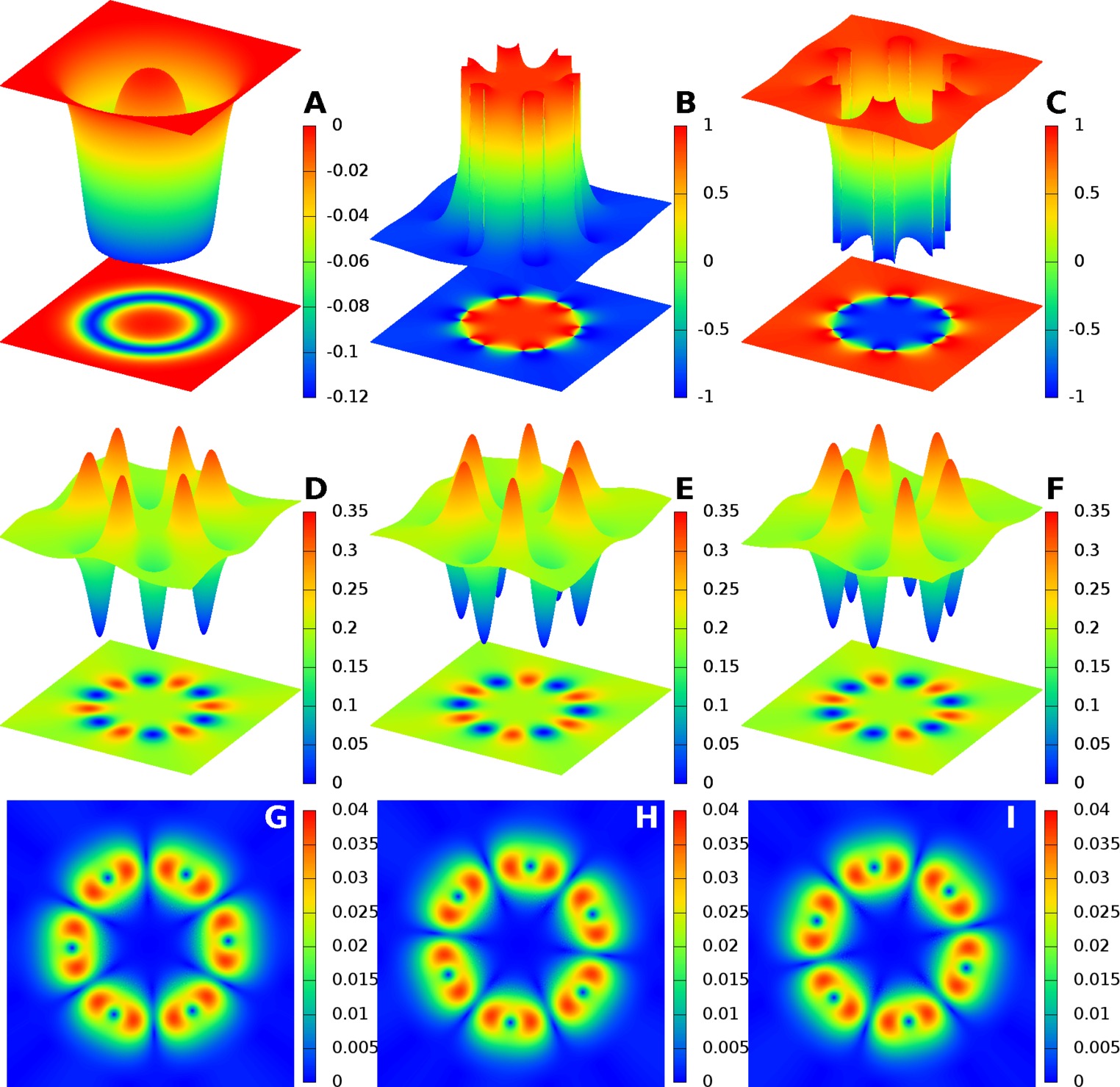}
 \hss}
\caption{
(Color online) -- 
A Skyrmion with  $\Q=6$ topological charge (which implies that 
it carries six flux quanta and consists of 18 fractional vortices). 
Displayed quantities are the magnetic flux $(\bf{A})$ and the sine 
of phase differences $\sin(\varphi_{12})$ $(\bf{B})$ 
$\sin(\varphi_{13})$  $(\bf{C})$. Condensate densities $|\psi^2_1|$, 
$(\bf{D})$, $|\psi^2_2|$, $(\bf{E})$ and $|\psi^2_3|$, $(\bf{F})$ 
are represented on the second line. The corresponding 
supercurrent densities $|J_1|$, $(\bf{G})$, $|J_2|$, $(\bf{H})$ and 
$|J_3|$, $(\bf{I})$ are displayed on the third line. Parameters are 
the same as in \Figref{Fig:Single1}. 
}
\label{Fig:Q=6}
\end{figure*}

When the magnetic field penetration length is sufficiently large 
($e$ small enough), the  repulsion between the fractional vortices 
confined on the domain wall can become strong enough to overcome 
the domain wall's tension.
It thus results in a formation of a topological soliton made up of $3N$ 
fractional vortices, stabilized by competing forces. Such `composite' 
topological solitons are thus made of a closed domain wall along 
which there are $N$ singularities in each condensate $|\psi_a|$. 
Around each singularity the phase $\varphi_a$ changes 
by $2\pi$. The total phase winding around the soliton is then 
$\oint \nabla\varphi_1 \dd\bs\ell= \oint \nabla\varphi_2 \dd\bs\ell 
=\oint\nabla \varphi_3 \dd\bs\ell=2\pi N$. Therefore it carries $N$ 
flux quanta. 
The \CPtwo topological invariant \Eqref{Topological_Charge} computed 
for such objects is found to be integer, whereas it is zero for ordinary 
composite vortices. As a result, the composite configuration made 
out of a domain wall between two $\Ztwo$ domains stabilized by 
repulsion between trapped vortices, is in fact a distinct topological 
defect: Chiral $GL^{(3)}$ skyrmion (chiral skyrmion for short). 

It was previously demonstrated that these topological defects exist 
and are indeed at least metastable \cite{Garaud.Carlstrom.ea:11}. Here we further 
investigate these objects. To investigate the existence and stability of 
the so-called chiral skyrmions, we use an energy minimization approach, 
using non-linear conjugate gradient algorithm. More details about the 
employed numerical schemes are provided in \Appref{Numerics}. 
The topological charge \Eqref{Topological_Charge} was computed 
numerically for all  configurations and was found to be integer within 
small numerical errors, less than $0.1\%$, thus providing an estimate of 
the accuracy of our solutions.

\begin{figure*}[!htb]
 \hbox to \linewidth{ \hss
  \includegraphics[width=0.75\linewidth]{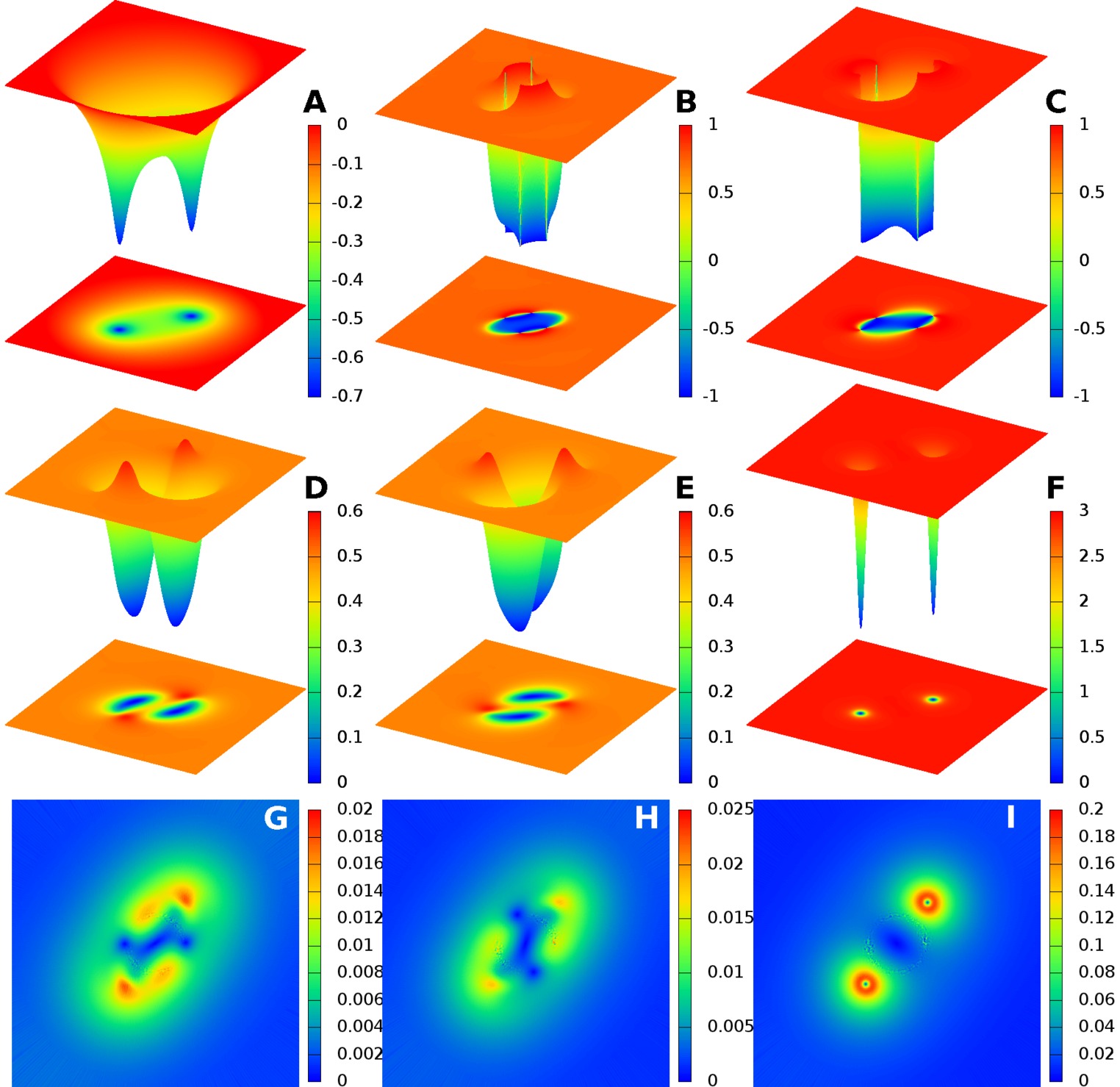}
  \hss}
\caption{
(Color online) -- A $\Q=2$ quantum soliton in a system with two 
identical passive bands $(\alpha_a,\beta_a)=(1,1)$ ($a=1,2$) coupled 
to a third active band with substantial disparity in the ground state 
densities $(\alpha_3,\beta_3)=(-2.75,1)$. Josephson coupling 
constants are $\eta_{12}=\eta_{13}=\eta_{23}=-3$. The 
system is in a strongly type-II regime $e=0.08$, the solutions here 
are stable even in the absence of bi-quadratic density interaction \ie 
$\gamma_{ab}=0$. 
Displayed quantities are the magnetic flux $(\bf{A})$ and the sine 
of phase differences $\sin(\varphi_{12})$  $(\bf{B})$ 
$\sin(\varphi_{13})$  $(\bf{C})$. Condensate densities $|\psi^2_1|$, 
$(\bf{D})$,  $|\psi^2_2|$, $(\bf{E})$ and $|\psi^2_3|$, $(\bf{F})$ are 
represented on the second line. The corresponding supercurrent 
densities $|J_1|$, $(\bf{G})$, $|J_2|$, $(\bf{H})$ and $|J_3|$, $(\bf{I})$ 
are displayed on the third line.
}
\label{Fig:Pair1}
\end{figure*}

\Figref{Fig:Single1} shows a $\Q=1$ chiral skyrmion in a 
superconductor with three passive bands (\ie the quadratic 
terms have positive prefactors $\alpha_a$). The fact that the bands 
are passive is  not important for the soliton's existence. It 
consists of three fractional vortices, each one carrying a fraction 
$|\psi_a|^2/\varrho^2$ of magnetic flux which adds up to a flux 
quantum $\Phi_0$. Since the fractional vortices are located quite 
close to each other they cannot be distinguished 
in the magnetic field profile in this case. 
Single charge skyrmions are more difficult to obtain than 
higher-charge skyrmions in this model. As will be explained later,
increasing the number of flux quanta $N$, usually makes the solution 
more stable (which contrasts with vortices where, in the type-II regime 
only $N=1$ vortices are stable). The bi-quadratic density interactions 
in the model \Eqref{freeEnergy} help to stabilize $\Q=1$ solutions. 
Single charge solitons are thus usually supported by bi-quadratic density 
interactions. Clearly, from the density plots (panels $(\bf{D}$--$\bf{F})$) 
in \Figref{Fig:Single1}, each component has a non-overlapping 
zero (the blue spots). A feature which can be observed in this regime 
is the strong density overshoot opposite to the cores (the red spots). 
 
Higher charge skyrmions are easily formed in many cases even 
when there is no bi-quadratic density interaction. There, the stability 
of the skyrmion against collapse of the domain wall is supported 
only by the electromagnetic repulsion and Josephson interactions. 
In different numerical simulations we quite easily constructed 
thousands of different skyrmionic configurations, for very different 
parameter sets. A sample of the various skyrmions is given in the 
Figures \ref{Fig:Single1}--\ref{Fig:Q=5-2}. More regimes are given 
in the appendix \Appref{Extra-graphics}. For \emph{all} such 
configurations the \CPtwo topological charge \Eqref{Topological_Charge} 
is integer with very good accuracy ( $|\Q/N-1|<10^{-3}$ ).

One key feature, in the Figures \ref{Fig:Single1}--\ref{Fig:Q=5-2}, 
is seen in the phase differences on panels $(\bf{B})$ and $(\bf{C})$. 
In each of these various regimes, the phase locking pattern 
`inside' the skyrmion is different from `outside', thus corresponding 
to either of the two $\Ztwo$ inequivalent ground states. As a result 
the  chiral skyrmions (in contrast to non-chiral) feature a 
domain wall separating the regions of different BTRS states. 
As discussed below \Partref{interaction}, the choice of one of 
the $\Ztwo$ ground states inside the skyrmion dictates a clockwise 
versus counter-clockwise arrangement of fractional vortices, thus 
motivating the terminology ``chiral" for these topological defects.

\begin{figure*}[!htb]
 \hbox to \linewidth{ \hss
  \includegraphics[width=0.75\linewidth]{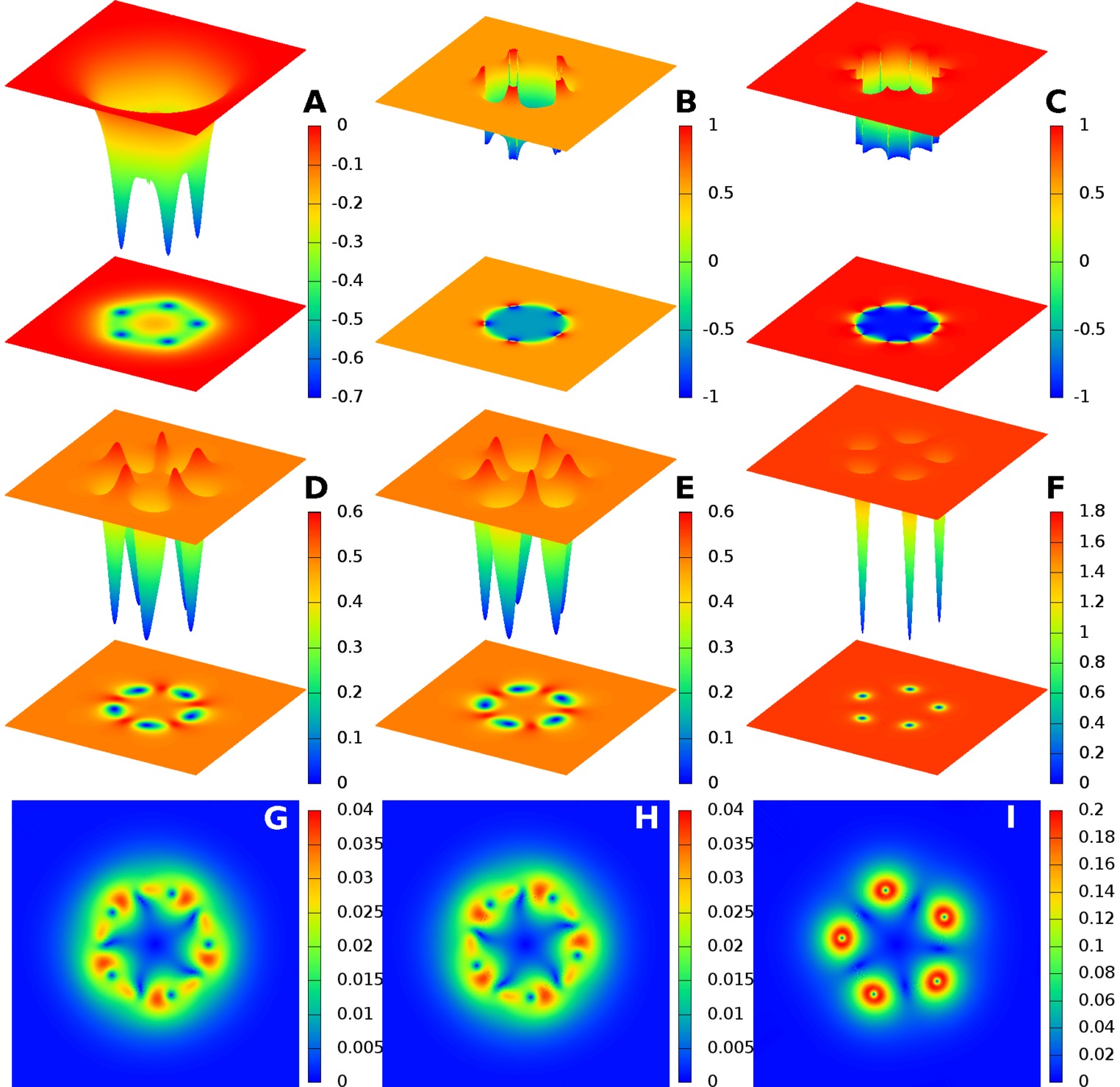}
  \hss}
\caption{
(Color online) -- A $\Q=5$ quantum soliton in a system with two 
identical passive bands as in \Figref{Fig:Pair1} coupled to a 
third active band with disparity in the ground state densities 
$(\alpha_3,\beta_3)=(-1.5,1)$. Josephson coupling constants 
are $\eta_{23}=-3$ and $\eta_{12}=\eta_{13}=1$. Here 
$e=0.2$ and there is no density-density interaction term 
$\gamma_{ab}=0$. 
The system is shaped as a pentagon  deformed  by the 
vortices of the strong active band carrying larger fractions 
of flux quantum. 
Displayed quantities are the same as in the previous pictures, 
\eg \Figref{Fig:Pair1}.
}
\label{Fig:Q=5-1}
\end{figure*}

Chiral skyrmions exhibit very unusual signatures of the 
magnetic field which can be seen from the panel $(\bf{A})$ 
in all of the Figures \ref{Fig:Single1}--\ref{Fig:Q=5-2} or 
in \Figref{Signature}. If the bands have similar density, 
each fractional vortex carries a similar fraction of flux quantum. 
As a result, the magnetic flux is almost uniformly spread along 
the domain wall, as in \Figref{Fig:Q=6}. On the other hand, when 
the condensates have quite different densities, the magnetic flux 
is carried non-uniformly by fractional vortices in different 
condensates. Consequently, the magnetic flux is inhomogeneously 
distributed along the soliton. This can be seen in \Figref{Fig:Pair1} 
where the third component carries a great fraction of the flux. 
The remaining fraction of flux is spread along the components 
having less density. The overall configuration can easily be mistaken 
for a vortex pair in such a superconductor.
For higher topological charge, the same system exhibits geometric 
structures (a pentagon as in \Figref{Fig:Q=5-1}) where the vertices 
are occupied by the fractional vortices of the band with bigger 
density. There again, geometrical arrangement of apparent vortices 
is a very typical signature of the chiral skyrmions. 

\begin{figure*}[!htb]
 \hbox to \linewidth{ \hss
  \includegraphics[width=0.75\linewidth]{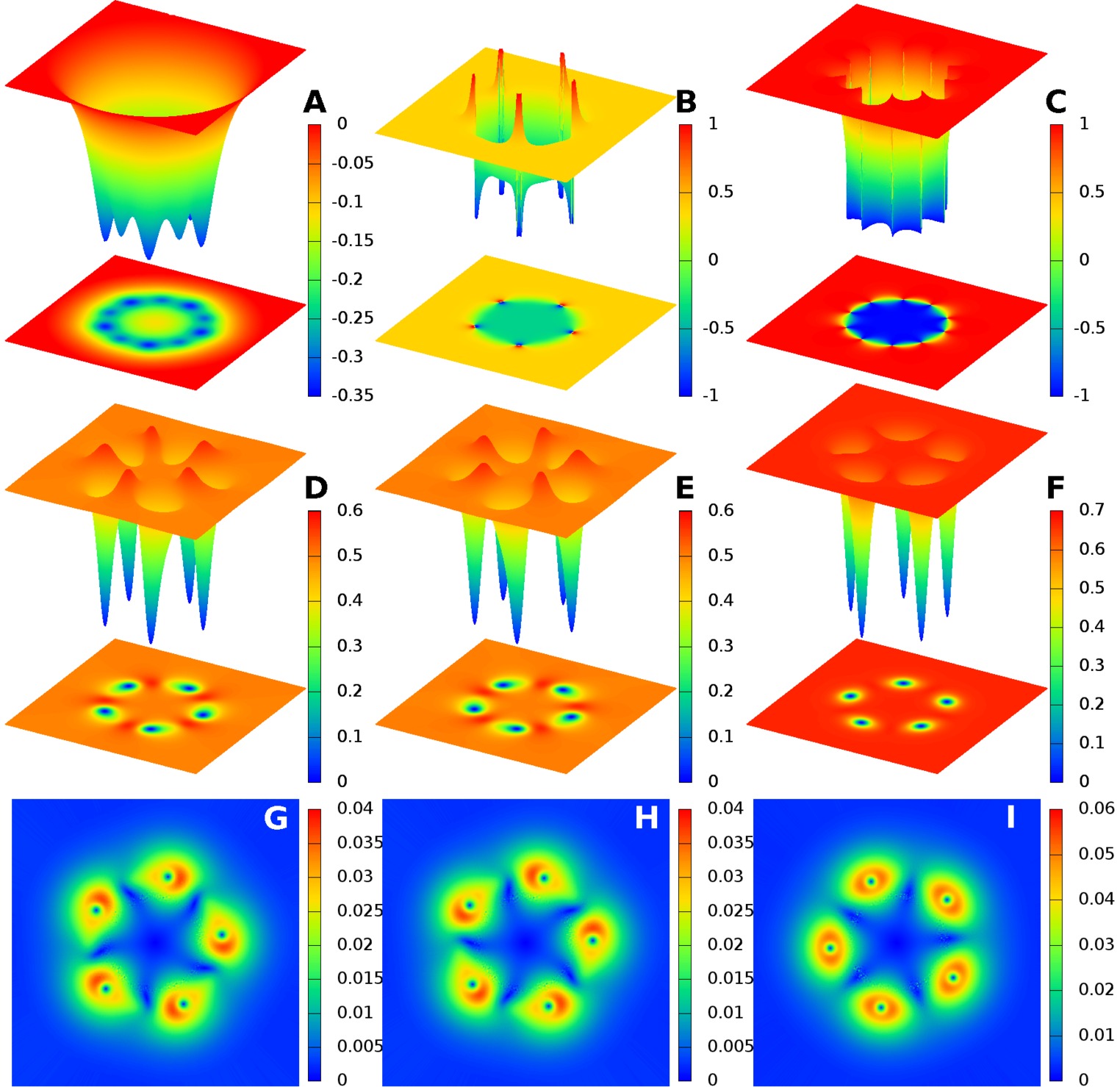}
  \hss}
\caption{
(Color online) -- A $\Q=5$ quantum soliton in a system with within 
the same parameter set as in \Figref{Fig:Q=5-1} apart from 
$(\alpha_3,\beta_3)=(-0.5,1)$.
Displayed quantities are the same as in the previous pictures, 
\eg \Figref{Fig:Pair1}.
}
\label{Fig:Q=5-2}
\end{figure*}

Among possible observable signatures of chiral skyrmions, is the 
varying fraction of magnetic flux carried by fractional vortices, 
as in \Figref{Fig:Q=5-2}. There, the magnetic field exhibits spots 
of different magnitude, larger spots associated to the two similar 
bands with more density while the small spots are associated with 
the active band. 

%%%%%%%%%%%%%%%%%%%%%%%%%%%%%%%%%%%%%%%%%%
\subsection{Chiral multi-skyrmions}

\begin{figure*}[!htb]
 \hbox to \linewidth{ \hss
\includegraphics[width=0.75\linewidth]{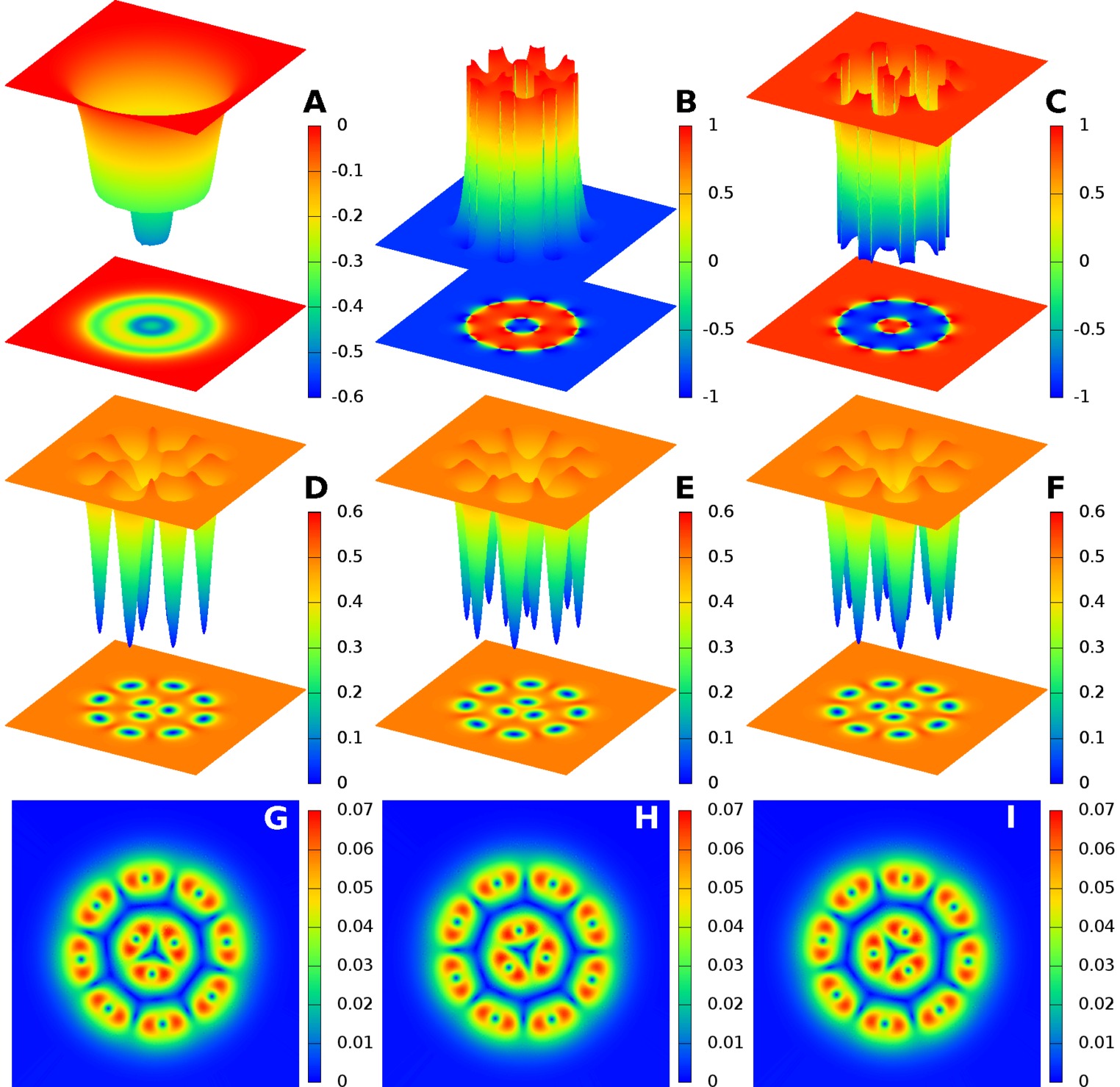}
  \hss}
\caption{
(Color online) -- A $\Q=11$ quantum \emph{multi}-soliton in a system 
with three identical passive bands as in \Figref{Fig:Mixed}. The current 
soliton is not made out of one but two stabilized domain walls thus 
being a homogeneous \emph{bi}-ring configuration. Panels $(\bf B)$ 
and  $(\bf C)$ clearly display the alternating  different ground-states. 
Since the three bands are identical, the magnetic field rather 
homogeneously spreads all along the solitons. Displayed quantities are 
the same as in the previous pictures, \eg \Figref{Fig:Pair1}. 
Note that while going counterclockwise along the outter ring, 
the fractional vortices have order band-"1,2,3". For the inner ring they 
are ordered as band-"1,3,2". The origin of this is discussed in \Partref{interaction}
}
\label{Fig:Bi-ring}
\end{figure*}

Besides having non trivial \CPtwo topological invariant 
\Eqref{Topological_Charge}, the chiral skyrmions in three component 
Ginzburg--Landau theory with BTRS have a given \emph{chirality}. 
Namely, there is a difference whether one or the other broken 
$\Ztwo$ state  is `inside'. Here we report bound states of chiral 
skyrmions with opposite chirality which can be called multi-skyrmions. 
More precisely a bound state of a skyrmion with a given chirality, 
carrying some topological charge say $\Q_1$ and a skyrmion with 
the opposite chirality carrying $\Q_2$, see \Figref{Fig:Bi-ring}. 
There the inner skyrmion has a smaller charge than the outer one, 
$\Q_1<\Q_2$ since the chiral skyrmion's size is controlled by the 
number of enclosed quanta. The bigger is the difference between 
$\Q_1$ and $\Q_2$, the weaker is the interaction between the two 
chiral skyrmions. Conversely, as $\Q_1\to\Q_2$ the chiral skyrmions 
interact progressively more strongly. For very close values of $\Q_1$ and 
$\Q_2$ the chiral skyrmions falls into each other's attractive basins 
and the domain walls annihilate. This allows decay to ordinary vortices.

Note that ``opposite chirality" should not be confused with opposite 
flux, \ie these objects have opposite chirality because they interpolate 
between two different $\Ztwo$ ground states. In that respect in the BTRS 
case, an additional $\Ztwo$ topological charge like those of ordinary 
domain walls can be attributed to skyrmions. However having opposite 
$\Ztwo$ topological charges does not mean that these objects represent a
skyrmion and an anti-skyrmion. This is because they have similar signs 
of $\Q_1$ and $\Q_2$ charges as well as similar signs of the total phase 
winding in the local $U(1)$ sector. That is, they carry magnetic flux in the 
same direction. For a given skyrmion one can construct an anti-skyrmion 
from similar number of anti-vortices. Using anti-vortices changes the overall 
phase winding and thus the direction of carried flux. As will be clear from the 
discussion below, an anti-Skyrmion with the same $\Ztwo$ charge as a Skyrmion 
will also have fractional vortices arranged in a different order.

\begin{figure*}[!htb]
 \hbox to \linewidth{ \hss
  \includegraphics[width=0.75\linewidth]{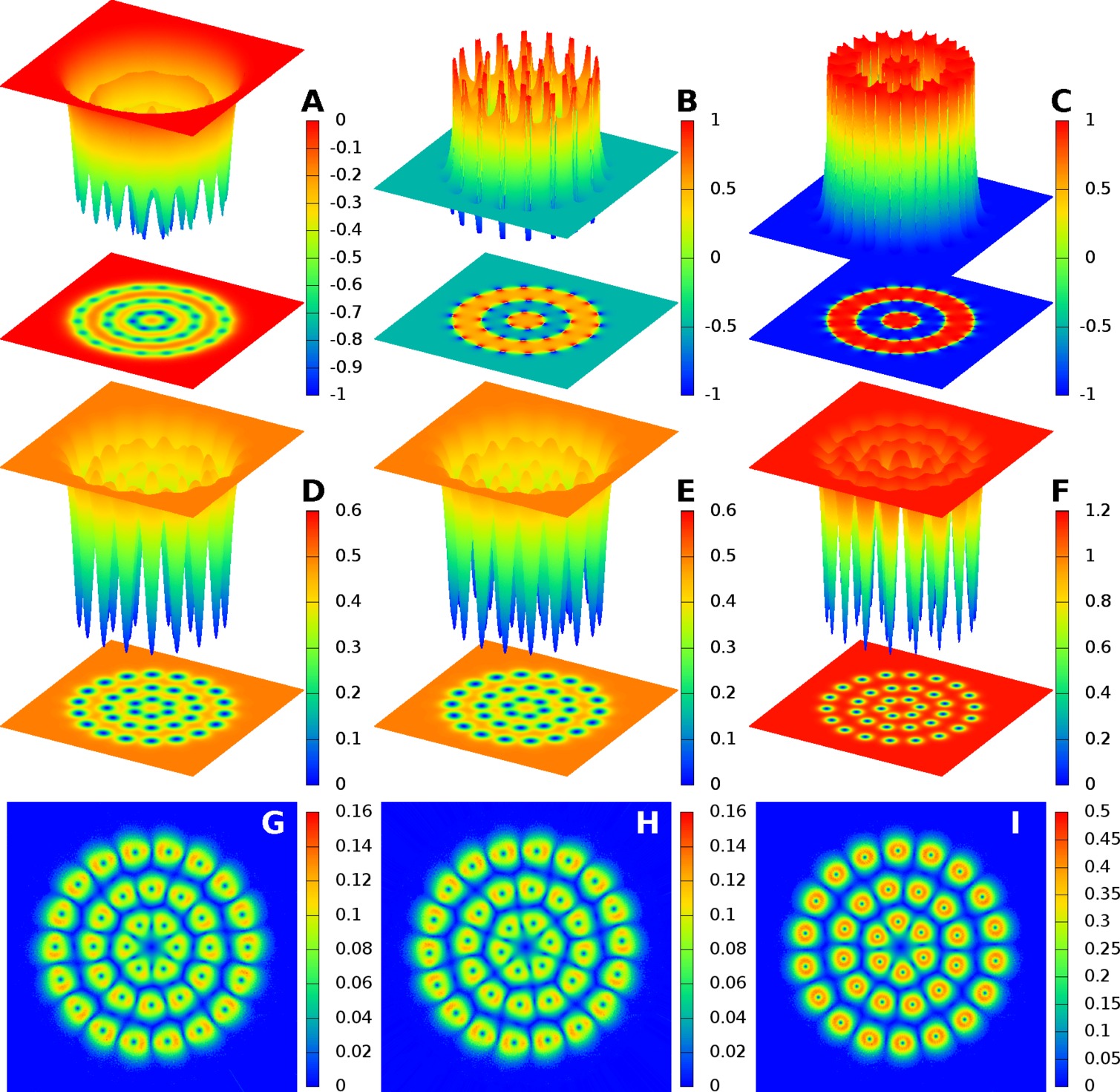}
  \hss}
\caption{
(Color online) -- A \emph{tri}-ring chiral skyrmion. The configuration 
carries total charge $\Q=36$  in a system with two identical passive 
bands $(\alpha_1,\beta_1)=(\alpha_2,\beta_2)=(1,1)$ coupled to a third 
active band  with $(\alpha_3,\beta_3)=(-1,1)$. Josephson coupling 
constants are $\eta_{23}=-3$ and $\eta_{12}=\eta_{13}=1$. $e=0.7$. 
Panels are the same as usual, \eg \Figref{Fig:Pair1}
}
\label{Fig:Triple}
\end{figure*}

Similarly, there exist also ``Russian nesting doll"-like  multi-skyrmions 
made of larger number of alternating skyrmions of opposite chiralities. 
Such a multiple skyrmion can be seen in \Figref{Fig:Triple} which shows 
\emph{tri}-ring solutions of skyrmion with alternating chiralities. 
This kind of numerical solution is quite easily obtained given a good 
initial guess. However this configuration can also spontaneously form 
from `collisional dynamics' of energy minimization of an initial configuration 
of  closely spaced ordinary vortices. This indicates that formation of 
multi-skyrmion solutions does not in general require fine tuning. Instead 
these solutions have a substantial ``attractive basin" in the GL energy 
landscape indicating they could also be observed in three component 
superconductors with Broken Time Reversal Symmetry.

%%%%%%%%%%%%%%%%%%%%%%%%%%%%%%%%%%%%%%%%%%%%%%%%%%%%%%%%%%%%%%%%%%%%%%
%%%%%%%%%%%%%%%%%%%%%%%%%%%%%%%%%%%%%%%%%%%%%%%%%%%%%%%%%%%%%%%%%%%%%%
\section{Physical properties of Chiral skyrmions}\label{Properties}

It is important to know the energetic properties of skyrmions 
compared to ordinary vortices, as well as their stability
 properties. Indeed if skyrmions are thermodynamically stable 
and form as the ground states in magnetic field, their experimental 
signatures are straightforward to detect. However, if they form as 
states with higher energy than \eg a vortex state, they are only 
metastable. When they are metastable states, skyrmions are 
protected against decay by an energy barrier. The height of this 
barrier depends non-trivially on the parameters of the potential 
and on the number of enclosed flux quanta. Metastable chiral 
skyrmions could be produced by quenching the system under 
applied magnetic field. In this section, we discuss these aspects.

%%%%%%%%%%%%%%%%%%%%%%%%%%%%%%%%%%%%%%%%%%%%%%%%%%%%%%%%%%%%%%%%%%%%%%
\subsection{Energy of Chiral skyrmions vs vortices} \label{Energy-ThStab}

For vanishing bi-quadratic density interaction couplings 
(\ie $\gamma_{ab}=0$), in all the regimes which we investigated, 
chiral skyrmions are always more expensive energetically than 
vortices. However, as suggested in 
Ref.~\onlinecite{Garaud.Carlstrom.ea:11}, bi-quadratic density 
interaction decreases the energy of chiral skyrmions relative to 
that of vortices. For sufficiently strong bi-quadratic density 
interaction chiral skyrmions are \emph{ground state} excitations 
\ie energetically cheaper than vortices and, for certain parameters, 
thermodynamically stable. 

\begin{figure*}[!htb]
\hbox to \linewidth{ \hss
   % \resizebox{150pt}{113pt}{ \input{figure10a} }
% GNUPLOT: LaTeX picture with Postscript
\begingroup
  \makeatletter
  \providecommand\color[2][]{%
    \GenericError{(gnuplot) \space\space\space\@spaces}{%
      Package color not loaded in conjunction with
      terminal option `colourtext'%
    }{See the gnuplot documentation for explanation.%
    }{Either use 'blacktext' in gnuplot or load the package
      color.sty in LaTeX.}%
    \renewcommand\color[2][]{}%
  }%
  \providecommand\includegraphics[2][]{%
    \GenericError{(gnuplot) \space\space\space\@spaces}{%
      Package graphicx or graphics not loaded%
    }{See the gnuplot documentation for explanation.%
    }{The gnuplot epslatex terminal needs graphicx.sty or graphics.sty.}%
    \renewcommand\includegraphics[2][]{}%
  }%
  \providecommand\rotatebox[2]{#2}%
  \@ifundefined{ifGPcolor}{%
    \newif\ifGPcolor
    \GPcolortrue
  }{}%
  \@ifundefined{ifGPblacktext}{%
    \newif\ifGPblacktext
    \GPblacktexttrue
  }{}%
  % define a \g@addto@macro without @ in the name:
  \let\gplgaddtomacro\g@addto@macro
  % define empty templates for all commands taking text:
  \gdef\gplbacktext{}%
  \gdef\gplfronttext{}%
  \makeatother
  \ifGPblacktext
    % no textcolor at all
    \def\colorrgb#1{}%
    \def\colorgray#1{}%
  \else
    % gray or color?
    \ifGPcolor
      \def\colorrgb#1{\color[rgb]{#1}}%
      \def\colorgray#1{\color[gray]{#1}}%
      \expandafter\def\csname LTw\endcsname{\color{white}}%
      \expandafter\def\csname LTb\endcsname{\color{black}}%
      \expandafter\def\csname LTa\endcsname{\color{black}}%
      \expandafter\def\csname LT0\endcsname{\color[rgb]{1,0,0}}%
      \expandafter\def\csname LT1\endcsname{\color[rgb]{0,1,0}}%
      \expandafter\def\csname LT2\endcsname{\color[rgb]{0,0,1}}%
      \expandafter\def\csname LT3\endcsname{\color[rgb]{1,0,1}}%
      \expandafter\def\csname LT4\endcsname{\color[rgb]{0,1,1}}%
      \expandafter\def\csname LT5\endcsname{\color[rgb]{1,1,0}}%
      \expandafter\def\csname LT6\endcsname{\color[rgb]{0,0,0}}%
      \expandafter\def\csname LT7\endcsname{\color[rgb]{1,0.3,0}}%
      \expandafter\def\csname LT8\endcsname{\color[rgb]{0.5,0.5,0.5}}%
    \else
      % gray
      \def\colorrgb#1{\color{black}}%
      \def\colorgray#1{\color[gray]{#1}}%
      \expandafter\def\csname LTw\endcsname{\color{white}}%
      \expandafter\def\csname LTb\endcsname{\color{black}}%
      \expandafter\def\csname LTa\endcsname{\color{black}}%
      \expandafter\def\csname LT0\endcsname{\color{black}}%
      \expandafter\def\csname LT1\endcsname{\color{black}}%
      \expandafter\def\csname LT2\endcsname{\color{black}}%
      \expandafter\def\csname LT3\endcsname{\color{black}}%
      \expandafter\def\csname LT4\endcsname{\color{black}}%
      \expandafter\def\csname LT5\endcsname{\color{black}}%
      \expandafter\def\csname LT6\endcsname{\color{black}}%
      \expandafter\def\csname LT7\endcsname{\color{black}}%
      \expandafter\def\csname LT8\endcsname{\color{black}}%
    \fi
  \fi
  \setlength{\unitlength}{0.0500bp}%
  \resizebox{150pt}{113pt}{ \begin{picture}(7200.00,5040.00)%
    \gplgaddtomacro\gplbacktext{%
      \csname LTb\endcsname%
      \put(1078,750){\makebox(0,0)[r]{\strut{} \huge 0.8}}%
      \put(1078,1680){\makebox(0,0)[r]{\strut{} \huge 0.9}}%
      \put(1078,2750){\makebox(0,0)[r]{\strut{} \huge 1}}%
      \put(1078,3750){\makebox(0,0)[r]{\strut{} \huge 1.1}}%
      \put(1078,4750){\makebox(0,0)[r]{\strut{} \huge 1.2}}%
      \put(1210,404){\makebox(0,0){\strut{} \huge 1}}%
      \put(2009,404){\makebox(0,0){\strut{} \huge 2}}%
      \put(2808,404){\makebox(0,0){\strut{} \huge 3}}%
      \put(3607,404){\makebox(0,0){\strut{} \huge 4}}%
      \put(4406,404){\makebox(0,0){\strut{} \huge 5}}%
      \put(5205,404){\makebox(0,0){\strut{} \huge 6}}%
      \put(6004,404){\makebox(0,0){\strut{} \huge 7}}%
      \put(6803,404){\makebox(0,0){\strut{} \huge 8}}%
      \put(-150,2739){\rotatebox{-270}{\makebox(0,0){\strut{} \huge $\frac{E(N)}{NE(N=1)}$}}}%
      \put(4006,154){\makebox(0,0){\strut{} \huge $N$}}%
      \put(5800,2500){\makebox(0,0)[r]{\strut{} \huge $0.4$}}%
      \put(5100,1900){\makebox(0,0)[r]{\strut{} \huge $0.6$}}%
      \put(5200,1000){\makebox(0,0)[r]{\strut{} \huge $\gamma_{ab}=0.8$}}%
 \put(6650,4300){\makebox(0,0)[r]{\strut{} \Huge $\bf{(A)}$}}%
  }%
    \gplgaddtomacro\gplfronttext{%
    }%
    \gplbacktext
   \put(0,0){\includegraphics{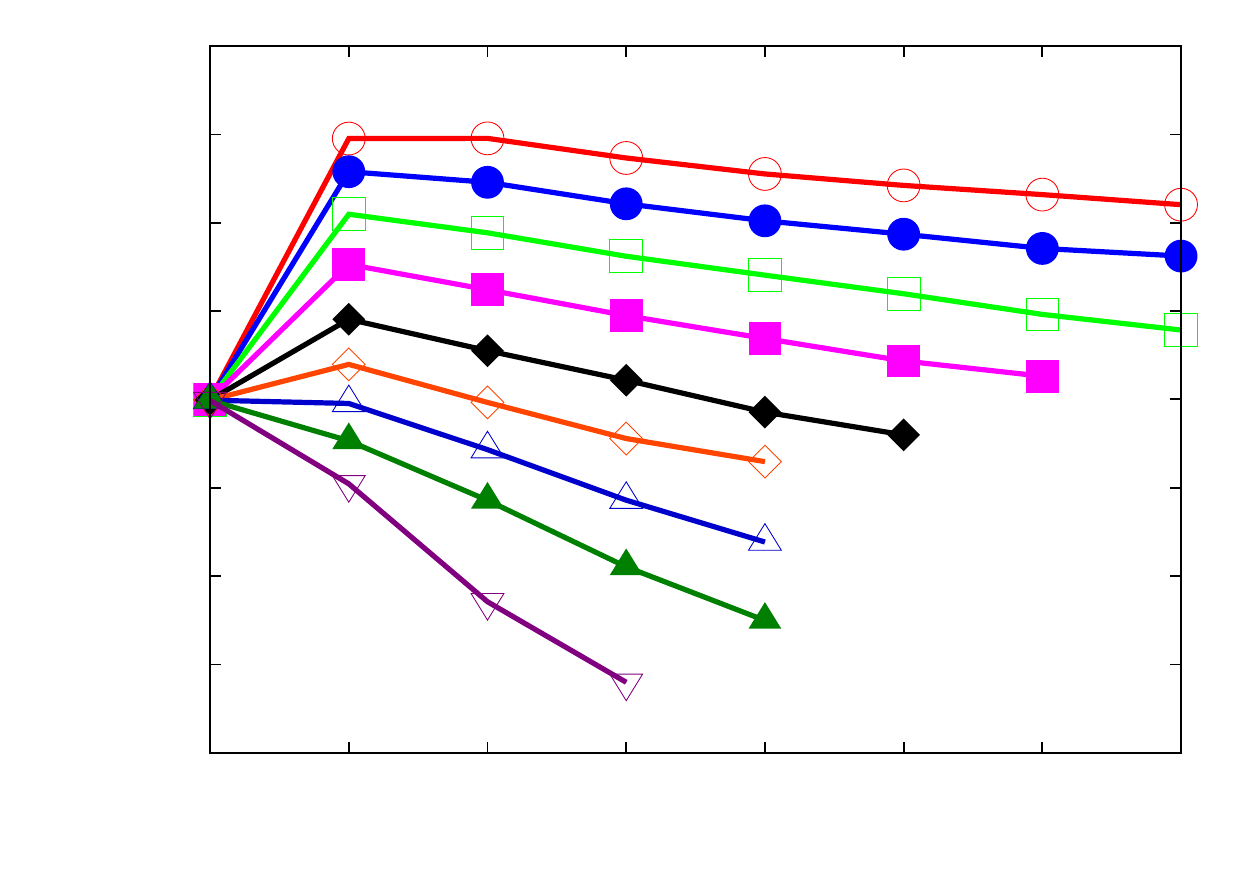}}%
     \gplfronttext
  \end{picture}%
}
\endgroup
%%%%%%%%%%%%%%%%%%%%%%%%%%%%%%%%%%%%%%%%%%%%%%%%%%%%%%%%%%%%%%%%%%%%%%
\hspace{0.15cm} 
%   \resizebox{150pt}{113pt}{ \input{figure10b} }
  % GNUPLOT: LaTeX picture with Postscript
\begingroup
  \makeatletter
  \providecommand\color[2][]{%
    \GenericError{(gnuplot) \space\space\space\@spaces}{%
      Package color not loaded in conjunction with
      terminal option `colourtext'%
    }{See the gnuplot documentation for explanation.%
    }{Either use 'blacktext' in gnuplot or load the package
      color.sty in LaTeX.}%
    \renewcommand\color[2][]{}%
  }%
  \providecommand\includegraphics[2][]{%
    \GenericError{(gnuplot) \space\space\space\@spaces}{%
      Package graphicx or graphics not loaded%
    }{See the gnuplot documentation for explanation.%
    }{The gnuplot epslatex terminal needs graphicx.sty or graphics.sty.}%
    \renewcommand\includegraphics[2][]{}%
  }%
  \providecommand\rotatebox[2]{#2}%
  \@ifundefined{ifGPcolor}{%
    \newif\ifGPcolor
    \GPcolortrue
  }{}%
  \@ifundefined{ifGPblacktext}{%
    \newif\ifGPblacktext
    \GPblacktexttrue
  }{}%
  % define a \g@addto@macro without @ in the name:
  \let\gplgaddtomacro\g@addto@macro
  % define empty templates for all commands taking text:
  \gdef\gplbacktext{}%
  \gdef\gplfronttext{}%
  \makeatother
  \ifGPblacktext
    % no textcolor at all
    \def\colorrgb#1{}%
    \def\colorgray#1{}%
  \else
    % gray or color?
    \ifGPcolor
      \def\colorrgb#1{\color[rgb]{#1}}%
      \def\colorgray#1{\color[gray]{#1}}%
      \expandafter\def\csname LTw\endcsname{\color{white}}%
      \expandafter\def\csname LTb\endcsname{\color{black}}%
      \expandafter\def\csname LTa\endcsname{\color{black}}%
      \expandafter\def\csname LT0\endcsname{\color[rgb]{1,0,0}}%
      \expandafter\def\csname LT1\endcsname{\color[rgb]{0,1,0}}%
      \expandafter\def\csname LT2\endcsname{\color[rgb]{0,0,1}}%
      \expandafter\def\csname LT3\endcsname{\color[rgb]{1,0,1}}%
      \expandafter\def\csname LT4\endcsname{\color[rgb]{0,1,1}}%
      \expandafter\def\csname LT5\endcsname{\color[rgb]{1,1,0}}%
      \expandafter\def\csname LT6\endcsname{\color[rgb]{0,0,0}}%
      \expandafter\def\csname LT7\endcsname{\color[rgb]{1,0.3,0}}%
      \expandafter\def\csname LT8\endcsname{\color[rgb]{0.5,0.5,0.5}}%
    \else
      % gray
      \def\colorrgb#1{\color{black}}%
      \def\colorgray#1{\color[gray]{#1}}%
      \expandafter\def\csname LTw\endcsname{\color{white}}%
      \expandafter\def\csname LTb\endcsname{\color{black}}%
      \expandafter\def\csname LTa\endcsname{\color{black}}%
      \expandafter\def\csname LT0\endcsname{\color{black}}%
      \expandafter\def\csname LT1\endcsname{\color{black}}%
      \expandafter\def\csname LT2\endcsname{\color{black}}%
      \expandafter\def\csname LT3\endcsname{\color{black}}%
      \expandafter\def\csname LT4\endcsname{\color{black}}%
      \expandafter\def\csname LT5\endcsname{\color{black}}%
      \expandafter\def\csname LT6\endcsname{\color{black}}%
      \expandafter\def\csname LT7\endcsname{\color{black}}%
      \expandafter\def\csname LT8\endcsname{\color{black}}%
    \fi
  \fi
  \setlength{\unitlength}{0.0500bp}%
   \resizebox{150pt}{113pt}{\begin{picture}(7200.00,5040.00)%
    \gplgaddtomacro\gplbacktext{%
      \csname LTb\endcsname%
      \put(946,704){\makebox(0,0)[r]{\strut{} \huge  0.6}}%
      \put(946,1867){\makebox(0,0)[r]{\strut{} \huge  0.8}}%
      \put(946,3030){\makebox(0,0)[r]{\strut{} \huge  1}}%
      \put(946,4193){\makebox(0,0)[r]{\strut{} \huge  1.2}}%
      \put(1896,404){\makebox(0,0){\strut{} \huge  2}}%
      \put(2714,404){\makebox(0,0){\strut{} \huge  3}}%
      \put(3532,404){\makebox(0,0){\strut{} \huge  4}}%
      \put(4349,404){\makebox(0,0){\strut{} \huge  5}}%
      \put(5167,404){\makebox(0,0){\strut{} \huge  6}}%
      \put(5985,404){\makebox(0,0){\strut{} \huge  7}}%
      \put(6803,404){\makebox(0,0){\strut{} \huge  8}}%
      \put(-120,2739){\rotatebox{-270}{\makebox(0,0){\strut{} \huge $\frac{E(N)-E(N-1)}{E(N=1)}$}}}%
      \put(3940,154){\makebox(0,0){\strut{} \huge $N$}}%
     \put(4200,4250){\makebox(0,0)[r]{\strut{} \huge $\gamma_{ab}=0.0$}}%
     \put(6000,1200){\makebox(0,0)[r]{\strut{} \huge $\gamma_{ab}=0.7$}}%
    \put(6650,4300){\makebox(0,0)[r]{\strut{} \Huge $\bf{(B)}$}}%
   }%
    \gplgaddtomacro\gplfronttext{%
    }%
    \gplbacktext
    \put(0,0){\includegraphics{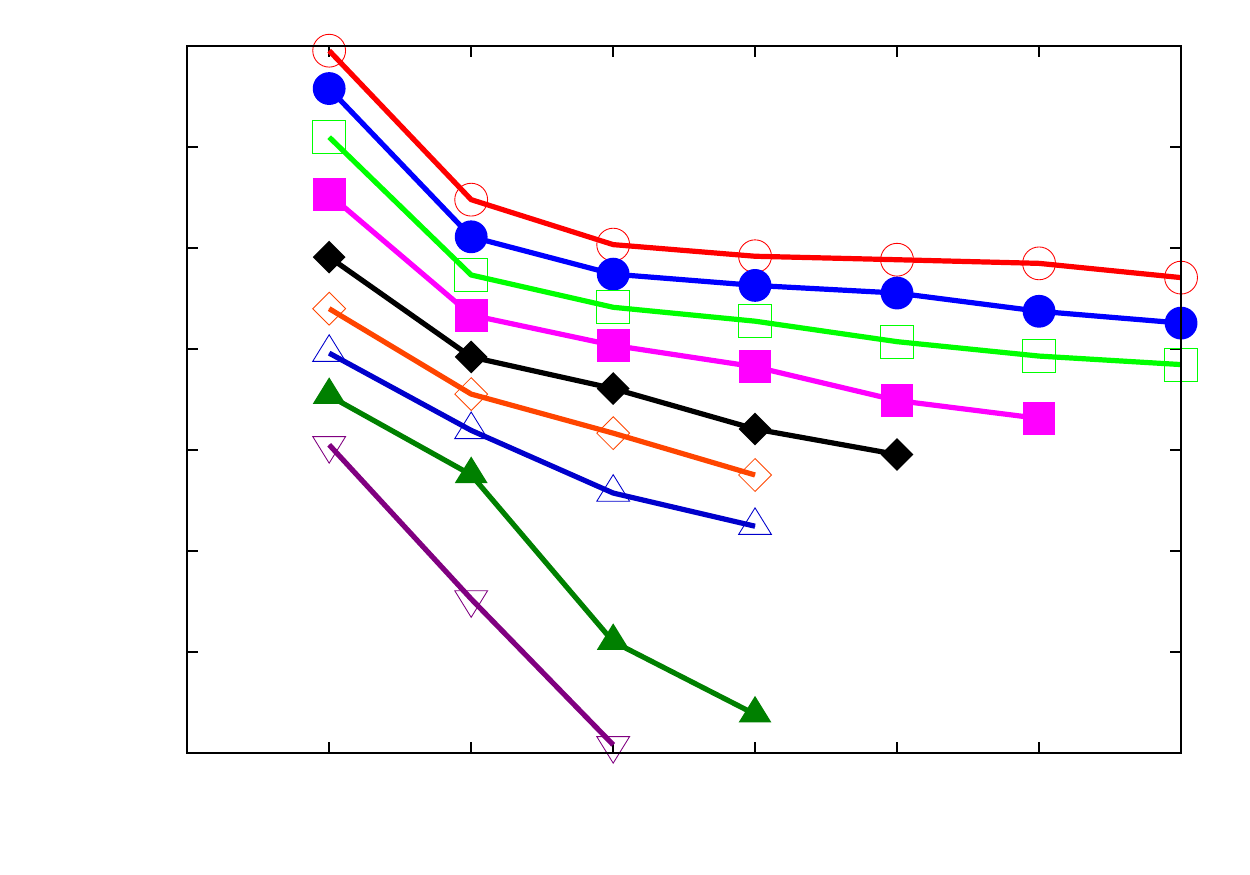}}%
    \gplfronttext
  \end{picture}%
}
\endgroup
%%%%%%%%%%%%%%%%%%%%%%%%%%%%%%%%%%%%%%%%%%%%%%%%%%%%%%%%%%%%%%%%%%%%%%
% \resizebox{150pt}{113pt}{ \input{figure10c}   }
% GNUPLOT: LaTeX picture with Postscript
\begingroup
  \makeatletter
  \providecommand\color[2][]{%
    \GenericError{(gnuplot) \space\space\space\@spaces}{%
      Package color not loaded in conjunction with
      terminal option `colourtext'%
    }{See the gnuplot documentation for explanation.%
    }{Either use 'blacktext' in gnuplot or load the package
      color.sty in LaTeX.}%
    \renewcommand\color[2][]{}%
  }%
  \providecommand\includegraphics[2][]{%
    \GenericError{(gnuplot) \space\space\space\@spaces}{%
      Package graphicx or graphics not loaded%
    }{See the gnuplot documentation for explanation.%
    }{The gnuplot epslatex terminal needs graphicx.sty or graphics.sty.}%
    \renewcommand\includegraphics[2][]{}%
  }%
  \providecommand\rotatebox[2]{#2}%
  \@ifundefined{ifGPcolor}{%
    \newif\ifGPcolor
    \GPcolortrue
  }{}%
  \@ifundefined{ifGPblacktext}{%
    \newif\ifGPblacktext
    \GPblacktexttrue
  }{}%
  % define a \g@addto@macro without @ in the name:
  \let\gplgaddtomacro\g@addto@macro
  % define empty templates for all commands taking text:
  \gdef\gplbacktext{}%
  \gdef\gplfronttext{}%
  \makeatother
  \ifGPblacktext
    % no textcolor at all
    \def\colorrgb#1{}%
    \def\colorgray#1{}%
  \else
    % gray or color?
    \ifGPcolor
      \def\colorrgb#1{\color[rgb]{#1}}%
      \def\colorgray#1{\color[gray]{#1}}%
      \expandafter\def\csname LTw\endcsname{\color{white}}%
      \expandafter\def\csname LTb\endcsname{\color{black}}%
      \expandafter\def\csname LTa\endcsname{\color{black}}%
      \expandafter\def\csname LT0\endcsname{\color[rgb]{1,0,0}}%
      \expandafter\def\csname LT1\endcsname{\color[rgb]{0,1,0}}%
      \expandafter\def\csname LT2\endcsname{\color[rgb]{0,0,1}}%
      \expandafter\def\csname LT3\endcsname{\color[rgb]{1,0,1}}%
      \expandafter\def\csname LT4\endcsname{\color[rgb]{0,1,1}}%
      \expandafter\def\csname LT5\endcsname{\color[rgb]{1,1,0}}%
      \expandafter\def\csname LT6\endcsname{\color[rgb]{0,0,0}}%
      \expandafter\def\csname LT7\endcsname{\color[rgb]{1,0.3,0}}%
      \expandafter\def\csname LT8\endcsname{\color[rgb]{0.5,0.5,0.5}}%
    \else
      % gray
      \def\colorrgb#1{\color{black}}%
      \def\colorgray#1{\color[gray]{#1}}%
      \expandafter\def\csname LTw\endcsname{\color{white}}%
      \expandafter\def\csname LTb\endcsname{\color{black}}%
      \expandafter\def\csname LTa\endcsname{\color{black}}%
      \expandafter\def\csname LT0\endcsname{\color{black}}%
      \expandafter\def\csname LT1\endcsname{\color{black}}%
      \expandafter\def\csname LT2\endcsname{\color{black}}%
      \expandafter\def\csname LT3\endcsname{\color{black}}%
      \expandafter\def\csname LT4\endcsname{\color{black}}%
      \expandafter\def\csname LT5\endcsname{\color{black}}%
      \expandafter\def\csname LT6\endcsname{\color{black}}%
      \expandafter\def\csname LT7\endcsname{\color{black}}%
      \expandafter\def\csname LT8\endcsname{\color{black}}%
    \fi
  \fi
  \setlength{\unitlength}{0.0500bp}%
 \resizebox{150pt}{113pt}{  \begin{picture}(7200.00,5040.00)%
    \gplgaddtomacro\gplbacktext{%
      \csname LTb\endcsname%
      \put(1078,704){\makebox(0,0)[r]{\strut{} \huge-0.3}}%
      \put(1078,1609){\makebox(0,0)[r]{\strut{} \huge-0.26}}%
      \put(1078,2513){\makebox(0,0)[r]{\strut{} \huge-0.22}}%
      \put(1078,3418){\makebox(0,0)[r]{\strut{} \huge-0.18}}%
      \put(1078,4323){\makebox(0,0)[r]{\strut{} \huge-0.14}}%
      \put(1210,404){\makebox(0,0){\strut{} \huge 1}}%
      \put(2009,404){\makebox(0,0){\strut{} \huge 2}}%
      \put(2808,404){\makebox(0,0){\strut{} \huge 3}}%
      \put(3607,404){\makebox(0,0){\strut{} \huge 4}}%
      \put(4406,404){\makebox(0,0){\strut{} \huge 5}}%
      \put(5205,404){\makebox(0,0){\strut{} \huge 6}}%
      \put(6004,404){\makebox(0,0){\strut{} \huge 7}}%
      \put(6803,404){\makebox(0,0){\strut{} \huge 8}}%
      \put(7200,2739){\rotatebox{-270}{\makebox(0,0){\strut{} \huge $\frac{1}{2}H_{c1}^2-|\F_{\mbox{\normalsize GS}}|$}}}%
      \put(4006,154){\makebox(0,0){\strut{} \huge $N$}}%
      \put(5050,3850){\makebox(0,0)[r]{\strut{} \huge $0.5$}}%
      \put(6700,3200){\makebox(0,0)[r]{\strut{} \huge $0.3$}}%
      \put(5000,2450){\makebox(0,0)[r]{\strut{} \huge $0.2$}}%
      \put(6500,1770){\makebox(0,0)[r]{\strut{} \huge $\gamma_{ab}=0.1$}}%
      \put(6650,4300){\makebox(0,0)[r]{\strut{} \Huge $\bf{(C)}$}}%

    }%
    \gplgaddtomacro\gplfronttext{%
    }%
    \gplbacktext
    \put(0,0){\includegraphics{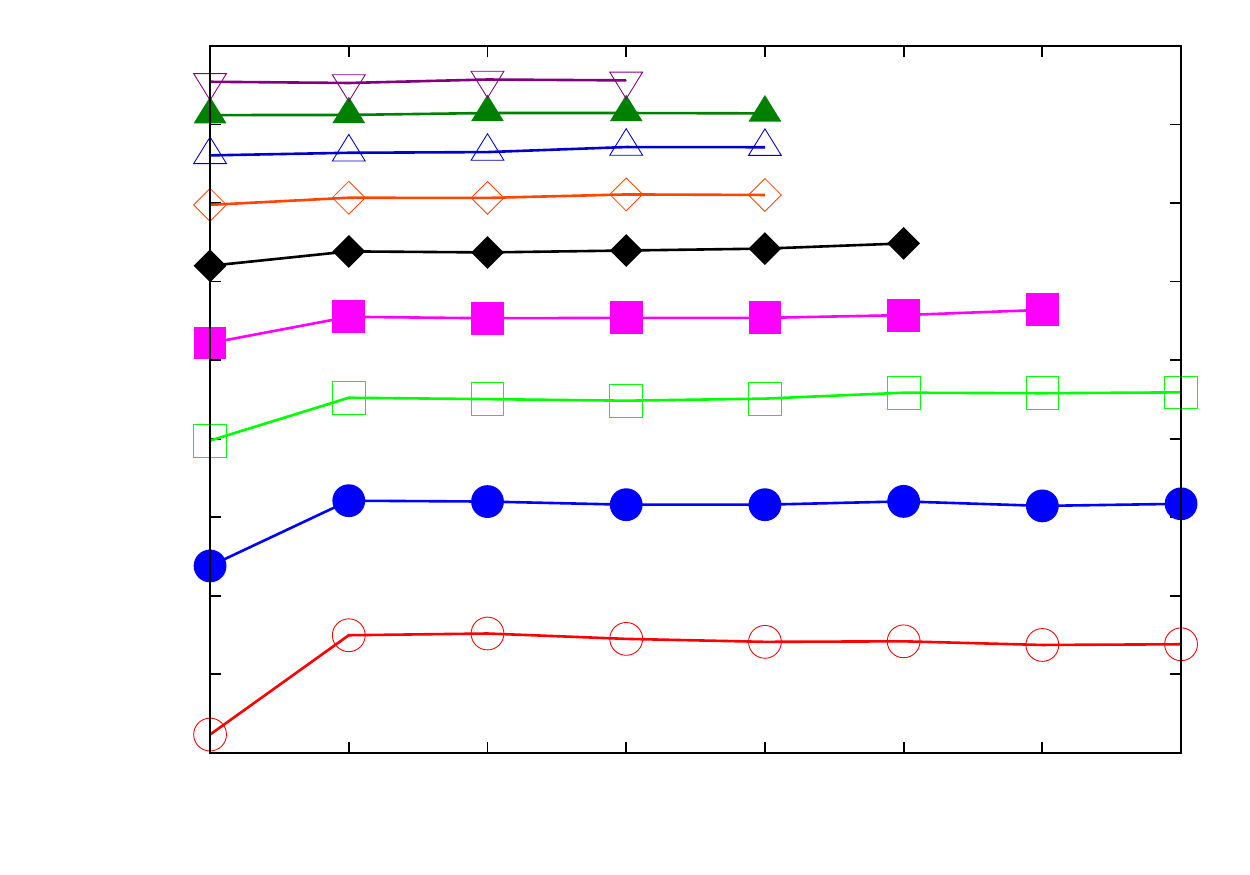}}%
    \gplfronttext
  \end{picture}%
}
\endgroup
%%%%%%%%%%%%%%%%%%%%%%%%%%%%%%%%%%%%%%%%%%%%%%%%%%%%%%%%%%%%%%%%%%%%%%
\hss}
\caption{
(Color online) -- 
Energies  per flux quantum of the skyrmions carrying $N$ flux 
quanta. The energy is given in units of the energy of the 
energetically cheapest  (either vortex or skyrmion) single 
quantum excitation $(\bf{A})$. 
Middle panel $(\bf{B})$ shows  $\frac{E(N)-E(N-1)}{E(N=1)}$ 
as a function of the number of flux quanta. When this quantity 
is less than one, it is energetically preferred to have a 
$N$-quantum skyrmion than having a $(N-1)$-skyrmion plus 
one isolated vortex.
The criterion for thermodynamical stability 
$\frac{H_{c1}^2}{2}-|F_{\mbox{\tiny GS}}|$, where the 
condensation energy is 
$F_{\mbox{\tiny GS}}\equiv F(\langle\psi_a\rangle,0)$, is 
shown on the right panel $(\bf{C})$. 
Here, the dependence of the solutions on $\gamma$ and 
$N$ is investigated, while the gauge coupling is fixed at 
$e=0.3$. Other parameters are $(\alpha_a,\beta_a)=(1,1)$ 
and $\eta_{ab}=-3$. 
Colors and symbols associated to different values of 
$\gamma_{ab}$ (shown on the picture) are the same over 
three panels.
Note that the reason why curves with high $\gamma_{ab}$ 
terminates for smaller $N$, is that the size of the skyrmion 
becomes comparable to the size of the numerical domain. 
To avoid any finite size effect, we chose to skip the 
corresponding points.
}
\label{Fig:Sk1}
\end{figure*}

The energy properties of the chiral skyrmions are displayed on 
the left panels of Figures~\ref{Fig:Sk1}-\ref{Fig:Sk2}. There, 
the energy per flux quantum of a given configuration is given 
in units of the single quantum flux carrying ground state. Namely 
$E(N)/[NE(N=1)]$ is represented as a function of $N$, the number 
of flux quanta. The corresponding energies are sublinear functions 
of enclosed flux quanta  for all solutions with $N>2$. This means 
that the energy cost per flux quantum decreases as $N$ grows.

Two different regimes can be distinguished. If a configuration has 
$E(N)/[NE(N=1)]>1$ (where $E(N=1)$ is the energy of a single vortex), 
then it is energetically preferable to have $N$ isolated type-II integer 
flux vortices. As discussed below, there, skyrmions should be 
understood as metastable objects. That is, they can decay into 
type-II (composite) vortices, \eg in case of  strong enough perturbations. 
On the other hand, when $E(N)/[NE(N=1)]<1$, then isolated vortices 
are no longer energetically preferred over a skyrmion. In the first case, 
(corresponding to the upper curves of \Figref{Fig:Sk1}), chiral 
skyrmions can exist as meta-stable excitations. In the second 
situation (the lower curves of \Figref{Fig:Sk1}), chiral skyrmions 
could form as true ground state topological excitations.
Note also that there is a regime where lower charge skyrmions 
are more expensive than type-II integer vortices, while higher 
charge ones are cheaper (see \Figref{Fig:Sk1}). In the regimes 
where there is  density-density interaction, even the smallest 
skyrmions  with $\Q=N=1$ can  be energetically cheaper than 
vortices. 

The relative cost of including an additional flux quantum into a 
chiral skyrmion is evaluated by computing 
$\frac{E(N)-E(N-1)}{E(N=1)}$. When this quantity is less than 
one, it is globally beneficial to merge an additional 
flux-quantum-carrying object with a skyrmion. It is displayed 
on panel $(\bf{B})$ of Figures~\ref{Fig:Sk1}-\ref{Fig:Sk2}. 
Note that it does not tell about the real work the system 
has to provide for bringing the isolated single quantum defect 
from infinity into the skyrmion, but only on global cost or benefit.

\subsection{Thermodynamical stability of Chiral skyrmions} \label{Energy-ThStab2}

\begin{figure*}[!htb]
\hbox to \linewidth{ \hss
 %  \resizebox{150pt}{113pt}{ \input{figure11a} }
% GNUPLOT: LaTeX picture with Postscript
\begingroup
  \makeatletter
  \providecommand\color[2][]{%
    \GenericError{(gnuplot) \space\space\space\@spaces}{%
      Package color not loaded in conjunction with
      terminal option `colourtext'%
    }{See the gnuplot documentation for explanation.%
    }{Either use 'blacktext' in gnuplot or load the package
      color.sty in LaTeX.}%
    \renewcommand\color[2][]{}%
  }%
  \providecommand\includegraphics[2][]{%
    \GenericError{(gnuplot) \space\space\space\@spaces}{%
      Package graphicx or graphics not loaded%
    }{See the gnuplot documentation for explanation.%
    }{The gnuplot epslatex terminal needs graphicx.sty or graphics.sty.}%
    \renewcommand\includegraphics[2][]{}%
  }%
  \providecommand\rotatebox[2]{#2}%
  \@ifundefined{ifGPcolor}{%
    \newif\ifGPcolor
    \GPcolortrue
  }{}%
  \@ifundefined{ifGPblacktext}{%
    \newif\ifGPblacktext
    \GPblacktexttrue
  }{}%
  % define a \g@addto@macro without @ in the name:
  \let\gplgaddtomacro\g@addto@macro
  % define empty templates for all commands taking text:
  \gdef\gplbacktext{}%
  \gdef\gplfronttext{}%
  \makeatother
  \ifGPblacktext
    % no textcolor at all
    \def\colorrgb#1{}%
    \def\colorgray#1{}%
  \else
    % gray or color?
    \ifGPcolor
      \def\colorrgb#1{\color[rgb]{#1}}%
      \def\colorgray#1{\color[gray]{#1}}%
      \expandafter\def\csname LTw\endcsname{\color{white}}%
      \expandafter\def\csname LTb\endcsname{\color{black}}%
      \expandafter\def\csname LTa\endcsname{\color{black}}%
      \expandafter\def\csname LT0\endcsname{\color[rgb]{1,0,0}}%
      \expandafter\def\csname LT1\endcsname{\color[rgb]{0,1,0}}%
      \expandafter\def\csname LT2\endcsname{\color[rgb]{0,0,1}}%
      \expandafter\def\csname LT3\endcsname{\color[rgb]{1,0,1}}%
      \expandafter\def\csname LT4\endcsname{\color[rgb]{0,1,1}}%
      \expandafter\def\csname LT5\endcsname{\color[rgb]{1,1,0}}%
      \expandafter\def\csname LT6\endcsname{\color[rgb]{0,0,0}}%
      \expandafter\def\csname LT7\endcsname{\color[rgb]{1,0.3,0}}%
      \expandafter\def\csname LT8\endcsname{\color[rgb]{0.5,0.5,0.5}}%
    \else
      % gray
      \def\colorrgb#1{\color{black}}%
      \def\colorgray#1{\color[gray]{#1}}%
      \expandafter\def\csname LTw\endcsname{\color{white}}%
      \expandafter\def\csname LTb\endcsname{\color{black}}%
      \expandafter\def\csname LTa\endcsname{\color{black}}%
      \expandafter\def\csname LT0\endcsname{\color{black}}%
      \expandafter\def\csname LT1\endcsname{\color{black}}%
      \expandafter\def\csname LT2\endcsname{\color{black}}%
      \expandafter\def\csname LT3\endcsname{\color{black}}%
      \expandafter\def\csname LT4\endcsname{\color{black}}%
      \expandafter\def\csname LT5\endcsname{\color{black}}%
      \expandafter\def\csname LT6\endcsname{\color{black}}%
      \expandafter\def\csname LT7\endcsname{\color{black}}%
      \expandafter\def\csname LT8\endcsname{\color{black}}%
    \fi
  \fi
  \setlength{\unitlength}{0.0500bp}%
 \resizebox{150pt}{113pt}{  \begin{picture}(7200.00,5040.00)%
    \gplgaddtomacro\gplbacktext{%
      \csname LTb\endcsname%
      \put(1078,704){\makebox(0,0)[r]{\strut{} \huge 0.88}}%
      \put(1078,1383){\makebox(0,0)[r]{\strut{} \huge 0.9}}%
      \put(1078,2061){\makebox(0,0)[r]{\strut{} \huge 0.92}}%
      \put(1078,2740){\makebox(0,0)[r]{\strut{} \huge 0.94}}%
      \put(1078,3418){\makebox(0,0)[r]{\strut{} \huge 0.96}}%
      \put(1078,4097){\makebox(0,0)[r]{\strut{} \huge 0.98}}%
      \put(1078,4775){\makebox(0,0)[r]{\strut{} \huge 1}}%
      \put(1210,404){\makebox(0,0){\strut{} \huge 1}}%
      \put(2009,404){\makebox(0,0){\strut{} \huge 2}}%
      \put(2808,404){\makebox(0,0){\strut{} \huge 3}}%
      \put(3607,404){\makebox(0,0){\strut{} \huge 4}}%
      \put(4406,404){\makebox(0,0){\strut{} \huge 5}}%
      \put(5205,404){\makebox(0,0){\strut{} \huge 6}}%
      \put(6004,404){\makebox(0,0){\strut{} \huge 7}}%
      \put(6803,404){\makebox(0,0){\strut{} \huge 8}}%
      \put(-150,2739){\rotatebox{-270}{\makebox(0,0){\strut{} \huge $\frac{E(N)}{NE(N=1)}$}}}%
      \put(4006,154){\makebox(0,0){\strut{} \huge $N$}}%
      \put(2800,950){\makebox(0,0)[r]{\strut{} \huge $e=0.3$}}%
      \put(6000,2100){\makebox(0,0)[r]{\strut{} \huge $e=0.8$}}%
      \put(6650,4300){\makebox(0,0)[r]{\strut{} \Huge $\bf{(A)}$}}%
    }%
    \gplgaddtomacro\gplfronttext{%
    }%
    \gplbacktext
    \put(0,0){\includegraphics{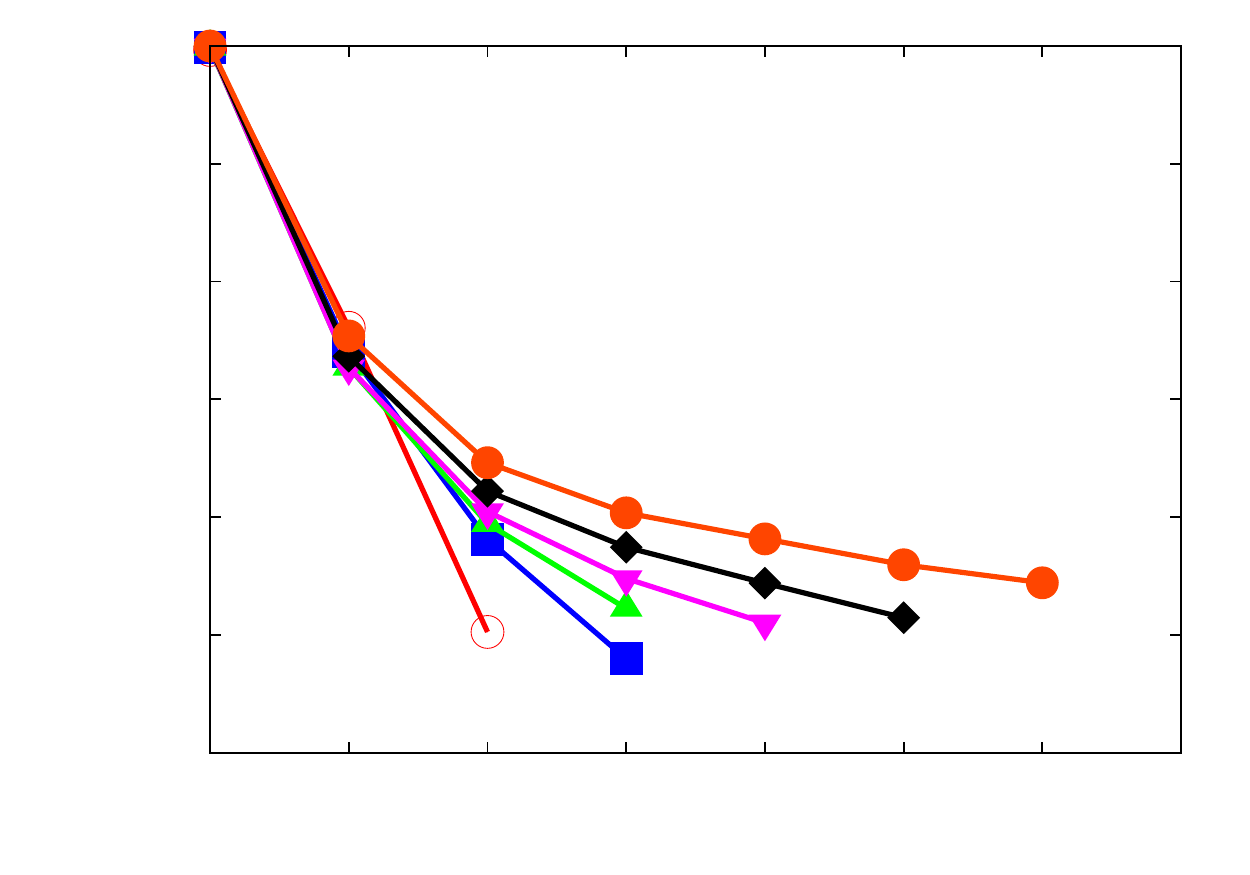}}%
    \gplfronttext
  \end{picture}%
}
\endgroup
%%%%%%%%%%%%%%%%%%%%%%%%%%%%%%%%%%%%%%%%%%%%%%%%%%%%%%%%%%%%%%%%%%%%%%
\hspace{0.15cm} 
  % \resizebox{150pt}{113pt}{ \input{figure11b} }
% GNUPLOT: LaTeX picture with Postscript
\begingroup
  \makeatletter
  \providecommand\color[2][]{%
    \GenericError{(gnuplot) \space\space\space\@spaces}{%
      Package color not loaded in conjunction with
      terminal option `colourtext'%
    }{See the gnuplot documentation for explanation.%
    }{Either use 'blacktext' in gnuplot or load the package
      color.sty in LaTeX.}%
    \renewcommand\color[2][]{}%
  }%
  \providecommand\includegraphics[2][]{%
    \GenericError{(gnuplot) \space\space\space\@spaces}{%
      Package graphicx or graphics not loaded%
    }{See the gnuplot documentation for explanation.%
    }{The gnuplot epslatex terminal needs graphicx.sty or graphics.sty.}%
    \renewcommand\includegraphics[2][]{}%
  }%
  \providecommand\rotatebox[2]{#2}%
  \@ifundefined{ifGPcolor}{%
    \newif\ifGPcolor
    \GPcolortrue
  }{}%
  \@ifundefined{ifGPblacktext}{%
    \newif\ifGPblacktext
    \GPblacktexttrue
  }{}%
  % define a \g@addto@macro without @ in the name:
  \let\gplgaddtomacro\g@addto@macro
  % define empty templates for all commands taking text:
  \gdef\gplbacktext{}%
  \gdef\gplfronttext{}%
  \makeatother
  \ifGPblacktext
    % no textcolor at all
    \def\colorrgb#1{}%
    \def\colorgray#1{}%
  \else
    % gray or color?
    \ifGPcolor
      \def\colorrgb#1{\color[rgb]{#1}}%
      \def\colorgray#1{\color[gray]{#1}}%
      \expandafter\def\csname LTw\endcsname{\color{white}}%
      \expandafter\def\csname LTb\endcsname{\color{black}}%
      \expandafter\def\csname LTa\endcsname{\color{black}}%
      \expandafter\def\csname LT0\endcsname{\color[rgb]{1,0,0}}%
      \expandafter\def\csname LT1\endcsname{\color[rgb]{0,1,0}}%
      \expandafter\def\csname LT2\endcsname{\color[rgb]{0,0,1}}%
      \expandafter\def\csname LT3\endcsname{\color[rgb]{1,0,1}}%
      \expandafter\def\csname LT4\endcsname{\color[rgb]{0,1,1}}%
      \expandafter\def\csname LT5\endcsname{\color[rgb]{1,1,0}}%
      \expandafter\def\csname LT6\endcsname{\color[rgb]{0,0,0}}%
      \expandafter\def\csname LT7\endcsname{\color[rgb]{1,0.3,0}}%
      \expandafter\def\csname LT8\endcsname{\color[rgb]{0.5,0.5,0.5}}%
    \else
      % gray
      \def\colorrgb#1{\color{black}}%
      \def\colorgray#1{\color[gray]{#1}}%
      \expandafter\def\csname LTw\endcsname{\color{white}}%
      \expandafter\def\csname LTb\endcsname{\color{black}}%
      \expandafter\def\csname LTa\endcsname{\color{black}}%
      \expandafter\def\csname LT0\endcsname{\color{black}}%
      \expandafter\def\csname LT1\endcsname{\color{black}}%
      \expandafter\def\csname LT2\endcsname{\color{black}}%
      \expandafter\def\csname LT3\endcsname{\color{black}}%
      \expandafter\def\csname LT4\endcsname{\color{black}}%
      \expandafter\def\csname LT5\endcsname{\color{black}}%
      \expandafter\def\csname LT6\endcsname{\color{black}}%
      \expandafter\def\csname LT7\endcsname{\color{black}}%
      \expandafter\def\csname LT8\endcsname{\color{black}}%
    \fi
  \fi
  \setlength{\unitlength}{0.0500bp}%
  \resizebox{150pt}{113pt}{\begin{picture}(7200.00,5040.00)%
    \gplgaddtomacro\gplbacktext{%
      \csname LTb\endcsname%
      \put(1078,704){\makebox(0,0)[r]{\strut{} \huge  0.76}}%
      \put(1078,1722){\makebox(0,0)[r]{\strut{} \huge  0.8}}%
      \put(1078,2740){\makebox(0,0)[r]{\strut{} \huge  0.84}}%
      \put(1078,3757){\makebox(0,0)[r]{\strut{} \huge  0.88}}%
      \put(1078,4775){\makebox(0,0)[r]{\strut{} \huge  0.92}}%
      \put(2009,404){\makebox(0,0){\strut{} \huge  2}}%
      \put(2808,404){\makebox(0,0){\strut{} \huge  3}}%
      \put(3607,404){\makebox(0,0){\strut{} \huge  4}}%
      \put(4406,404){\makebox(0,0){\strut{} \huge  5}}%
      \put(5205,404){\makebox(0,0){\strut{} \huge  6}}%
      \put(6004,404){\makebox(0,0){\strut{} \huge  7}}%
      \put(6803,404){\makebox(0,0){\strut{} \huge  8}}%
      \put(-120,2739){\rotatebox{-270}{\makebox(0,0){\strut{} \huge $\frac{E(N)-E(N-1)}{E(N=1)}$}}}%
      \put(4006,154){\makebox(0,0){\strut{}\huge $N$}}%
      \put(4600,1400){\makebox(0,0)[r]{\strut{} \huge $e=0.4$}}%
      \put(6650,4300){\makebox(0,0)[r]{\strut{} \Huge $\bf{(B)}$}}%
   }%
    \gplgaddtomacro\gplfronttext{%
    }%
    \gplbacktext
    \put(0,0){\includegraphics{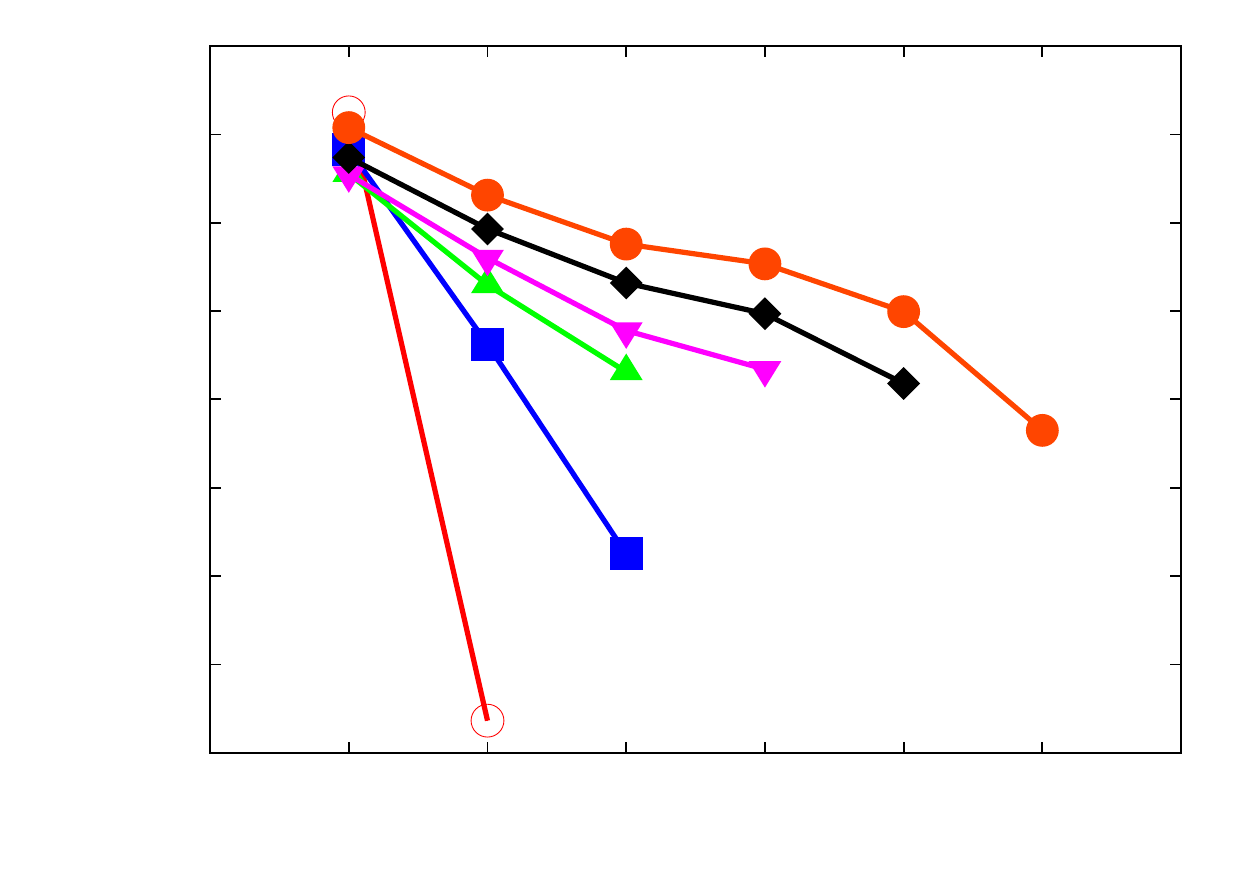}}%
    \gplfronttext
  \end{picture}%
}
\endgroup
 %%%%%%%%%%%%%%%%%%%%%%%%%%%%%%%%%%%%%%%%%%%%%%%%%%%%%%%%%%%%%%%%%%%%%%
 %\resizebox{150pt}{113pt}{ \input{figure11c}   }
% GNUPLOT: LaTeX picture with Postscript
\begingroup
  \makeatletter
  \providecommand\color[2][]{%
    \GenericError{(gnuplot) \space\space\space\@spaces}{%
      Package color not loaded in conjunction with
      terminal option `colourtext'%
    }{See the gnuplot documentation for explanation.%
    }{Either use 'blacktext' in gnuplot or load the package
      color.sty in LaTeX.}%
    \renewcommand\color[2][]{}%
  }%
  \providecommand\includegraphics[2][]{%
    \GenericError{(gnuplot) \space\space\space\@spaces}{%
      Package graphicx or graphics not loaded%
    }{See the gnuplot documentation for explanation.%
    }{The gnuplot epslatex terminal needs graphicx.sty or graphics.sty.}%
    \renewcommand\includegraphics[2][]{}%
  }%
  \providecommand\rotatebox[2]{#2}%
  \@ifundefined{ifGPcolor}{%
    \newif\ifGPcolor
    \GPcolortrue
  }{}%
  \@ifundefined{ifGPblacktext}{%
    \newif\ifGPblacktext
    \GPblacktexttrue
  }{}%
  % define a \g@addto@macro without @ in the name:
  \let\gplgaddtomacro\g@addto@macro
  % define empty templates for all commands taking text:
  \gdef\gplbacktext{}%
  \gdef\gplfronttext{}%
  \makeatother
  \ifGPblacktext
    % no textcolor at all
    \def\colorrgb#1{}%
    \def\colorgray#1{}%
  \else
    % gray or color?
    \ifGPcolor
      \def\colorrgb#1{\color[rgb]{#1}}%
      \def\colorgray#1{\color[gray]{#1}}%
      \expandafter\def\csname LTw\endcsname{\color{white}}%
      \expandafter\def\csname LTb\endcsname{\color{black}}%
      \expandafter\def\csname LTa\endcsname{\color{black}}%
      \expandafter\def\csname LT0\endcsname{\color[rgb]{1,0,0}}%
      \expandafter\def\csname LT1\endcsname{\color[rgb]{0,1,0}}%
      \expandafter\def\csname LT2\endcsname{\color[rgb]{0,0,1}}%
      \expandafter\def\csname LT3\endcsname{\color[rgb]{1,0,1}}%
      \expandafter\def\csname LT4\endcsname{\color[rgb]{0,1,1}}%
      \expandafter\def\csname LT5\endcsname{\color[rgb]{1,1,0}}%
      \expandafter\def\csname LT6\endcsname{\color[rgb]{0,0,0}}%
      \expandafter\def\csname LT7\endcsname{\color[rgb]{1,0.3,0}}%
      \expandafter\def\csname LT8\endcsname{\color[rgb]{0.5,0.5,0.5}}%
    \else
      % gray
      \def\colorrgb#1{\color{black}}%
      \def\colorgray#1{\color[gray]{#1}}%
      \expandafter\def\csname LTw\endcsname{\color{white}}%
      \expandafter\def\csname LTb\endcsname{\color{black}}%
      \expandafter\def\csname LTa\endcsname{\color{black}}%
      \expandafter\def\csname LT0\endcsname{\color{black}}%
      \expandafter\def\csname LT1\endcsname{\color{black}}%
      \expandafter\def\csname LT2\endcsname{\color{black}}%
      \expandafter\def\csname LT3\endcsname{\color{black}}%
      \expandafter\def\csname LT4\endcsname{\color{black}}%
      \expandafter\def\csname LT5\endcsname{\color{black}}%
      \expandafter\def\csname LT6\endcsname{\color{black}}%
      \expandafter\def\csname LT7\endcsname{\color{black}}%
      \expandafter\def\csname LT8\endcsname{\color{black}}%
    \fi
  \fi
  \setlength{\unitlength}{0.0500bp}%
 \resizebox{150pt}{113pt}{ \begin{picture}(7200.00,5040.00)%
    \gplgaddtomacro\gplbacktext{%
      \csname LTb\endcsname%
      \put(1210,704){\makebox(0,0)[r]{\strut{} \huge-0.13}}%
      \put(1210,1722){\makebox(0,0)[r]{\strut{} \huge-0.12}}%
      \put(1210,2740){\makebox(0,0)[r]{\strut{} \huge-0.11}}%
      \put(1210,3757){\makebox(0,0)[r]{\strut{} \huge-0.1}}%
      \put(1210,4775){\makebox(0,0)[r]{\strut{} \huge-0.09}}%
      \put(1342,404){\makebox(0,0){\strut{} \huge 1}}%
      \put(2122,404){\makebox(0,0){\strut{} \huge 2}}%
      \put(2902,404){\makebox(0,0){\strut{} \huge 3}}%
      \put(3682,404){\makebox(0,0){\strut{} \huge 4}}%
      \put(4463,404){\makebox(0,0){\strut{} \huge 5}}%
      \put(5243,404){\makebox(0,0){\strut{} \huge 6}}%
      \put(6023,404){\makebox(0,0){\strut{} \huge 7}}%
      \put(6803,404){\makebox(0,0){\strut{} \huge 8}}%
      \put(7200,2739){\rotatebox{-270}{\makebox(0,0){\strut{} \huge $\frac{1}{2}H_{c1}^2-|\F_{\mbox{\normalsize GS}}|$}}}%
      \put(4072,154){\makebox(0,0){\strut{} \huge $N$}}%
      \put(5900,2850){\makebox(0,0)[r]{\strut{} \huge $0.7$}}%
      \put(5100,2200){\makebox(0,0)[r]{\strut{} \huge $0.6$}}%
      \put(5100,1600){\makebox(0,0)[r]{\strut{} \huge $e=0.5$}}%
    \put(6650,4300){\makebox(0,0)[r]{\strut{} \Huge $\bf{(C)}$}}%
  }%
    \gplgaddtomacro\gplfronttext{%
    }%
    \gplbacktext
    \put(0,0){\includegraphics{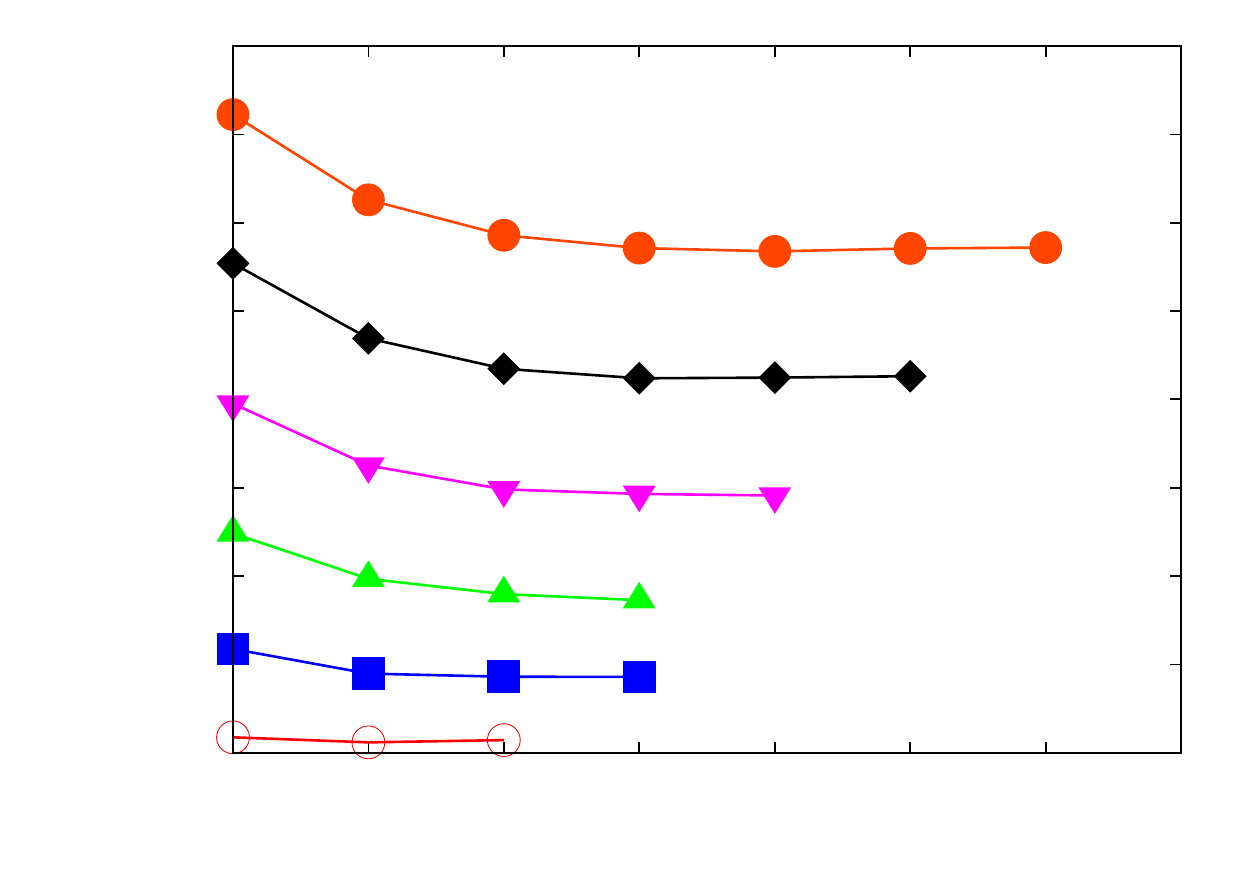}}%
    \gplfronttext
  \end{picture}%
}
\endgroup
%%%%%%%%%%%%%%%%%%%%%%%%%%%%%%%%%%%%%%%%%%%%%%%%%%%%%%%%%%%%%%%%%%%%%%
\hss}
\caption{
(Color online) -- 
Energies per flux quantum of the chiral skyrmions, in the units 
of the energy of the energetically cheapest (either vortex or skyrmion) 
single quantum excitation $(\bf{A})$. Curves with same color
 and symbols on different panels have same parameters. The 
middle panel $(\bf{B})$ shows that it is always beneficial 
(within a parameter range) to have a higher charge skyrmion 
than a lower charge one plus an isolated one quantum vortex. 
The criterion for thermodynamical stability of $N$-quantum 
solitons $\frac{H_{c1}^2}{2}-|\F_{\mbox{\tiny GS}}|$, where 
$\F_{\mbox{\tiny GS}}\equiv \F(\langle\psi_a\rangle,0)$ is the 
condensation energy $(\bf{C})$. The dependence of the solutions 
on $e$ and $N$ is investigated, for a strength of the density-density 
interactions $\gamma_{ab}=0.8$. Other parameters are the same 
as in \Figref{Fig:Sk1}. Here again curves are truncated when the 
soliton's size becomes comparable to the numerical domain.
}
\label{Fig:Sk2}
\end{figure*}

The first critical field is defined as the applied magnetic field 
at which the formation of a single flux carrying defect (vortex 
or skyrmion) becomes energetically favorable. It is defined in 
analogy with the first critical field for ordinary vortices 
$H_{c1}= E_d/\Phi_d$, where $E_d$ and $\Phi_d$ are the 
energy and magnetic flux of the topological defect 
$E_d=\int (\F(\psi_a,\bs A)-\F_{\mbox{\tiny GS}})$ and 
$\F_{\mbox{\tiny GS}}\equiv \F(\langle\psi_a\rangle,0)$ is the 
ground state energy. \Ie it is energetically preferred to form a 
topological defect carrying flux $\Phi_d$ in external field $H_0$ 
if the Gibbs free energy $E_d  - \Phi_d H_0 <0$. The external 
field $H_0$ should be smaller than the thermodynamical 
critical magnetic field $H_{ct}=2\sqrt{\F(0,0) - \F_{\mbox{\tiny GS}} }$.  
The criterion for thermodynamical stability is investigated 
on the right panels $(\bf{C})$ of Figures~\ref{Fig:Sk1}-\ref{Fig:Sk2}. 
For all these regimes, $\frac{H_{c1}^2}{2}-|\F_{\mbox{\tiny GS}}|<0$. 
In all displayed cases, skyrmions satisfy this criterion. That means 
that under certain conditions they can be induced by an applied 
external field.

\subsection{Perturbative stability of Chiral skyrmions} \label{Stability}

%%%%%%%%%%%%%%%%%%%%%%%%%%%%%%%%%%%%%%%%%%%%%%%%%%%%%%%%%%%%%%%%%%%%%%
\begin{figure*}[!htb]
  \hbox to \linewidth{ \hss
  \includegraphics[width=\linewidth]{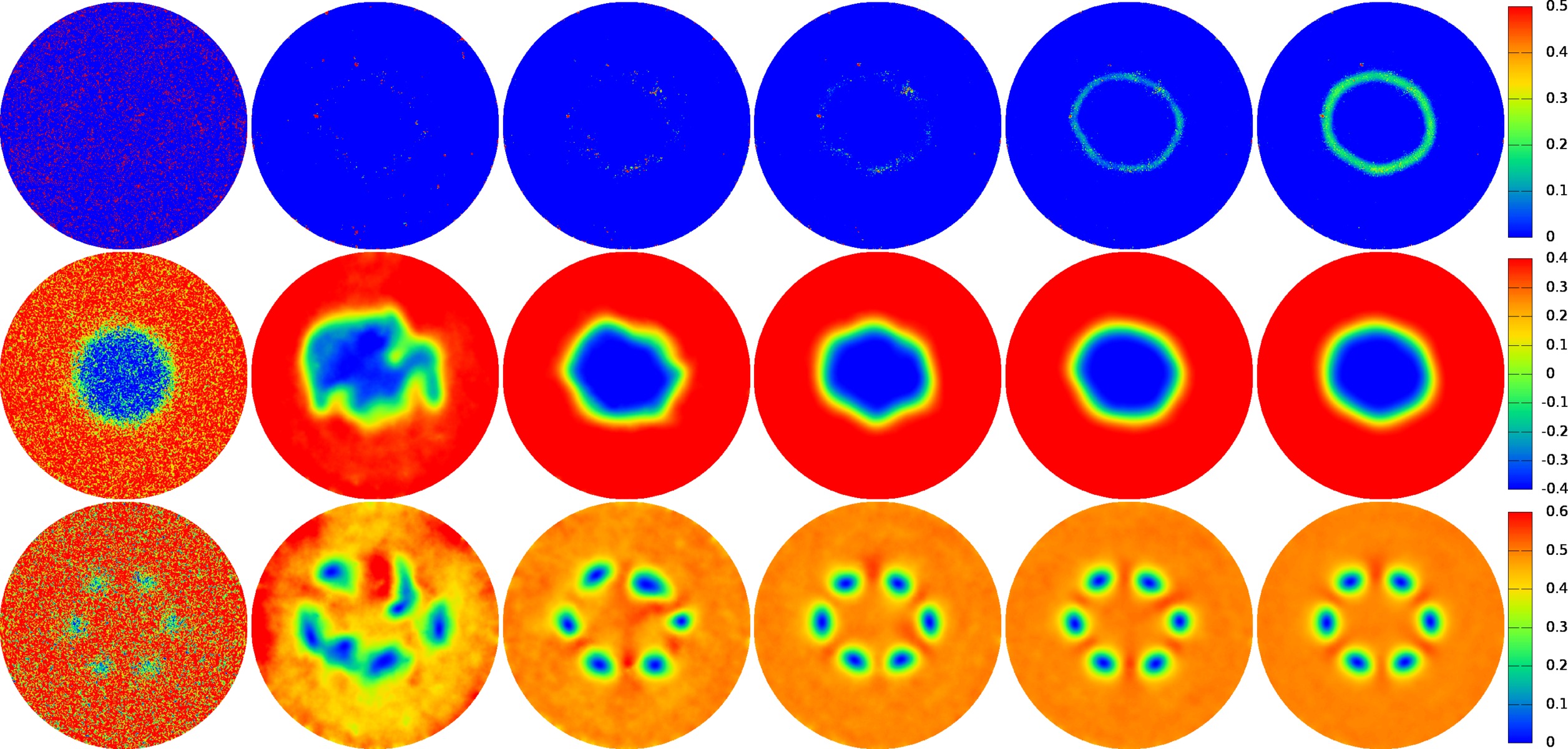}
\hss}
%\hline
\vspace{0.2cm}{\hrule height0.8pt}\vspace{0.2cm}
  \hbox to \linewidth{ \hss
  \includegraphics[width=\linewidth]{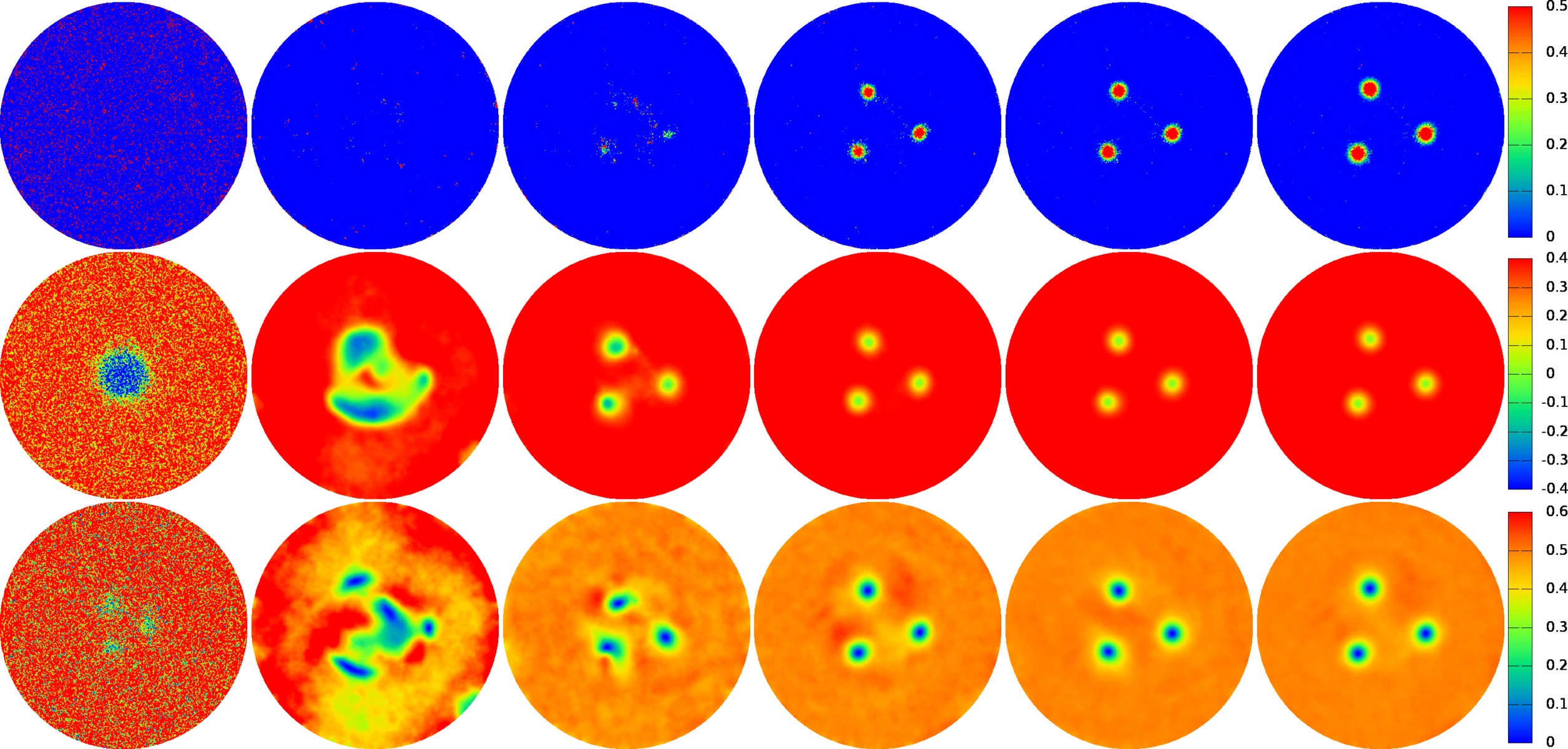}
  \hss}
\caption{
(Color online) -- 
Relaxation of a randomly perturbed chiral skyrmion. Displayed 
quantities are the energy density, $\Im(\psi_1^*\psi_2)$ and 
$|\psi_1|^2$. The parameters are the same as in \Figref{Fig:Single1}, 
but vanishing bi-quadratic density interactions $\gamma_{ab}=0$ 
and  $e=0.3$. Thus it is only meta-stable. 
The snapshots show the state of the system at different stages of 
the energy minimization algorithm after the applied perturbation. 
On the top panel, a $\Q=6$ chiral skyrmion with initial white noise 
of 70 \% of the ground state values. The configurations relaxes to 
a chiral skyrmion. 
On the bottom panel, a perturbation of a metastable charge 
$\Q=3$ soliton with an initial noise $P=0.8$. Here, the noise is 
strong enough to break up the domain wall. The soliton thus relaxes 
to ordinary type-II vortices (one can clearly see the disappearance 
of the domain wall between blue and red area in the middle row). 
The last snapshot in the lower configuration does not represent a 
stationary configuration: the vortices repel each other and are 
in process of drifting apart.
}
\label{Stability-fig1}
\end{figure*}

Chiral skyrmions can appear as thermodynamically stable ground 
states or metastable states in superconductors with Broken Time 
Reversal Symmetry. In this work, they are obtained by minimizing 
the energy. Consequently, they are always minima (at least local) 
of the free energy landscape. When the chiral skyrmions are 
metastable states  they are protected against decay into type-II 
vortices by a finite energy barrier. The analysis carried out in this 
subsection concerns the metastable solutions. In all the regimes 
which we considered, metastable chiral skyrmions are found to be 
very robust. They are easily formed during the energy minimization, 
\eg in closely spaced groups of vortices. The energy barrier 
preventing them from decay to type-II vortices is typically quite high. 
Although difficult to quantify, it is interesting to have a qualitative 
insight into the behaviour of metastable skyrmions against fluctuations. 

One possible approach to study the stability of skyrmions is the 
linear stability analysis which consists of applying infinitesimally 
small perturbation to the fields, and investigating the eigenvalue 
spectrum of the (linear) perturbation operator, on the background 
of a given solution. When the background solution is (meta) stable 
all infinitesimally small perturbations are positive modes and thus 
can only increase the energy. As a result linear stability analysis 
cannot tell anything especially interesting about the properties of 
skyrmions.
A strong perturbation should cause a decay of a metastable chiral 
skyrmion to ordinary vortices. Here, the stability is investigated 
numerically by perturbing the chiral skyrmion by white noise. This 
allows one to investigate the full non-linear response where the 
meaningful information belongs.
The white noise applied to all degrees of freedom, is generated 
as follows 
\Align{Perturbations}{
  \psi_a &= \psi_a^\oz 
  + P\mathrm{max}(|\psi|)\mu_a^{\mbox{\tiny $\psi$}}(x,y)\, , \nonumber \\
  A_i&= A_i^\oz
  +P\mathrm{max}(|\bs A|)\mu_i^{\mbox{\tiny $\bs A$}}(x,y)	\,.
}
Here $^\oz$ denotes the fields of the initial skyrmionic state, $P$ 
is a ratio giving the relative magnitude of the perturbation with 
respect to the maximal amplitude of a given field of the initial state. 
$\mu_a^{\mbox{\tiny $\psi$}}(x,y)$, and $\mu_i^{\mbox{\tiny $\bs A$}}(x,y)$ 
are (independent) random functions of the space. They satisfy 
$|\mu_a^{\mbox{\tiny $\psi$}}|<1$ and $|\mu_i^{\mbox{\tiny $\bs A$}}|<1$. 
As a result all fields initially receive a noise whose relative 
amplitude is $P$. The perturbation has very large field gradients 
since it is applied locally on the mesh. After applying noise the 
system is then relaxed using the same minimization scheme as 
for constructing the skyrmions. Despite the strong field gradients, 
if the white noise does not exceed a certain threshold, the 
configuration relaxes back to the initial chiral skyrmion solution. 
This can be seen from the upper panel of \Figref{Stability-fig1}. 
The noise was gradually increased, confirming that indeed, a 
sufficiently strong perturbation drives the metastable solution 
over the barrier, in the energy landscape,  thus leading to its decay 
to ordinary vortex solutions as shown on the bottom panel of 
\Figref{Stability-fig1}. The precise value of the relative amplitude 
required to destabilize a given chiral skyrmion, obviously depends 
on the parameters of the Ginzburg--Landau functional and on the 
number of flux quanta of the solution.

As expected, if a perturbation is strong enough, the metastable 
chiral skyrmion decays to the configuration with less energy, \ie 
isolated type-II vortices. The observed behaviour confirms the 
expectations from energy arguments \Partref{Energy-ThStab}. 
Moreover, the deeper in the type-II regime, the less breakable 
are the skyrmions. One of the easiest ways for a skyrmion to 
decay is to deform it enough so that the domain wall self 
intersects. The configuration then can decay to skyrmionic 
configurations with lower $\Q$ which are less stable and can 
further decay into integer vortices.

%%%%%%%%%%%%%%%%%%%%%%%%%%%%%%%%%%%%%%%%%%%%%%%%%%%%%%%%%%%%%%%%%%%%%%
%%%%%%%%%%%%%%%%%%%%%%%%%%%%%%%%%%%%%%%%%%%%%%%%%%%%%%%%%%%%%%%%%%%%%%
\section{Interactions of Chiral skyrmions}\label{Interactions}

The analysis of the energetic properties of chiral skyrmions suggests 
they should have quite non trivial interactions. 
Generally, the energy 
per flux quantum \emph{decreases} with the topological charge (see \eg 
\Figref{Fig:Sk1}). In some cases it is also preferable to absorb isolated 
vortices into a skyrmion, \ie the energy of an $N$-quantum vortex is less 
than that of an $(N-1)$-quantum vortex and an isolated vortex.
In those cases, the interaction at short range should be 
attractive. On the other hand, they exist in regimes where vortices usually 
exhibit repulsive interaction (type-II or even type-1.5). 
Moreover, the lack of axial symmetry and complicated internal structure 
featuring fractional vortices can provide very non-trivial contribution to the 
interaction of skyrmions in BTRS superconductors.

%%%%%%%%%%%%%%%%%%%%%%%%%%%%%%%%%%%%%%%%%%%%%%%%%%%%%%%%%%%%%%%%%%%%%%
\subsection{Chiral skyrmion--vortex interaction}

\begin{figure}[!htb]
 \hbox to \linewidth{ \hss
  \includegraphics[width=\linewidth]{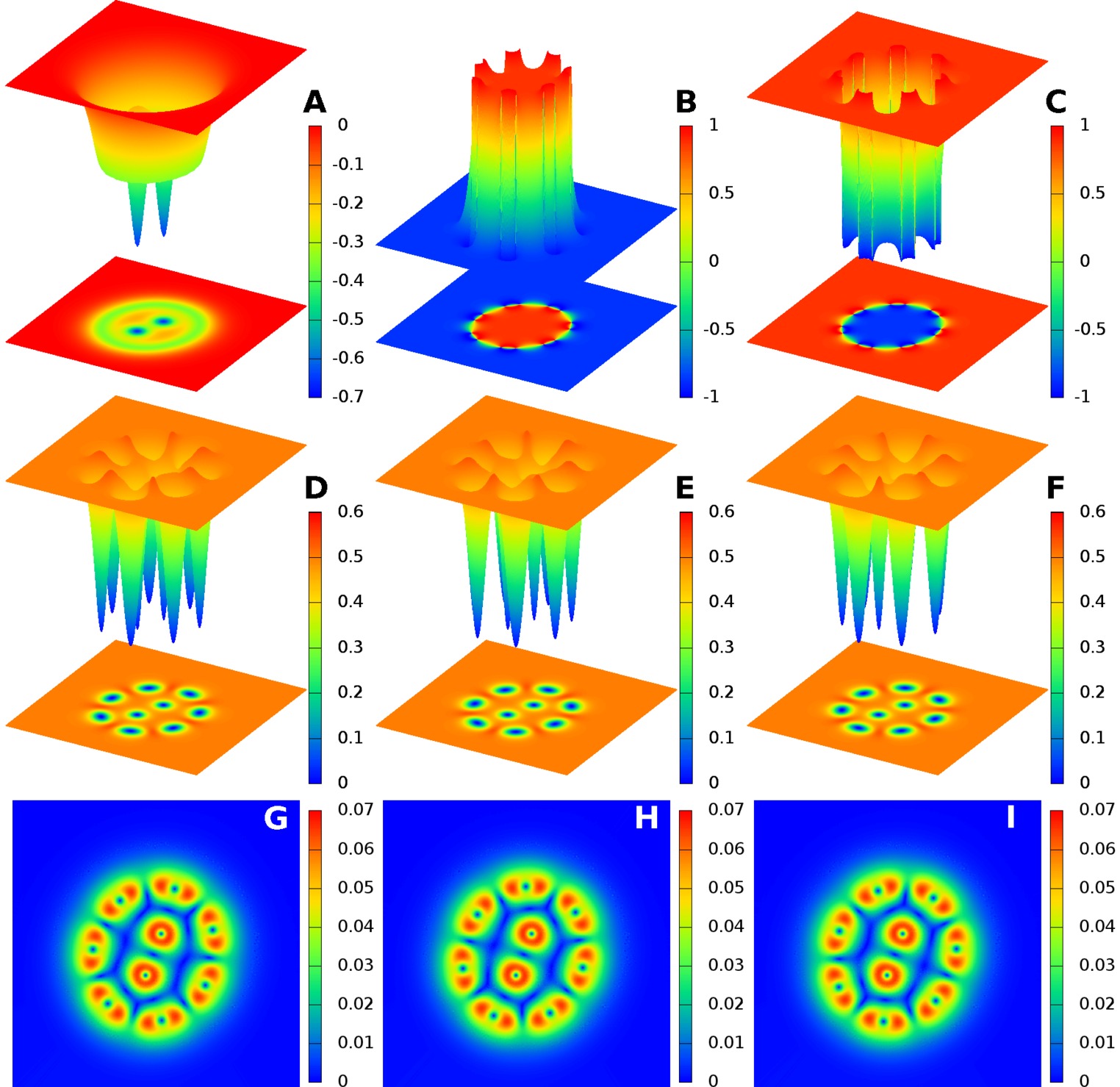}
  \hss}
\caption{
(Color online) -- A $\Q=9$ quantum configuration of \emph{mixed} 
vortices and skyrmions in a system with three identical passive 
bands as in \Figref{Stability-fig1}. This configuration is made 
out of a skyrmion surrounding two ordinary vortices. It is known, 
from energy considerations that interaction is short range 
attractive. Interaction with vortices deforms the skyrmions. 
This shows that it is long range repulsive.
}
\label{Fig:Mixed}
\end{figure}

Chiral skyrmions can have very non trivial, non-monotonic 
interaction with vortices. As seen from the numerically 
obtained solutions shown on \Figref{Fig:Sk1} and 
\Figref{Fig:Sk2}, in applied field, chiral skyrmions can be 
either ground states (for a given phase winding) or represent 
metastable states.
For some regimes, as seen from the middle panels of 
\Figref{Fig:Sk1} and \Figref{Fig:Sk2}, a vortex placed sufficiently 
close to a chiral skyrmion should be absorbed in the domain wall 
and split into fractional vortices, thus increasing the charge of 
the skyrmion and then decreasing its energy per flux quantum. 
Consequently, the interaction is expected to be attractive at 
short range. Indeed, as we observe in numerical calculations, 
if vortices are placed close enough to a domain wall, they are 
easily trapped to form a skyrmion of larger topological charge. 
However the long range forces between skyrmions and vortices 
can be repulsive. This is clearly seen from the existence of 
stable configurations where a number of integer flux vortices 
are confined within a chiral skyrmion, as shown on \Figref{Fig:Mixed}. 
That figure demonstrates that there is a repulsion between inner 
``ordinary vortices'', and the fractional vortices comprising  
the chiral skyrmion, which follows  from (i) the stability of the 
configuration and (ii) the fact that the type-II vortices visibly 
stretch the skyrmion.  Thus the interaction here is non-monotonic, 
being long range repulsive, but short range attractive. 

The repulsive long-range skyrmion-vortex interaction follows 
from the following considerations. In the ground state a vortex 
is an axially symmetric object with all phases winding around 
the same core. Thus in the type-II limit its energy and long-range 
interactions are dominated by the supercurrent $\bf J$  term in 
\Eqref{JPart}. At long separations when linearized theory applies, 
the interaction between a skyrmion and a vortex is dominated by 
this current-current $\bf J$-mediated interaction, resulting in repulsion. 
The attractive interaction at short distances is a nonlinear effect 
where split fractional vortices in a Skyrmion can deform a vortex 
by ``polarizing" it. \ie they can split its constituent fractional vortices 
thus inducing ``dipole''-like interactions. This interaction attracts 
the vortex so that it merges into the skyrmion.

%%%%%%%%%%%%%%%%%%%%%%%%%%%%%%%%%%%%%%%%%%%%%%%%%%%%%%%%%%%%%%%%%%%%%%
\subsection{Skyrmion--skyrmion interaction}\label{interaction}

In contrast to ordinary vortices in Ginzburg--Landau theory, chiral 
skyrmions do not exhibit rotational symmetry. An important 
consequence is that inter-soliton interactions should in general 
depend on the relative orientation of the solitons. First, note that 
the orientation and position of a soliton can be described by the 
position of the fractional vortices. The shape of a soliton, including 
the positions of the constituting fractional vortices is determined 
by energy minimization. The energy of the skyrmion is invariant 
under overall rotation and translation. 

\begin{figure}[!htb]
 \hbox to \linewidth{ \hss
  \includegraphics[width=\linewidth]{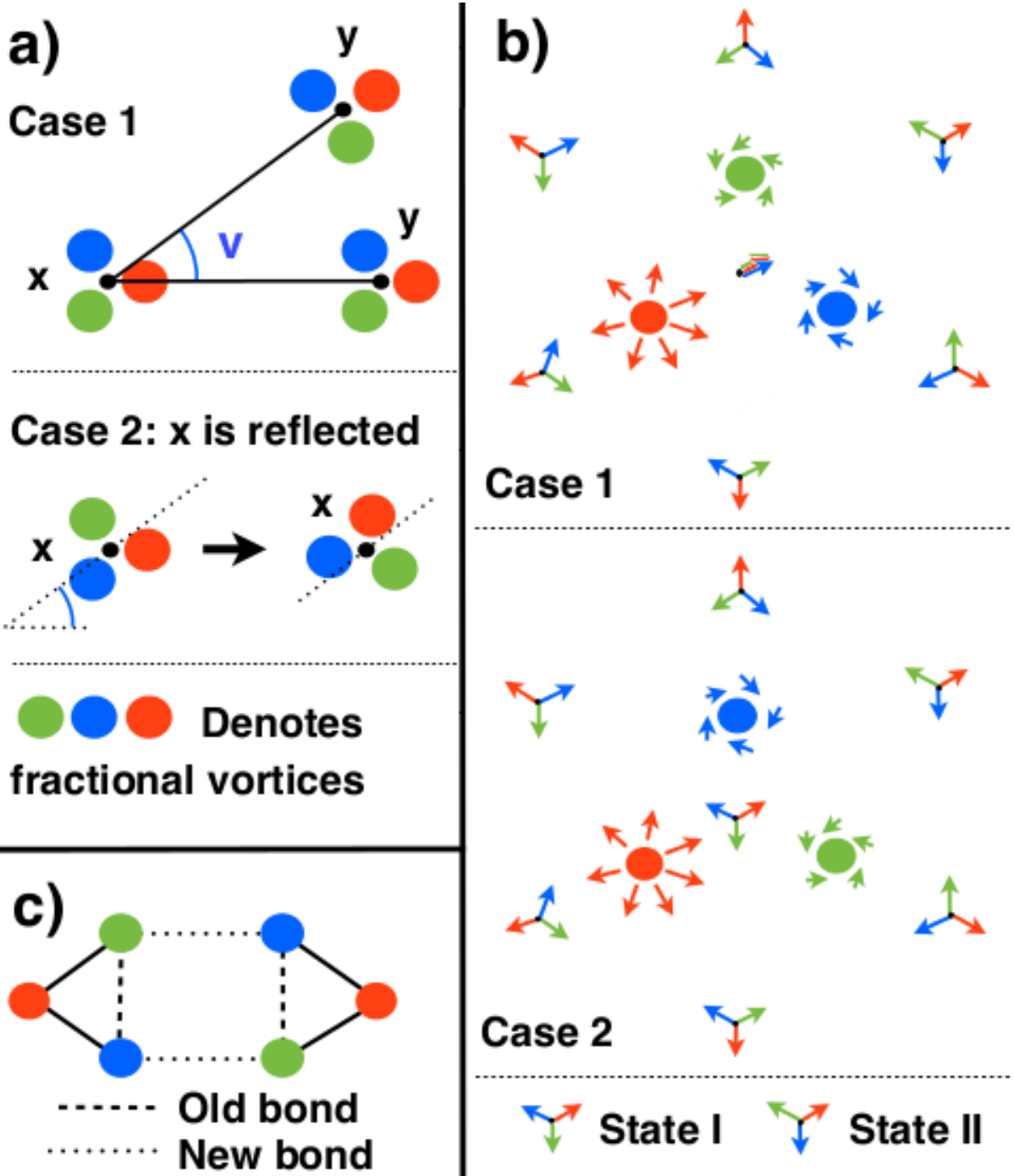}
  \hss}
\caption{(Color online) -- 
Panel {\bf (a)} shows a schematic picture of how soliton 
interactions are computed. This figure shows the interaction 
between two single quanta solitons, each consisting of three 
fractional vortices shown in green, blue and red. This generalizes 
easily to larger solitons. One soliton ($\rm x$) is placed in the origin, 
while the second ($\rm y$) is placed at a distance $R$ at an angle 
$v$. Consequently, as $v$ is varied, the relative orientation of the 
solitons changes. Case 1 shows a system of two solitons with 
identical chiralities (same ordering $\order$), while case 2 shows 
two solitons with opposite chiralities (different $\order$), although 
the mirrored soliton is not necessarily stable. 
A schematic comparison of solitons with different ordering $\order$ 
is displayed on panel {\bf (b)}. In the case (2), the gradients in 
phase difference due to the fractional vortices naturally interpolate 
between two $\Ztwo$ states. For this reason, the case (2) is 
energetically preferable over (1) and it was verified numerically. 
Finally, panel {\bf (c)} gives a schematic view of the merging 
of two single quanta solitons. In order to merge, they should 
have same ordering but opposite orientation. 
}
\label{cartoon}
\end{figure}

Finally note that there are two orders in which the fractional 
vortices can be arranged. Going counter-clockwise along the 
domain wall, the vortices can be ordered $1,2,3$ or $1,3,2$. 
We denote this order $\order=\epsilon_{abc}$, $\epsilon$ being 
the Levi-Civita symbol and $a,b,c$ are the band indices of the 
fractional vortices. For a skyrmion carrying integer flux, 
$\order=\pm1$ (note that this ordering closely relates to the 
concept of chirality). As illustrated in \Figref{cartoon} {\bf (a)}, 
a system of two solitons is thus described by the distance 
between them $R$, their relative orientation $v$ together with 
the ordering (chirality) of each individual skyrmion.

\subsubsection{Chirality of skyrmions: inequivalence of left- and right- handed solutions}

In general, for a chiral skyrmion, the energy is not independent 
of the ordering $\order$. For a given $\Ztwo$ ground state 
outside of a skyrmion, the system allows \emph{only one 
particular ordering $\order$ of the fractional vortices in the 
skyrmion}. The mechanism that gives rise to this behaviour is 
illustrated in \Figref{cartoon} {\bf (b)}:  For a given external 
phase-locking pattern (a $\Ztwo$ state), only a particular 
ordering $\order$  gives the opposite $\Ztwo$ state inside. 
In the illustration the two solitons (case 1 and 2) differ in the 
ordering of the fractional vortices (represented by red blue 
and green dots with band index 1,2,3 respectively) -- the 
corresponding phase configurations are shown by the arrows. 
Thus, the ordering of the first one (case 1) is 
$\order=\epsilon_{132}=-1$ while the ordering of the second 
(case 2) is $\order=\epsilon_{123}=+1$. Now for a same 
given ground state outside both solitons, the phase-locking 
inside is determined consistently with the phase gradients of 
each fractional vortex. In the first case, it results in a phase 
arrangement inside the soliton that is not a ground state. 
However, in the second case, the state obtained inside is 
a different $\Ztwo$ ground-state. As a result, there is a 
synergy effect where the phase gradients due to the fractional 
vortices go from one $\Ztwo$ state to another. Therefore 
$\order=+1$ is energetically cheaper than $\order=-1$ for 
which the inner phase locking is the farthest from the ground 
state. This is indeed confirmed in our numerical simulations
 where a skyrmion $\order=-1$ decays into a skyrmion 
$\order=+1$. 
Thus the ordering of the fractional vortices does matter in 
BTRS superconductors. It results in the discrimination of one 
ordering. This further motivates the terminology \emph{chiral}.

\subsubsection{Numerical calculations on inter-skyrmion forces}

As illustrated in \Figref{cartoon} {\bf (a)}, inter-soliton forces 
are computed according to the following procedure. First, the 
structure of the soliton is determined by unconstrained energy 
minimization, thus determining the actual position of the 
fractional vortices constituting the skyrmion. Then two skyrmions 
($\rm x$ and $\rm y$ in \Figref{cartoon}) are place at a distance 
$R$ and a relative orientation $v$. There, the energy is 
minimized with respect to all degrees of freedom, except the 
position of the singularities of each fractional vortex. As shown 
in \Figref{cartoon} {\bf (a)}, the energy is computed for every 
distance and relative orientation $R$ and $v$. 
While allowing computation of long-range inter soliton forces, 
this procedure has an important limitation. It does not take into 
account one of the nonlinear effects: Deformation of interacting 
solitons in the form of changes of the position of the fractional 
vortices. However, this is primarily a problem at short separation, 
where the deformation is generally the strongest.
\begin{figure}[!htb]
 \hbox to \linewidth{ \hss
  \includegraphics[width=\linewidth]{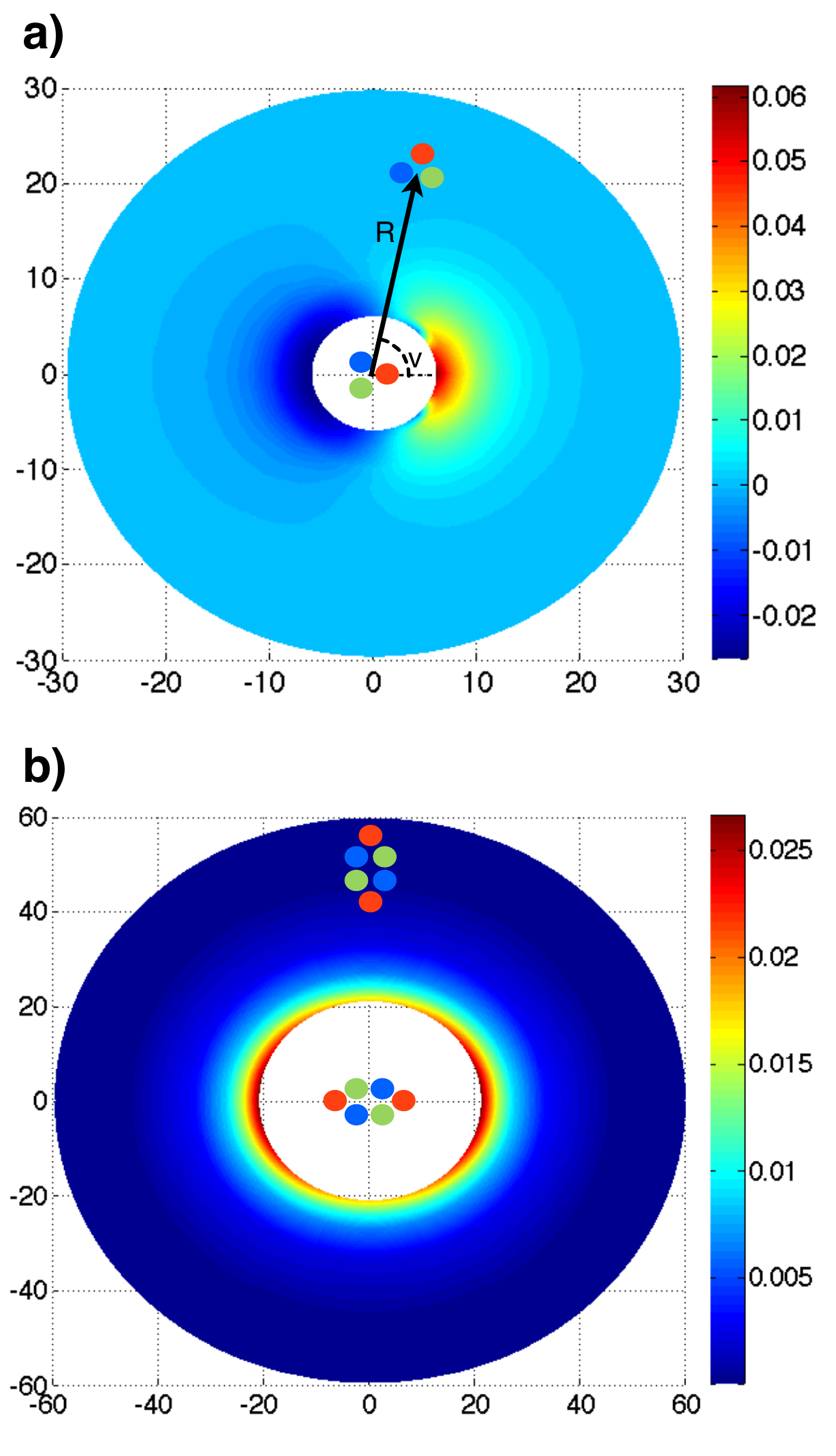}
  \hss}
\caption{(Color online) -- 
Panel {\bf (a)} displays the interaction energy of two single 
quantum skyrmions. One soliton is placed at the origin, the 
interaction energy is plotted as a function of the position and 
relative orientation of the second soliton. The interaction 
energy is maximal when $v=0$ while it is minimal for the 
opposite orientation,  $v=\pi$. The strength of the interaction 
decreases with the separation $R$. The model parameters 
are the same as in \Figref{Fig:Single1}. 
Panel {\bf (b)} shows the interaction energy of two $\Q=2$ 
quanta solitons, for the same parameters as in \Figref{Fig:Pair1}. 
Note that the skyrmion has two-fold symmetry (it is invariant 
under global rotations of $\pi$). The minimum energy is found 
for the relative orientation $v\pm\pi/2$. 
}
\label{interact}
\end{figure}

\Figref{interact} {\bf (a)} shows the interaction energy of two 
single quanta skyrmions, identical to the one in \Figref{Fig:Single1}. 
From \Figref{Fig:Sk2} it is clear that the energy per flux quanta 
decreases with the number of flux quanta. For the solitons to 
merge, they need to have opposite orientation, see 
\Figref{cartoon} {\bf (c)}. The computed interaction energy, 
\Figref{interact} {\bf (a)}, is indeed consistent with this picture. 
When the relative orientation, $v$ is not optimal \ie $v\not=\pi$, 
the solitons exert a torque on each other, so that they attain this 
optimal orientation. Then, an attractive channel opens in the 
potential, allowing them to get closer where nonlinear effects 
are strong, ultimately leading to a merger. 

The interaction energy of a slightly more complex soliton is 
shown in \Figref{interact} {\bf (b)}. There, each skyrmion 
carries two flux quanta (\ie their topological charge is $\Q=2$). 
The parameters are the one of \Figref{Fig:Pair1}, from which we 
know that superconducting components are not identical and 
that the skyrmion is more or less elliptic. Global orientation of 
the skyrmion is chosen so that when $v=0$ the major axis of 
both solitons lie along the horizontal axis. Note that these 
skyrmions are not only invariant under global rotation by $2\pi$, 
but also by $\pi$. Within the numerical accuracy, the inter-skyrmion 
interaction is always repulsive. Note that this approach can 
accurately determine the interaction only at sufficiently long 
distances. Indeed, by fixing the positions of the fractional vortices, 
it assumes that the skyrmions are almost-rigid bodies. The relative 
position of singularities in each fractional vortex is fixed once for 
all, but the fields can deform around this rigid `skeleton'. This 
neglects the possibility of mutually induced deformations of the 
`skeleton', which can open an attractive channel. Since our 
``almost-rigid body'' approximation holds only at large enough 
distances, short range data are irrelevant and not displayed 
in \Figref{interact}.
We also derive general long-range intersoliton forces in the more 
formal framework of \Partref{section:lrisf}. In \Partref{section:btrs} 
the formal long-range interactions are applied to the particular 
case of a BTRS superconductor. The predictions derived there are 
consistent with the numerical results presented in this section.

%%%%%%%%%%%%%%%%%%%%%%%%%%%%%%%%%%%%%%%%%%%%%%%%%%%%%%%%%%%%%%%%%%%%%%
%%%%%%%%%%%%%%%%%%%%%%%%%%%%%%%%%%%%%%%%%%%%%%%%%%%%%%%%%%%%%%%%%%%%%%
\section{Mathematical analysis of long-range intersoliton forces}\label{section:math}

The model considered in this paper has many properties that 
are interesting from a formal, mathematical point of view. 
In this section we show how, by re-writing the free energy in 
terms of gauge-invariant fields, we can identify a hidden 
topological charge, associated with the topology of the 
complex projective space $\CP^2$, and devise a 
mathematically satisfactory scheme for deducing the nature 
(attractive or repulsive) and range of the 
dominant force between well-separated solitons (either 
vortices or skyrmions). For generic parameter choices, the 
final step in this scheme (finding the spectrum of a symmetric 
real matrix) must be done numerically, but there are several 
symmetric cases and parametric limits where all calculations 
can be completed explicitly. After treating the general case, 
we consider two such special cases, both of potential 
phenomenological interest.

\subsection{\texorpdfstring{Reduction to a supercurrent coupled $\CP^{k-1}$ model}
{Reduction to a supercurrent coupled CP {k-1} model}}

In this section we consider a general $k$ component GL 
model, with no restriction on the potential terms $V$. 
The $k$ complex fields $\psi_a$ may be collected into 
a complex $k$-vector $\Psi:M\ra\C^k$, where $M=\R^2$ 
denotes physical space. It is convenient to use polar 
coordinates on $\C^k$ by defining
\beq\label{polcoo}
\Psi=:\rho Z
\eeq
where $\rho=\sqrt{\Psi^\dagger\Psi}\geq 0$ and 
$Z^\dagger Z=1$. Let $\pi:\C^k\less\{0\}\ra\CP^{k-1}$ 
denote the canonical projection which takes a point in 
$\C^k$ to the complex line through $0$ containing that 
point, and for any $X\in\C^k$, $X\neq 0$, denote by $[X]$ 
its projective equivalence class (so $[X]=\pi(X)$). By gauge 
invariance, the potential $V(\Psi)$ can actually depend only 
on $\rho$ and $[Z]\in\CP^{k-1}$, the projective equivalence 
class of $Z$, or, equivalently, of $\Psi$. Let $\Phi=\pi\circ\Psi$. 
This is a $\CP^{k-1}$-valued field which maps each $p\in M$ 
to $[\Psi(p)]=[Z(p)]\in\CP^{k-1}$. By construction it is, 
like $\rho$, gauge invariant.  We may rewrite the free energy 
entirely in terms of the gauge-invariant quantities $\rho, \Phi$ 
and $J=e\Im(\Psi^\dagger D\Psi)$, the total supercurrent. 
To do so, it is convenient to think of the gauge field $A$ and the 
supercurrent $J$ as one-forms rather than vector fields (so we 
use the metric on physical space $M=\R^2$ to ``lower the indices'' 
on vectors $A^i$ and $J^i$). In this language, the covariant 
derivative of $\Psi$ is, likewise, a one-form 
\beq
D\Psi=\d\Psi+ieA\Psi
\eeq
 with values in $\C^k$.

On $\C^k\less\{0\}$, let us define the real one-form
\beq
\nu=-{\rm Im}\frac{X^\dagger\d X}{|X|^2}.
\eeq
where $X=(X_1,\ldots,X_k)$ is a global coordinate on 
$\C^k\backslash\{0\}$ and $\d X=(\d X_1,\ldots, \d X_k)$ 
are the corresponding holomorphic one-forms. Then the 
total supercurrent is
\beq\label{sk}
J=e\rho^2\{eA-\Psi^*\nu\}
\eeq
where $\Psi^*\nu$ denotes the pullback of 
$\nu\in\Omega^1(\C^{k}\less\{0\})$ to $M$ by the map 
$\Psi:M\ra\C^{k}\less\{0\}$. In less compact notation, this 
is the one-form on $M$ whose $\d x^i$ component is 
$-\rho^{-2}{\rm Im}\Psi^\dagger\cd_i\Psi$. It follows that 
the magnetic field (thought of as a two-form) is
\beq
B=\d A=\frac1e\left(\d(\Psi^*\nu)-\frac{1}{e}\d\left(\frac{J}{\rho^2}\right)\right).
\eeq
It is a general fact that the exterior differential operator $\d$ 
commutes with pullback of differential forms, so 
$\d(\Psi^*\nu)=\Psi^*(\d\nu)$. Note that $\d\nu$ is a closed 
two-form on $\C^k\less\{0\}$. Let $h$ denote the Fubini-Study 
metric on $\CP^{k-1}$ with constant holomorphic sectional 
curvature $1$, and $\omega$ denote its associated k\"ahler 
form. Then the pullback of $\omega$ by 
$\pi:\C^k\less\{0\}\ra\CP^{k-1}$ is, like $\d\nu$, a closed 
two-form on $\C^k\less\{0\}$. In fact, $\omega$ is defined 
\cite{kobnom} by the requirement that
\beq
\pi^*\omega=2\d\nu.
\eeq
Hence
\Align{}{
\d(\Psi^*\nu)=\Psi^*(\d\nu)&=\frac12\Psi^*(\pi^*\omega) \nonumber\\
&=\frac12(\pi\circ\Psi)^*\omega=\frac12\Phi^*\omega,
}
and so
\beq\label{rheabl}
B=\frac1e\left(\frac12\Phi^*\omega-\frac{1}{e}
\d\left(\frac{J}{\rho^2}\right)\right).
\eeq

Similarly, we may rewrite $|D\Psi|^2$ entirely in terms of the 
gauge invariant quantities $\rho,\Phi$ and $J$. From \Eqref{sk}, 
we see that
\Align{}{
D\Psi &=\d\Psi+i\left(\Psi^*\nu-\frac{J}{e\rho^2}\right)\Psi  \nonumber \\
	 &=(\d\rho)Z+\rho\d Z+i\left(\Psi^*\nu-\frac{J}{e\rho^2}\right)\rho Z.
}
Let $e_1,e_2$ denote an orthonormal frame on $M$ (for example 
$e_i=\cd/\cd x^i$) and $X_i=\d Z(e_i)\in T_Z S^{2k-1}$. Then 
$\Re(Z^\dagger X_i)=0$ since $X_i$ is tangent to the unit sphere in 
$\C^k$ at $Z$. Hence
\Align{anke}{
|D\Psi|^2&=\sum_i(D\Psi(e_i))^\dagger D\Psi(e_i)\nonumber \\
&=\sum_i\Bigg\{(\d\rho(e_i))^2+\rho^2|X_i|^2  \nonumber\\ &
\quad+2\Im(X_i^\dagger Z)\rho^2\left(\frac{J(e_i)}{e\rho^2}-\Psi^*\nu(e_i)\right) \nonumber\\ &
\quad+\rho^2\left(\frac{J(e_i)}{e\rho^2}-\Psi^*\nu(e_i)\right)^2\Bigg\}\nonumber \\
&=|\d\rho|^2+\frac{1}{e^2\rho^2}|J|^2+\rho^2\sum_i(|X_i|^2-\nu(X_i)^2)
}
since $\Im(X_i^\dagger Z)=\nu(X_i)=(\Psi^*\nu)(e_i)$. Consider 
$\pi^*h$, the pullback by $\pi$ of the Fubini-Study metric on 
$\CP^{k-1}$ to $\C^{k}\less\{0\}$. Given any tangent vector 
$X\in T_ZS^{2k-1}$, 
\Align{}{
(\pi^*h)(X,X)&=h(\d\pi X, \d\pi X)
=\omega(\d\pi X, i\d\pi X)\nonumber \\
&=\omega(\d\pi X, \d\pi iX)
=\pi^*\omega(X,iX)\nonumber \\
&=2\d\nu(X,iX)
=4(|X|^2-\nu(X)^2)
}
where we have used the fact that $\pi:\C^k\less\{0\}\ra\CP^{k-1}$ 
is holomorphic (so $\d\pi$ commutes with $i$). Hence
\Align{}{
\sum_i(|X_i|^2&-(\Psi^*\nu)(e_i)^2)=
\frac14\sum_i\pi^*h(X_i,X_i)\nonumber \\
&=\frac14\sum_i\pi^*h(\d\Psi e_i,\d\Psi e_i)\nonumber \\
&=\frac14\sum_i h(\d\Phi e_i,\d\Phi e_i)
\label{ak}
=\frac14|\d\Phi|^2,
}
where $|\d\Phi|$ denotes the norm of the linear map 
$\d\Phi_p:T_pM\ra T_{\Phi(p)}\CP^{k-1}$ with respect 
to the metric $h$. Substituting \Eqref{ak} into \Eqref{anke}, 
one sees that
\beq
|D\Psi|^2=|\d\rho|^2+\frac{|J|^2}{e^2\rho^2}+\frac{\rho^2}{4}|\d\Phi|^2.
\eeq
Finally, we obtain an expression for the total free energy
\Align{angker}{
F=&\int_M\Bigg\{
\frac12|\d\rho|^2+\frac{\rho^2}{8}|\d\Phi|^2+\frac{|J|^2}{2e^2\rho^2}\nonumber \\
&+\frac{1}{2e^2}\left|\d\left(\frac{J}{e\rho^2}\right)-\frac12\Phi^*\omega\right|^2+V(\rho,\Phi)\Bigg\}.
}
The above expression for $F$ is valid for any number 
of condensates $k$, and for all field configurations where 
$\Psi^{-1}(0)\subset M$ has measure zero, \ie where the 
set of points in physical space at which the condensates 
$\psi_a$ all simultaneously vanish is negligible. This condition 
holds for skyrmions ($\Psi^{-1}(0)$ is empty), and for 
(multi-)vortices ($\Psi^{-1}(0)$ is finite), so we can use 
\Eqref{angker} for questions involving either type of soliton, 
though one should note that, for vortices, the $\CP^{k-1}$-valued 
field $\Phi$ is undefined at the finite collection of vortex positions.

In the special case $k=2$, we may identify $\CP^{k-1}$ with 
the unit two-sphere $S^2$ , by mapping $[Z_1,Z_2]\in\CP^1$ 
to the point on $S^2$ with stereographic coordinate $Z_2/Z_1$, 
so that $\Phi$ can be interpreted as being two-sphere valued. 
The k\"ahler form $\omega$  coincides with the area form on 
$S^2$ under this identification, so that the expression for $F$ 
\Eqref{angker} reduces to the  decomposition in Ref.~\onlinecite{bfn}.
In the general $k$ case (which was previously discussed, in 
somewhat different mathematical language, in context of an 
$SU(N)$ model in  Ref.~\onlinecite{Hindmarsh:93}), the field $\Phi$ takes 
values in $\CP^{k-1}$, which we cannot identify with any sphere. 

\subsection{Flux quantization and the topological charge}\label{section:quantization}

In order for a configuration on $M=\R^2$ to have finite total 
energy, $\Phi$ and $\rho$ should tend to constants 
$\Phi_0\in\CP^{k-1}$, $\rho_0\in(0.\infty)$, and $J$ should 
tend to $0$ as $|x|\ra\infty$. It follows, from \Eqref{rheabl} 
and Stokes's theorem, that the total magnetic flux of a finite 
energy configuration is
\beq\label{skonpihg}
\int_M B=\frac{1}{2e}\int_M\Phi^*\omega=:\frac{2\pi}{e}\Q(\Phi),
\eeq
which is a homotopy invariant of the map $\Phi:M\ra\CP^{k-1}$, 
because $\omega$ is closed. In the case $k=2$, $\Q$ is the 
winding number of the map $\Phi:M\ra S^2$. For $k>2$, $\Q$ 
is still an integer, but its geometric interpretation is more subtle: 
the image of $M$ under $\Phi$ is homologous to $\Q(\Phi)$ 
copies of the generator of $H_2(\CP^{k-1})$. This gives an 
alternative interpretation of $\Q$, to augment the physical 
interpretation, described in \Partref{Fractional-vortices}, of the 
magnetic flux being carried by an integer number of sets of $k$ 
fractional-flux vortices.

It is straightforward to give an integral formula for $\Q(\Phi)$ 
in terms of the original condensates $\Psi$, using the fact that 
$\pi^*\omega=2\d\nu$: 
\Align{}{
\Phi^*\omega&=(\pi\circ\Psi)^*\omega=\Psi^*(\pi^*\omega)=2\Psi^*\d\nu 
\nonumber\\
&=\frac{2}{i}\Psi^*\left(\frac{\d Z^\dagger\wedge\d Z}{|Z|^2}+
\frac{Z^\dagger\d Z\wedge \d Z^\dagger Z}{|Z|^4}\right)\nonumber \\
&=\frac{2}{i|\Psi|^4}\left(|\Psi|^2 \d\Psi^\dagger\wedge\d\Psi
+\Psi^\dagger\d\Psi\wedge\d\Psi^\dagger\Psi\right).
}
Hence
\Equation{}{
   \Q(\Psi)=\int_{\Real^2}\frac{i\epsilon_{ji}}{2\pi|\Psi|^4} \left[
   |\Psi|^2\partial_i\Psi^\dagger\partial_j\Psi
   +\Psi^\dagger\partial_i\Psi\partial_j\Psi^\dagger\Psi
   \right]\dd^2x\,.
}
One should note that the flux-quantization condition 
\Eqref{skonpihg} and the integral  formula for the topological 
charge $\Q$ above are valid only for field configurations for 
which $\Psi$ never vanishes. Note that flux is also quantized 
for ordinary vortices, for  which $\Psi$ vanishes, but then it is 
no longer associated with the  the topological charge $\Q$, 
but with a $U(1)$ topological charge associated with the total 
phase winding at spatial infinity. This expression for $\Q$ can 
be easily discretized for use on a numerical lattice. Comparing 
$\Q$ with the total number of flux quanta gives a convenient 
way of distinguishing between vortices and skyrmions numerically.

\subsection{Long-range intersoliton forces}\label{section:lrisf}

The key to understanding long-range forces between solitons 
is to identify the point sources which replicate, in the linearization 
of the field theory about the vacuum, the asymptotic fields 
of an isolated soliton \cite{spe_point_vortex}. Assuming that 
the vacuum is not $\Psi=0$, we can use the gauge-invariant 
variables $\rho,\Phi,J$, and expression \Eqref{angker} for this 
purpose. So, let the vacuum (\ie minimum of $V$) occur at 
$\rho=\rho_0$, $\Phi=\Phi_0$. To identify the linearization of the 
theory about this vacuum, we set $\rho=\rho_0+\sigma$, 
$\Phi=\Phi_0+Y$, where $Y\in T_{\Phi_0}\CP^{k-1}$, and expand 
$F$ to quadratic order in the small quantities $\sigma,Y$ and $J$:
\Align{flinhess}{
F_{lin}&=\int_M\bigg\{
\frac18\rho_0^2|\d Y|_{T_{\Phi_0}\CP^{k-1}}^2+\frac12|\d\sigma|^2\nonumber\\
&+\frac12\hess_{(\rho_0,\Phi_0)}((\sigma,Y),(\sigma,Y))
\nonumber\\
&+\frac{1}{2e^4\rho_0^4}(|\d J|^2+e^2\rho_0^2|J|^2)\bigg\}%\label{flinhess}
}
where $\hess_{(\rho_0,\Phi_0)}$ is the Hessian of the function 
$V:(0,\infty)\times\CP^{k-1}\ra\R$ about its minimum 
$(\rho_0,\Phi_0)$, which we now define. Let 
$P=(0,\infty)\times\CP^{k-1}$ and $p_0=(\rho_0,\Phi_0)$, so that 
$p_0$ is the minimum of $V:P\ra\R$. Let $p(t)$ be any smooth 
curve in $P$ with $p(0)=p_0$, and let $\dot{p}(0)=X\in T_{p_0}P$. 
Since $p_0$ is a critical point of $V$, $\d V_{p_0} X=(V\circ p)'(0)=0$. 
Now $\hess_{p_0}$ is, by definition, the unique symmetric bilinear 
form on $T_{p_0}P$ such that
\beq
\left.\frac{d^2 V(p(t))}{dt^2}\right|_{t=0}=\hess_{p_0}(X,X)
\eeq
for all curves $p(t)$. Since $p_0$ is a minimum of $V$, $\hess_{p_0}$ 
is non-negative, that is, $\hess_{p_0}(X,X)\geq 0$ for all $X$. 
The vector space $T_{p_0}P$ is equipped with an inner product, 
\beq\label{srue}
\ip{(\sigma,Y),(\sigma',Y')}_{T_{(\rho_0,\Phi_0)}P}
=\sigma\sigma'+\frac14\rho_0^2\ip{Y,Y'}_{T_{\Phi_0}\CP^{k-1}},
\eeq
so we can uniquely identify $\hess_{p_0}$ with a self-adjoint linear 
map ${\cal H}_{p_0}:T_{p_0}P\ra T_{p_0}P$ such that
\beq
\hess_{p_0}(X,X')=\ip{X,{\cal H}_{p_0}X'}.
\eeq
Let $E_i$, $i=1,\ldots,2k-1$ be an orthonormal basis of eigenvectors 
of ${\cal H}_{p_0}$ with corresponding eigenvalues $m_i^2\geq 0$. 
Then we can expand $(\sigma,Y)\in T_{p_0}P$ relative to this basis
\beq
(\sigma,Y)=\sum_{i=1}^{2k-1}\alpha_i E_i,
\eeq
whereupon we obtain
\Align{sruesiad}{
F_{lin}=\frac12\int_M\Bigg\{&
\frac{1}{e^4\rho_0^4}(|\d J|^2+e^2\rho_0^2|J|^2) \nonumber\\
&+\sum_{i=1}^{2k-1}(|\d\alpha_i|^2+m_i^2\alpha_i^2)\Bigg\}.
}
This is the energy functional of a set of decoupled fields, consisting 
of a Proca (vector boson) field $J$ of mass 
\beq
m_J=e\rho_0
\eeq
and $(2k-1)$ real Klein-Gordon (scalar boson) fields $\alpha_i$, 
of masses $m_i$. 

In general, the asymptotic fields of a soliton will have all these 
degrees of freedom non-zero, and the dominant force between 
well-separated solitons will be mediated by whichever mode has 
longest range, that is, lowest mass. So the first task in predicting 
long range intersoliton forces is to compute the spectrum of the 
self-adjoint linear map ${\cal H}_{(\rho_0,\Phi_0)}$. For a generic 
choice of $V$ in the family we are considering \Eqref{freeEnergy},
it is not possible to compute even the vacuum $(\rho_0,\Phi_0)$ 
explicitly, so the matrix ${\cal H}_{(\rho_0,\Phi_0)}$, and hence 
its spectrum, is perforce known only numerically. There are, 
however, some interesting cases where explicit analytic progress 
is possible.

\begin{figure*}[!htb]
 \hbox to \linewidth{ \hss
 \includegraphics[width=0.75\linewidth]{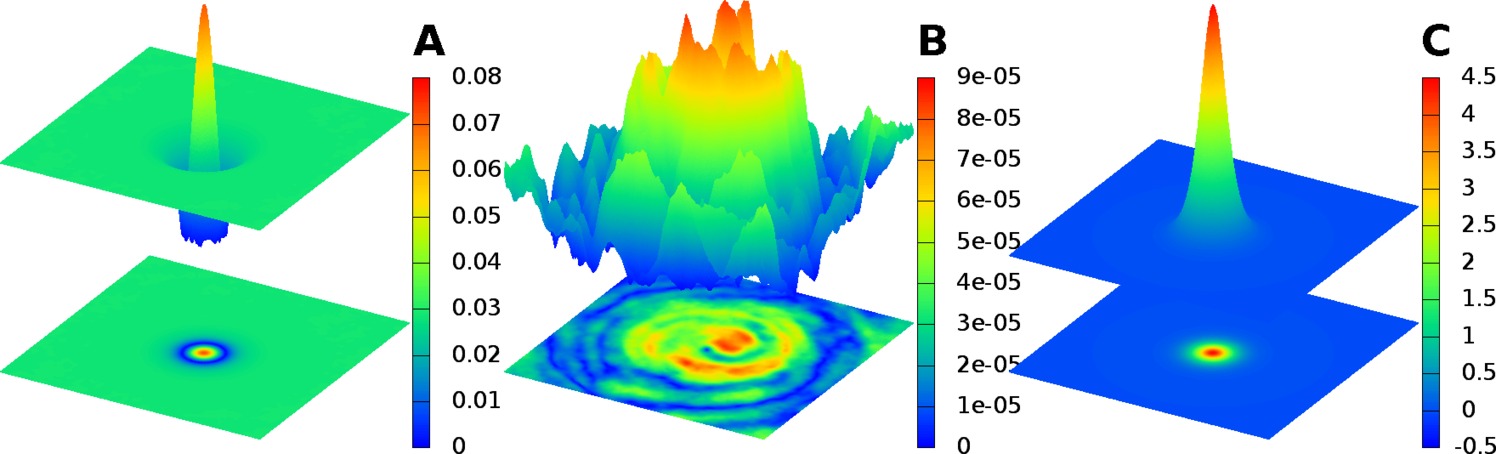}
 \hspace{0.1cm}
 \includegraphics[width=0.225\linewidth]{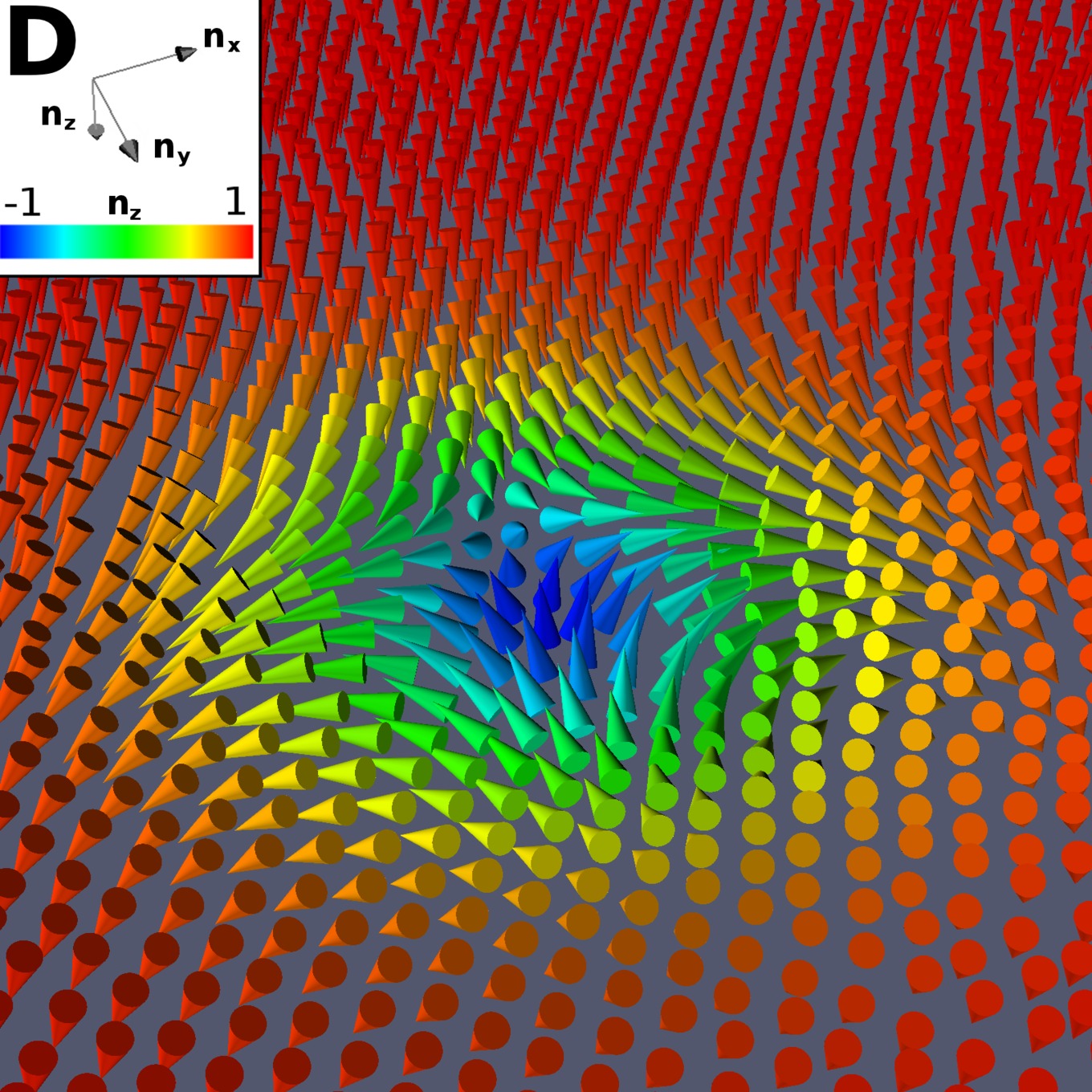}
 \hss}
\caption{(Color online) -- 
The $\Q=1$ soliton for the $U(3)$ symmetric model broken by 
Josephson interactions of the form \Eqref{rheablelc}, with 
$\Lambda=20$ and $\eta_0=1$. The quantities $|1-|\psi|^2|$ 
(panel $\bf A$), and $|e_3^\dagger\Psi|$ (panel $\bf B$), 
measure the deviation from the $\sigma$-model. They converge 
to zero as $\Lambda$ is increased. Panel $\bf C$ shows the 
energy density of the skyrmion.The fourth panel (panel $\bf D$) 
displays the texture of the field $\nvec$, which is similar to 
that of a baby skyrmion.
}
\label{rheablel}
\end{figure*}
\subsection{The sigma model limit}  \label{sigma-model}

In this section we
consider the $k$-component GL model with potential
\beq
V=\frac12\Lambda(1-|\Psi|^2)^2-\frac12\Psi^\dagger\eta\Psi
\eeq
in the limit $\Lambda\ra\infty$, where $\eta$ is a real-symmetric 
$k\times k$ matrix, with zero diagonal, parametrizing a general 
collection of Josephson interactions. In the notation of 
\Partref{section:theoretical_framework} this is the case 
$-\alpha_a=\beta_a=\gamma_{ab}=\Lambda$ for all $a,b$. 
The special case where $\eta=0$, $\Lambda\ra\infty$ and 
$e\ra\infty$, which reduces to a pure sigma model, was 
considered in Refs.~\onlinecite{golo,dadda}. It is possible to 
find explicit formulae for the topological solitons in that case.
The case of finite $\Lambda$ and $e$, with $\eta=0$, has also 
been treated previously 
\cite{achucarro1,Hindmarsh:92,Achucarro.Vachaspati:00}. 
The field equations for the model \Eqref{angker} in the sigma 
model limit (in fact, in the case where $\Phi$ is valued in any 
compact k\"ahler manifold) were studied in detail, from a 
geometric viewpoint, in Ref.~\onlinecite{spe}. Our focus 
here is on the new phenomena introduced by the Josephson 
terms $\eta$.

In terms of the polar coordinates $\rho,Z$, the limit 
$\Lambda\ra\infty$ amounts to the constraint $\rho\equiv 1$, 
and the potential $V$ reduces, in this limit, to
\beq\label{rh}
V([Z])=-\frac12 Z^\dagger\eta Z=-\frac12
\frac{Z^\dagger\eta Z}{|Z|^2}.
\eeq
We have included the factor of $|Z|^2$ in the denominator 
of this expression (which, of course, equals $1$ since $|Z|=1$ 
by definition) so that the right hand side is manifestly a function 
of the projective equivalence class of $Z$ only, not $Z$ per se, 
that is, $V([cZ])=V([Z])$ for all $c\in\C\less\{0\}$. This is convenient 
when one comes to compute the Hessian of $V$. Since $\eta$ is 
real symmetric, it has a unitary basis of eigenvectors 
$e_1,e_2,\ldots,e_k$, with corresponding real eigenvalues 
$\lambda_1\geq\lambda_2\geq\cdots\geq \lambda_k$. Expanding 
$Z$ relative to this basis
\beq
Z=\sum_{i=1}^k\chi_i e_i,\qquad \chi\in\C^k,\quad |\chi|=1,
\eeq
we see that
\beq
V=-\frac1{2}\sum_{i=1}^k\lambda_i|\chi_i|^2.
\eeq
Hence, the $U(k)$ symmetry of the model, which is preserved 
by the sigma-model limit, is broken by $\eta$ generically to 
$U(1)^k$. In the case where the spectrum of $\eta$ is degenerate, 
the breaking may be partial. For example, if $\lambda_1=\lambda_2$ 
and all other $\lambda_i$ are distinct, the free energy remains 
invariant under $U(2)\times U(1)^{k-1}$, where $U(2)$ acts in 
the obvious way on the span of $\{e_1,e_2\}$.

Clearly, $V:\CP^{k-1}\ra\R$ attains its minimum at $[Z]=[e_1]$, 
and this minimum is unique if $\lambda_1\neq\lambda_2$.  If 
$\lambda_1=\lambda_2=\cdots=\lambda_j>\lambda_{j+1}
\geq\cdots\geq\lambda_k$, 
then any $Z$ in the span of $\{e_1,\ldots,e_j\}$ minimizes $V$, 
so the set of minima of $V$ is a $\CP^{j-1}$ submanifold of 
$\CP^{k-1}$. In this case, there can be no energy minimizer on 
$\R^2$ with $\Q\neq 0$, by Derrick's scaling argument \cite{der}, 
(\ie solitons are unstable against expanding indefinitely) so let us 
assume, henceforth, that $\lambda_1\neq \lambda_2$, so that the 
vacuum of the model, $[e_1]$, is unique. If the field 
$\Phi=\pi\circ\Psi:\R^2\ra\CP^{k-1}$ has topological charge $\Q=1$ 
then it wraps $\R^2$ once around some submanifold homologous 
to $\CP^1$ in $\CP^{k-1}$.  In order to minimize the contribution 
of $V$, it should be the $\CP^1$ on which $\Psi$ lies in the span of 
$\{e_1,e_2\}$, the sum of the two highest eigenspaces of $\eta$. 
So we predict that
\beq
\Psi\approx \chi_1 e_1+\chi_2 e_2
\eeq
everywhere, where 
$\chi_i=e_i^\dagger\Psi$ are complex valued functions on $\R^2$. 
From the pair $(\chi_1,\chi_2)$ we can construct a $S^2$-valued 
field using the usual identification of $\CP^1$ with $S^2$, that is
\beq
\nvec=(\ol{\chi}_1\:\: \ol{\chi}_2)\tauvec\Vector{\chi_1\\ \chi_2}
\eeq
where $\tauvec=(\tau_1,\tau_2,\tau_3)$ are the Pauli spin matrices. 
In this way, a $\Q=1$ energy minimizer can, conjecturally, be 
identified with a degree 1 texture $\nvec:\R^2\ra S^2$. Since 
$\Psi$ is parallel to $e_1$ at $|x|=\infty$, we see that 
$\chi_2(\infty)=0$, and hence $\nvec(\infty)=(0,0,1)^T$. 

We present numerical evidence in favor of this conjecture in 
\Figref{rheablel}, in the case $k=3$,
\beq\label{rheablelc} 
\eta=\eta_0\left(\begin{array}{ccc}0&-1&-1\\-1&0&-2\\-1&-2&0\end{array}\right)
\eeq
$\eta_0=1$ and $\Lambda=20$. It is found that $\Psi$ 
approximately satisfies the sigma-model constraint, more precisely, 
$\eps_1=\max_{x\in\R^2}|1-|\Psi|^2|<0.04$. For this choice of 
$\eta$,
\Align{}{
\lambda_1=2\eta_0\,,
   e_1&=\frac1{\sqrt{2}}\Vector{0\\1\\-1}	\nonumber\\
\lambda_2=(\sqrt{3}-1)\eta_0\,,
   e_2&=\frac1{\sqrt{6+2\sqrt{3}}}\Vector{\sqrt{3}+1\\-1\\-1}\nonumber\\
\lambda_3=-(\sqrt{3}+1)\eta_0\,,
   e_3&=\frac1{\sqrt{6-2\sqrt{3}}}\Vector{\sqrt{3}-1\\1\\1}.\nonumber
}

We expect the $\Q=1$ energy minimizer to have $\Psi$ in the 
span of $\{e_1,e_2\}$ which, since the eigenvectors form a 
unitary frame, is equivalent to satisfying $e_3^\dagger\Psi=0$. 
Again, this turns out to be approximately true: 
$\eps_2=\max_{x\in\R^2}|e_3^\dagger\Psi|<0.03$. We find 
that both the errors $\eps_1$ and $\eps_2$ become smaller 
as $\Lambda$ increases with $\eta_0$ held fixed. This indicates 
that the sigma model limit is well founded and should be a 
reliable approximation for $\Lambda$ large but finite. Qualitatively, 
in this special case of the 3-component model, the $\nvec$ field 
we find numerically is  similar to the field of a so-called 
baby-skyrmion \cite{pieschzak}.

\begin{figure*}[!htb]
\hbox to \linewidth{ \hss
 \includegraphics[width=0.25\linewidth]{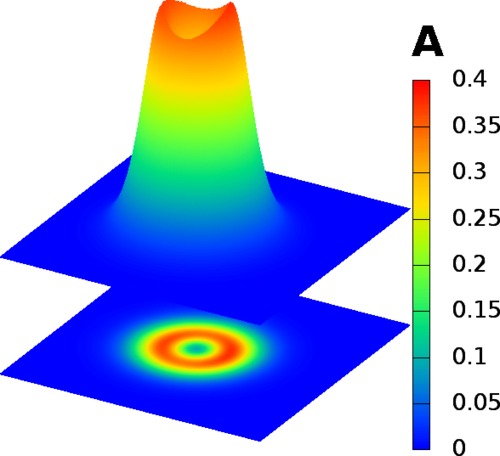}
 \includegraphics[width=0.25\linewidth]{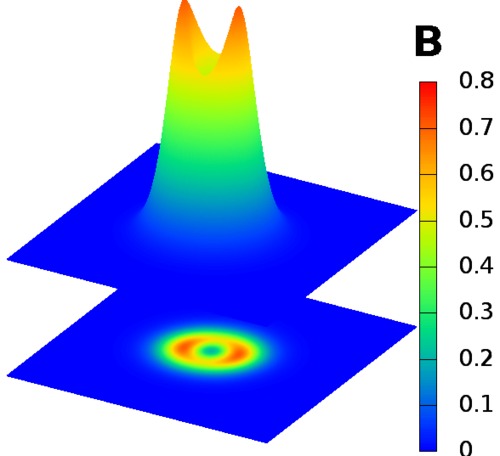}
 \includegraphics[width=0.25\linewidth]{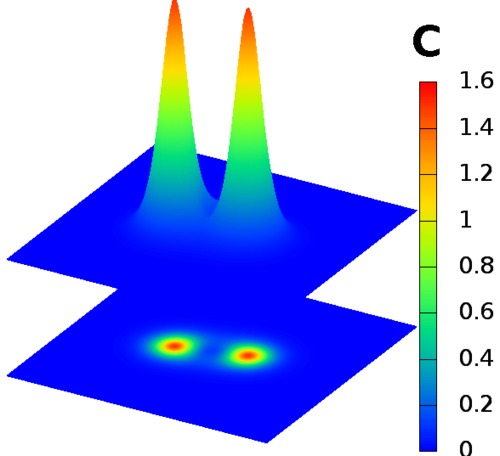}
 \includegraphics[width=0.25\linewidth]{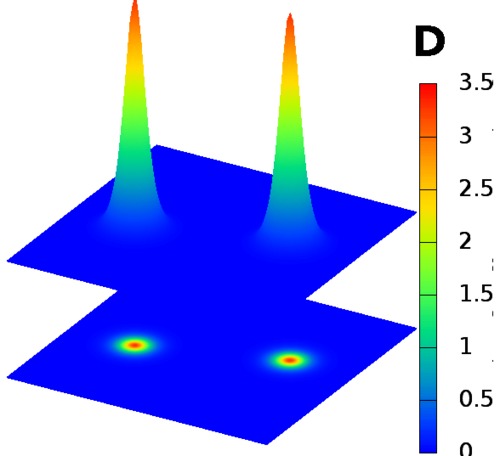}
\hss}
\vspace{0.2cm}
 \hbox to \linewidth{ \hss
 \includegraphics[width=0.5\linewidth]{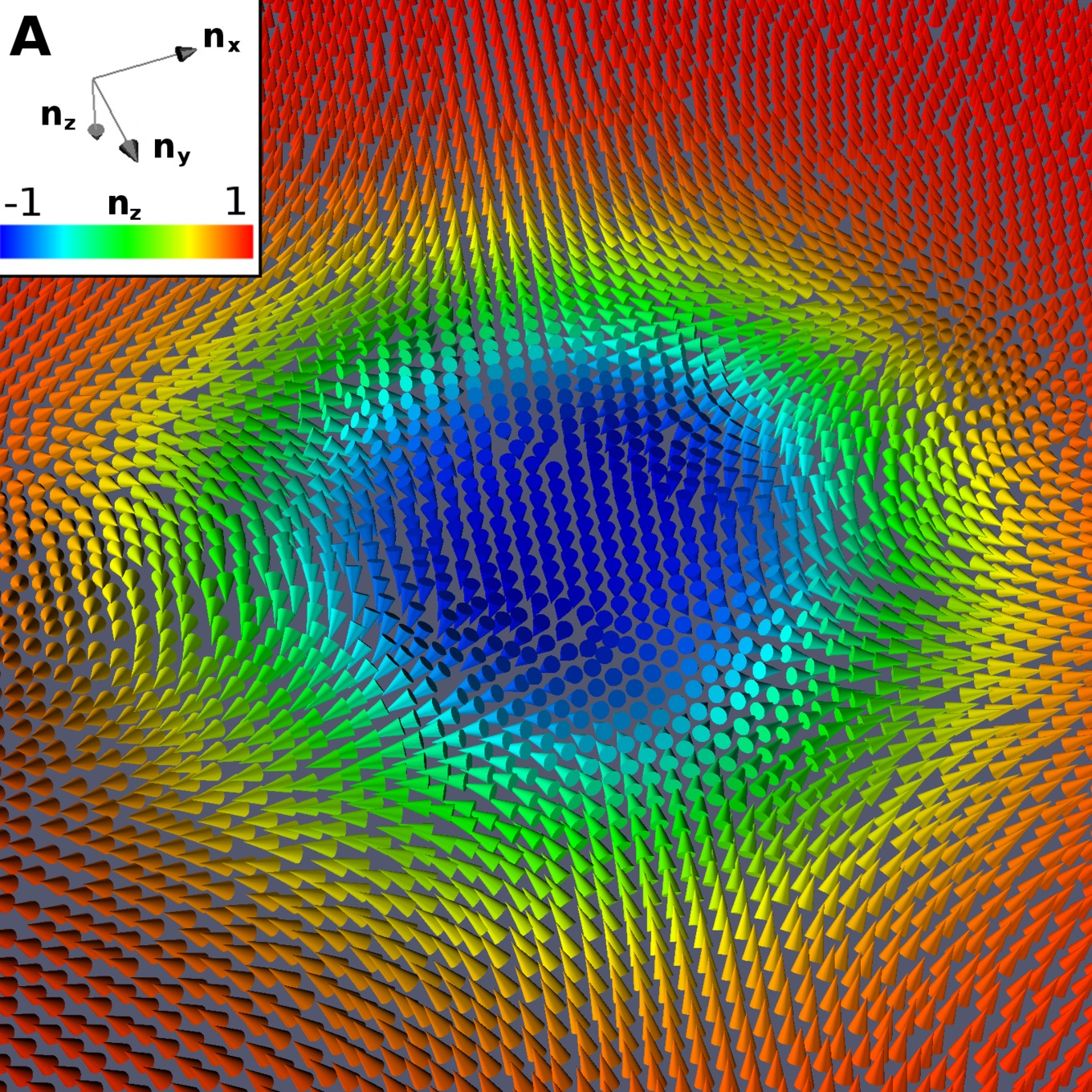}
 \includegraphics[width=0.5\linewidth]{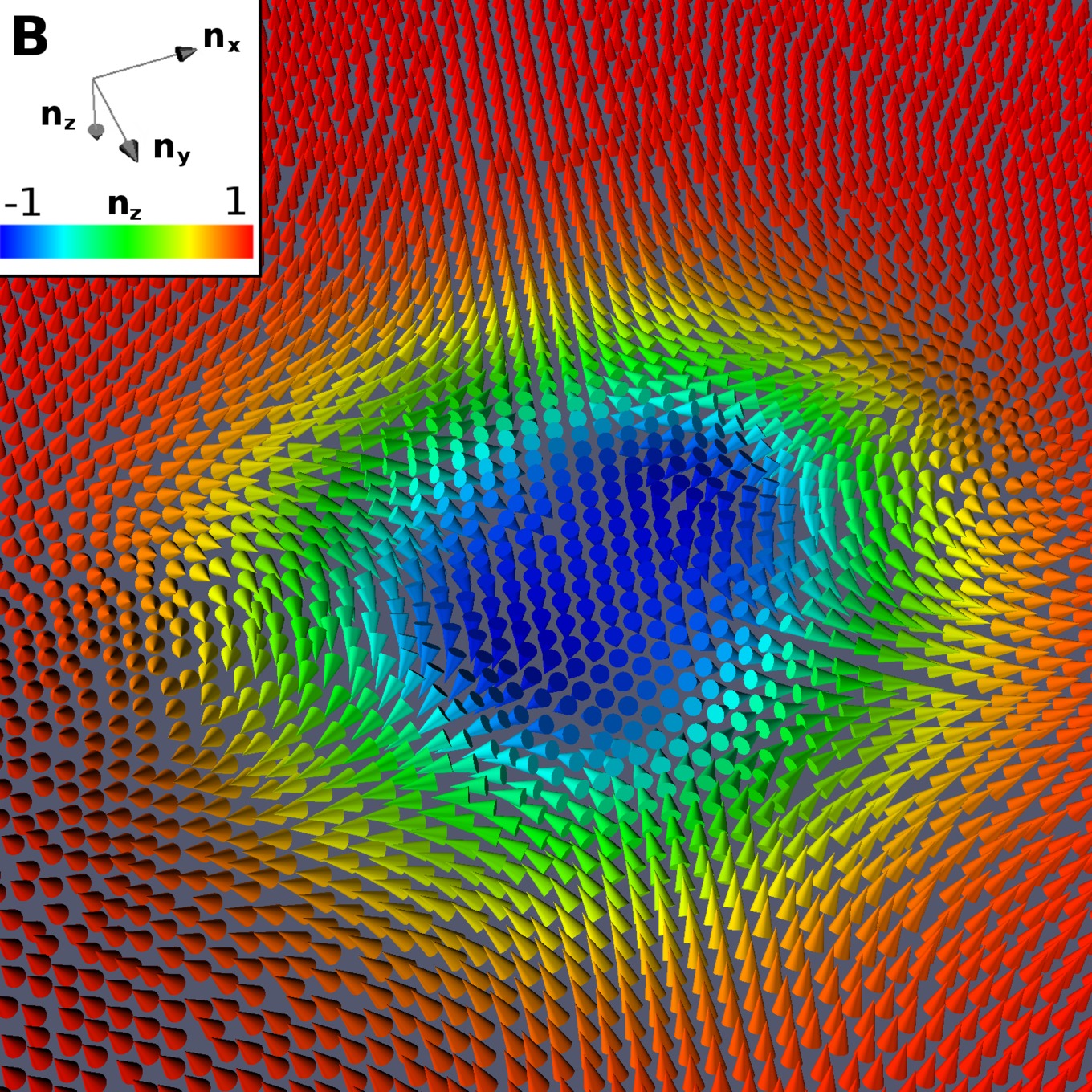}
\hss}
\vspace{0.025cm}
 \hbox to \linewidth{ \hss
 \includegraphics[width=0.5\linewidth]{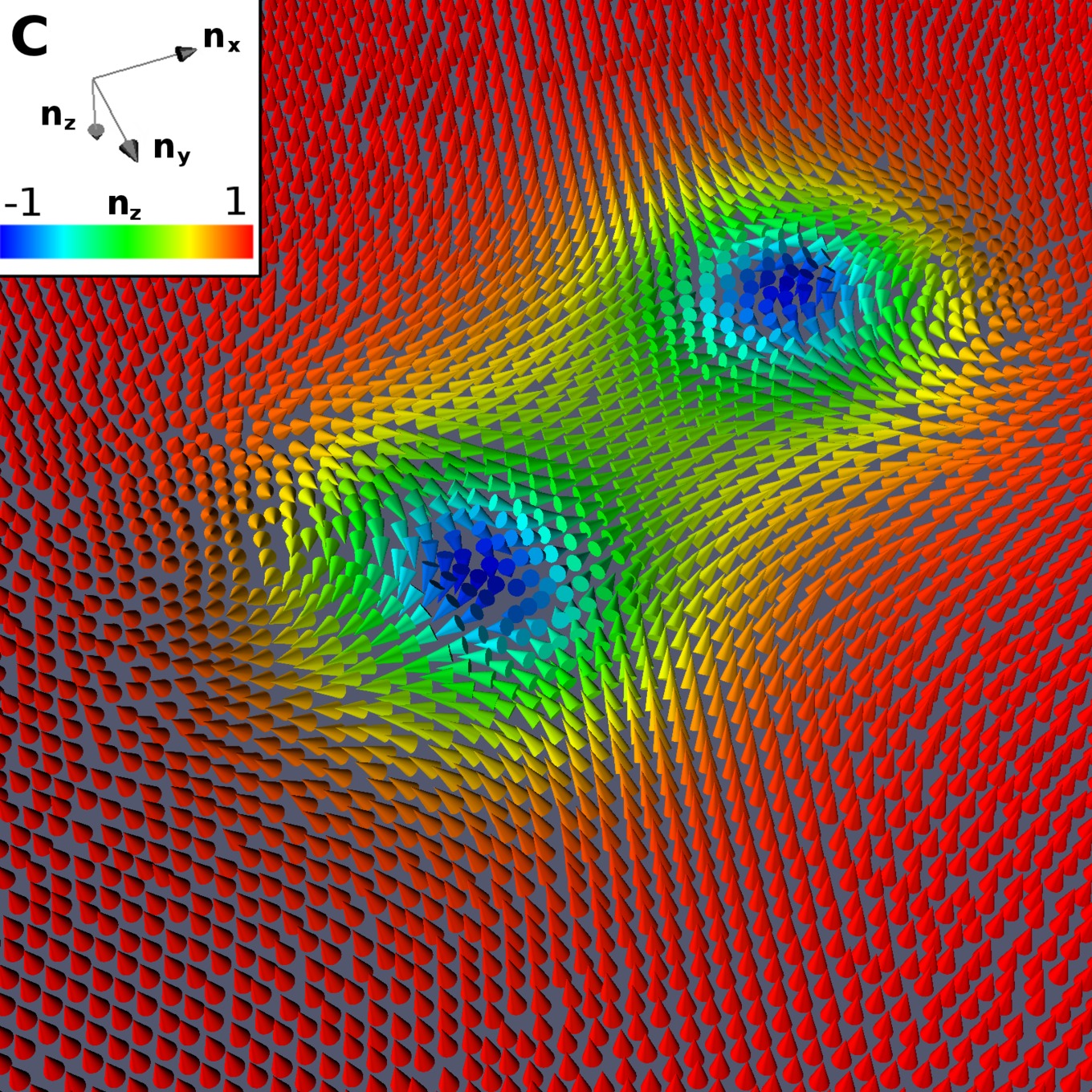}
 \includegraphics[width=0.5\linewidth]{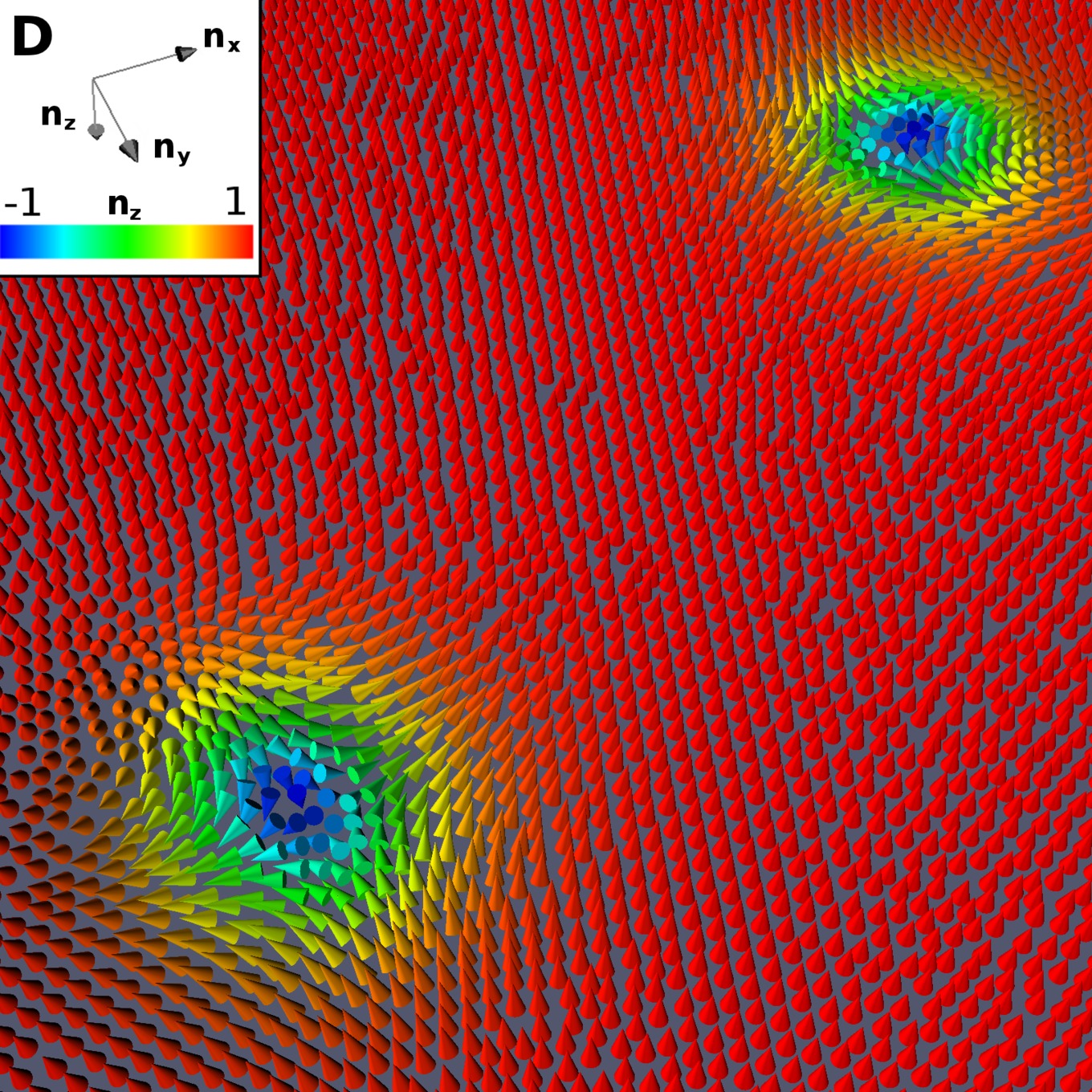}
\hss}
\caption{(Color online) -- 
The results of energy minimization with charge $\Q=2$ for 
$\Lambda=20$, and $e=1.0$ and varying $\eta_0$. First row 
shows the energy density while the second and third row displays 
the corresponding texture field.Panels ($\bf A$, $\eta_0=0.1$), 
($\bf B$, $\eta_0=0.2$)), ($\bf C$, $\eta_0=0.3$)), are for 
$\eta_0<\eta_0^*$ where interaction between skyrmions is 
attractive. Two $\Q=1$ skyrmions coalesce into either one 
$\Q=2$ skyrmion ($\bf A$, $\bf{B}$ and $\bf C$). Configuration 
displayed on panel $\bf{C}$ resemble bound state of 
two $\Q=1$ skyrmions.
Panel ($\bf D$, $\eta_0=0.8$)) has $\eta_0>\eta_0^*$, then in 
the repulsive channel. Here the two $\Q=1$ skyrmions are  
repelling each other. So the snapshot on panel $\bf D$ shows 
a late but unconverged iteration (\ie it represents a fairly 
converged pair of individual skyrmions which are, however, 
still drifting apart).
}
\label{rhea}
\end{figure*}
 
\begin{figure}[!htb]
 \hbox to \linewidth{ \hss
  \includegraphics[width=\linewidth]{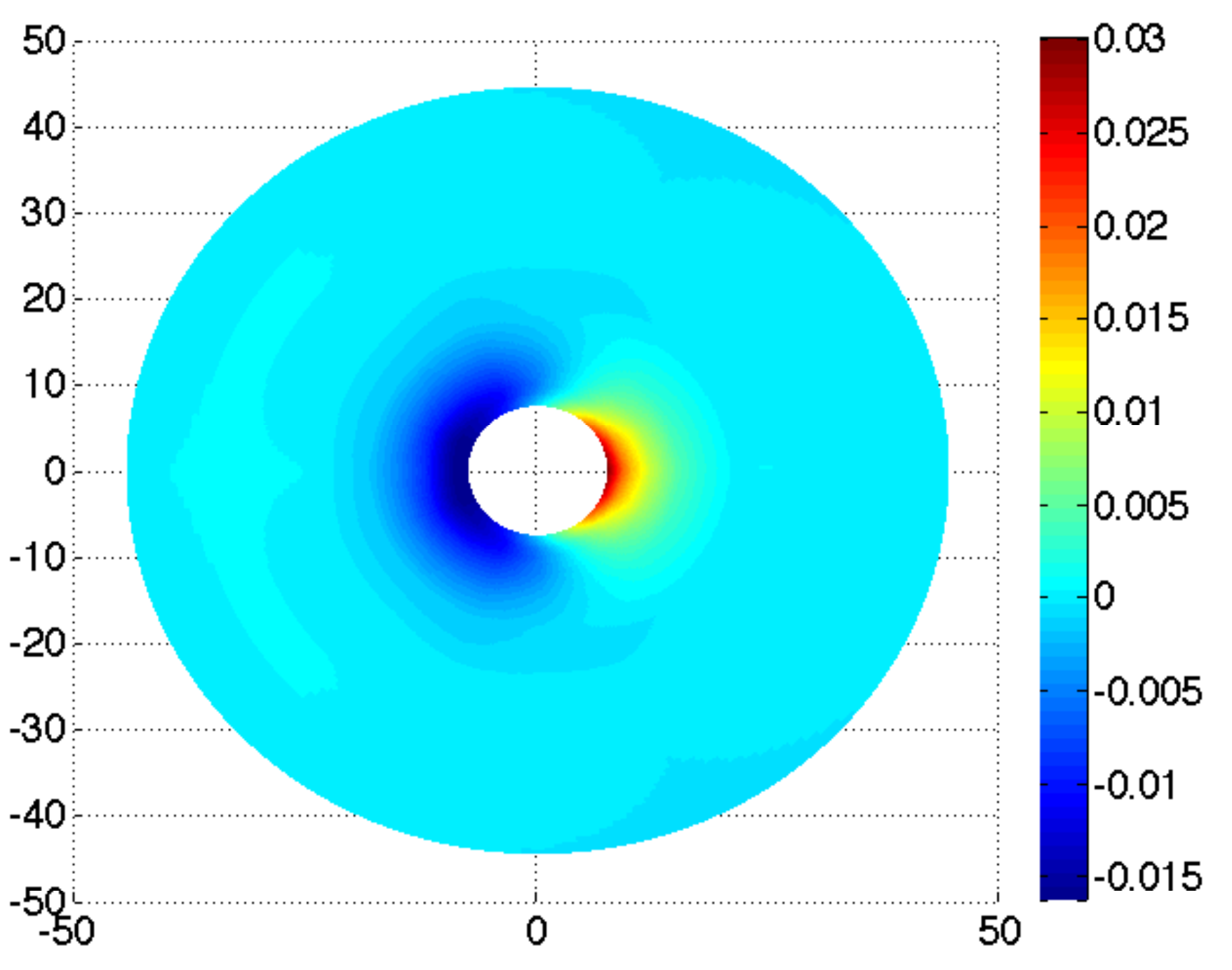}
  \hss}
\caption{(Color online)--
Interaction energy of two single quantum solitons. The the GL 
parameters are $\alpha_a=-20$, $\beta_a=20$, 
$\gamma_{ab}=20$, $\eta_{ab}=-1$ and $e=1$. Thus, 
the potential part of the free energy density can be written 
as $U=\lambda(1-|\psi_1|^2-|\psi_2|^2-|\psi_3|^2)^2 
-\eta_{ab}|\psi_a||\psi_b|\cos(\varphi_a-\varphi_b)$ 
with $\lambda=10$, \ie the Hamiltonian features $SU(3)$ 
symmetry broken by Josephson interaction term. 
}
\label{su3_10}
\end{figure}

If we place two $\Q=1$ energy minimizers a long distance 
apart and allow the system to relax, do they repel one another 
and escape to infinity, or do they attract one another and 
coalesce into a $\Q=2$ bound state? To predict this, we need 
to compute the spectrum of the Hessian of $V$ about 
$\Phi_0=[e_1]$, as described in \Partref{section:lrisf}. In this 
case, $\rho$ is frozen by the constraint, so $P=\CP^{k-1}$. 
It is useful to identify the tangent space $T_{[e_1]}\CP^{k-1}$ 
with the $(k-1)$-dimensional complex vector space
\beq
\VV=\{Y\in\C^k\: :\: e_1^\dagger Y=0\}.
\eeq
Then the natural metric on $T_{p_0}P$ \Eqref{srue} reduces to
\beq
\ip{Y,Y'}_\VV=\Re(Y^\dagger Y')
\eeq
the restriction of the Euclidean metric on $\C^k$ to $\VV$.
 To compute the Hessian of $V$ about $[e_1]$, we consider a 
curve $Z(t)$ in $\C^k$ with $Z(0)=e_1$ and $\dot{Z}(0)=Y\in\VV$. 
Then
\Align{}{
\hess_{[e_1]}(Y,Y)&=\left.\frac{d^2\: }{dt^2}\right|_{t=0}V(\Psi(t)) 
    \nonumber\\
&=Y^\dagger[\lambda_1\I_k-\eta]Y	\nonumber\\
&=\ip{Y,(\lambda_1\I_k-\eta)Y}_{\VV}
}
where we have used the fact that $\lambda_1\I_k-\eta$ is self adjoint 
with $e_1$ in its kernel, so that $e_1^\dagger(\lambda\I_k-\eta)Y=0$. 
Hence, the associated self-adjoint linear map ${\cal H}_{[e_1]}:\VV\ra\VV$ 
is the restriction to $\VV$ of $\lambda_1\I_k-\eta$. It follows that the 
eigenvalues of ${\cal H}_{[e_1]}$ are $\lambda_1-\lambda_i$, $i=2,\ldots,k$, 
each of multiplicity $2$, and that the corresponding eigenspaces are 
two real-dimensional, spanned by $\{e_i,ie_i\}$, $i=2,\ldots,k$. So there 
are $2k-2$ real scalar bosons in this model, occurring in pairs, having mass
\beq
m_i=\sqrt{\lambda_1-\lambda_i}.
\eeq
This should be compared with the mass of the supercurrent field, 
\ie the inverse London penetration length,
\beq
m_J=e.
\eeq

Numerics suggest that the supercurrent of a $\Q=1$ energy minimizer 
is, at large $|x|$, similar to that of a vortex, while the lightest (complex) 
Klein-Gordon mode  $\chi_2$ is similar to the asymptotic field of a 
baby-skyrmion. Hence, we expect $J$ to mediate a repulsive force of 
range $1/e$ and $\chi_2$ to mediate a short-range scalar dipole-dipole 
force. The range of this force is  $1/\sqrt{\lambda_1-\lambda_2}$. 
The latter force is attractive provided the two solitons are appropriately 
aligned; see the discussion of baby-Skyrme models \cite{jayspesut} 
for a detailed analysis. The dipole like interaction is also natural from 
the viewpoint of the fractional-vortex picture of skyrmions (see discussion 
in \Partref{Interactions} and in Refs.~\onlinecite{frac,smiseth}). Hence, 
we predict that a pair of $\Q=1$ solitons, in the model which we consider 
in this subsection, always repel (for all relative orientations) if 
$e^2<\lambda_1-\lambda_2$, so higher $\Q$ bound states cannot form. 
On the other hand, if $e^2>\lambda_1-\lambda_2$, well-separated 
solitons have an attractive channel, and we predict that they can coalesce 
into higher $\Q$ bound states. Numerical evidence of this predicted 
dichotomy in the three component case is presented in \Figref{rhea} 
and direct numerical evidence of dipolar interaction of two $\Q=1$ 
skyrmions is presented in \Figref{su3_10}. 
 
\subsection{A symmetric case with BTRS}\label{section:btrs}

In this section we consider the GL$^{(3)}$ model with three 
identical active bands, coupled through identical Josephson terms. 
The potential is
\beq\label{sgl3}
V(\Psi)=\frac\lambda8\sum_{a=1}^3(1-|\psi_a|^2)^2
+\frac\eta 2\Psi^\dagger N\Psi,
\eeq
where $N$ denotes the symmetric coupling matrix
\beq
N=\left(\begin{array}{ccc}
0&1&1\\
1&0&1\\
1&1&0\end{array}\right).
\eeq
Note that, in contrast to section \ref{sigma-model}, $\eta$ 
denotes a real parameter here, not a matrix. In terms of the 
notation of section \ref{section:theoretical_framework}, this is 
the special case $\alpha_1=\alpha_2=\alpha_3=-\frac\lambda4$, 
$\beta_1=\beta_2=\beta_3=\frac\lambda4$, $\gamma_{ab}=0$ 
and $\eta_{12}=\eta_{13}=\eta_{23}=-\eta$. The vacuum 
manifold for this potential is a disjoint union of two circles, 
the gauge orbits of
\beq\label{vac}
\Psi=\rho_0 v_0\qquad\mbox{and}\qquad\Psi=\rho_0 v_1
\eeq
where
\beq
\rho_0=\sqrt{3+\frac{6\eta}{\lambda}},
\eeq
and (with $\xi=e^{2\pi i/3}$)
\beq
v_0=\frac{1}{\sqrt{3}}\Vector{1 \\\xi \\ \xi^2},~
v_1=\frac{1}{\sqrt{3}}\Vector{1 \\\xi^2 \\ \xi},~
v_2=\frac{1}{\sqrt{3}}\Vector{1 \\ 1\\ 1},
\eeq
are simultaneous unit eigenvectors of the symmetric coupling 
matrix $N$ and the permutation matrix $P$,
\beq
P=\left(\begin{array}{ccc}
0&0&1\\
1&0&0\\
0&1&0\end{array}\right).
\eeq
Note that
\bea
&&Nv_0=-v_0,\quad Nv_1=-v_1,\quad Nv_2=2v_2,\\
&&Pv_0=\xi v_0,\quad Pv_1=\xi^2v_1,\quad Pv_2=v_2.
\eea
We shall, without loss of generality, choose the vacuum 
$\rho_0v_0$ (rather than $\rho_0 v_1$). Since $[\ol\Psi]\neq[\Psi]$ 
for this vacuum, the model has broken time-reversal symmetry. 

There are axially symmetric vortex solutions which interpolate 
between $(0,0,0)$ at $r=0$ and the above vacuum at $r=\infty$. 
To construct them, one only needs to solve a {\em single} 
component GL model:
\beq\label{scm}
F_*=\int_{\R^2}\Bigg\{\frac12|\d A|^2+\frac12|D\phi|^2+\frac{\lambda}{24}
(\rho_0^2-|\phi|^2)^2\Bigg\}.
\eeq
Given a vortex solution $(\phi,A)$ of \Eqref{scm},
\beq\label{embvor}
\Psi=\phi v_0,\qquad A
\eeq
is a vortex solution of the symmetric GL$^{(3)}$ model \Eqref{sgl3}. 
The numerical results of \Partref{section:theoretical_framework} 
strongly suggest that \Eqref{sgl3} also supports skyrmion solutions, 
at least for  $\Q$ and $1/e$ sufficiently large. 

Once again, we wish to compute the spectrum for the Hessian of 
$V$ about the vacuum $(\rho_0,[v_0])$. The potential is, in polar 
coordinates \Eqref{polcoo},
\Align{}{
V(\rho,Z)&=\frac{\lambda}{8}\sum_{a=1}^3(1-\rho^2{|Z_a|^2})^2
  +\frac\eta2\rho^2{Z^\dagger N Z}\\
&=\frac{3\lambda}{8}-\frac\lambda4\rho^2
  +\frac\lambda8\rho^4U([Z])+\frac\eta2\wt{U}([Z])
}
where 
\Align{Udef}{
U([Z])&=\frac{1}{|Z|^4}\sum_{a=1}^3|Z_a|^4\\
\wt{U}([Z])&=\frac{Z^\dagger NZ}{|Z|^2}.
}
We have  included the factors of $|Z|^2$ in the denominators of 
these expressions (which, of course, equals $1$ by definition) so 
that the right hand sides are manifestly functions $[Z]$ only. Recall 
that $\hess$ is a symmetric bilinear form on the tangent space to 
$(0,\infty)\times\CP^2$ at the vacuum $(\rho_0,[v_0])$. In general, 
there is no reason why this bilinear form should not couple the 
direction tangent to $(0,\infty)$ with directions tangent to $\CP^2$. 
We shall see that in this case permutation symmetry prevents such 
coupling.

First, we note that $[v_0]$ is a fixed point of the permutation map
\beq
\PP:\CP^2\ra\CP^2,\qquad[Z]\mapsto [PZ],
\eeq
 and that $\d\PP_{[v_0]}:T_{[v_0]}\CP^2\ra T_{[v_0]}\CP^2$
has maximal rank, so it follows that $[v_0]$ is a critical point of 
{\em any} function $\CP^2\ra\R$ invariant under $\PP$. In particular,
\beq\label{sruesi}
\d U_{[v_0]}=\d\wt{U}_{[v_0]}=0.
\eeq
 Consider now a two-parameter variation $p(s,t)=(\rho(s),[Z(t)])$ 
through $p_0=(\rho_0,[v_0])$ in $P=(0,\infty)\times\CP^2$, with 
$\cd_sp(0,0)=(\sigma,0)$ and $\cd_tp(0,0)=(0,Y)$. Then
\Align{}{
\hess_{p_0}((\sigma,0),(0,Y))&=
\left.\frac{\cd^2 V(p(s,t))}{\cd s\cd t}\right|_{s=t=0} \nonumber\\
&=\frac\lambda4\rho_0^3\sigma\d U_{[v_0]}Y
+\eta\rho_0\sigma\d\wt{U}_{[v_0]}Y \nonumber\\
&=0
}
by \Eqref{sruesi}. Hence
\beq\label{hessbtrs}
\hess=(\lambda+2\eta)\d\rho_{\rho_0}^2+\frac\lambda8\rho_0^4H
  +\frac\eta2\rho_0^2\wt{H}
\eeq
where $H,\wt{H}:T_{[v_0]}\CP^2\times T_{[v_0]}\CP^2\ra\R$ are 
the Hessians of the functions $U,\wt{U}$ respectively. It follows that 
one of the real scalar bosons $\alpha_i$ in \Eqref{sruesiad} is just 
$\sigma$ (the linearization of $\rho$ about $\rho_0$) and that this 
has mass
\beq
m_\rho=\sqrt{\lambda+2\eta}.
\eeq

It remains to compute $H$ and $\wt{H}$. For this purpose, we 
identify the tangent space $T_{[v_0]}\CP^2$ with the two 
dimensional complex vector space
\beq
\VV=\{Y\in\C^3\: :\: v_0^\dagger Y=0\}
\eeq
which is spanned by $\{v_1,v_2\}$, and give $\VV$ the induced 
Euclidean metric
\beq
\ip{X,Y}_\VV=\frac12(X^\dagger Y+Y^\dagger X)
  =\frac14\ip{X,Y}_{FS}
\eeq
where $\ip{\cdot ,\cdot}_{FS}$ denotes the Fubini-Study metric, 
used to compute $|\d Y|^2 $ in equation \Eqref{angker}.

In fact, we already know $\wt{H}$, since this is a special case 
of the general Josephson coupling matrix considered in section 
\ref{sigma-model}:
\beq
\wt{H}(X,Y)=2\ip{X,(N+1)Y}_{\VV}
\eeq
It is convenient to expand $X,Y$ relative to the unitary 
(for $\ip{\cdot ,\cdot}_{\VV}$) basis $v_1,v_2$, which are 
eigenvectors of $N$. Namely, if
\Align{}{
X&=(x_1+ix_2)v_1+(x_3+ix_4)v_2,\nonumber \\
Y&=(y_1+iy_2)v_1+(y_3+iy_4)v_2
}
then
\beq\label{Htildemat}
\wt{H}(X,Y)=6x^T\left(\begin{array}{cccc}
0&0&0&0\\
0&0&0&0\\
0&0&1&0\\
0&0&0&1
\end{array}
\right)y.
\eeq
Note that this is a hermitian bilinear form on
$\VV$, that is $\wt{H}(iX,iY)=\wt{H}(X,Y)$. 

Turning to $H$, one should not expect it to be hermitian, 
because $U$ contains terms like $Z_1^2\ol{Z}_1^2$. 
Consider a two-parameter variation $Z_{s,t}$ in 
$S^{5}\subset\C^3$ with $Z_{0,0}=v_0$ and 
$\cd_s Z_{s,t}|_{0,0}=X\in\VV$, $\cd_t Z_{s,t}|_{0,0}=\VV$. 
By definition,
\beq
H(X,Y)=\left.\frac{\cd^2 U(Z_{s,t})}{\cd s\cd t}\right|_{s=t=0}.
\eeq
Using the explicit formula \Eqref{Udef} for $U(Z)$, we find that
\Equation{}{
H(X,Y)=2\sum_{a=1}^3(\ol{X}_aZ_a+\ol{Z}_aX_a)(\ol{Y}_aZ_a+\ol{Z}_aY_a)~~
}
where $Z=v_0$. Note this is {\em not} Hermitian because, 
for example $H(iv_1,iv_2)=-H(v_1,v_2)$. Again, we can 
express this as a $4\times 4$ real matrix, by expanding 
$X,Y$ relative to $v_1,v_2$. One finds that
\beq\label{Hmat}
H(X,Y)=\frac43x^T\left(\begin{array}{cccc}
1&0&1&0\\
0&1&0&-1\\
1&0&1&0\\
0&-1&0&1
\end{array}\right)y.
\eeq
Substituting \Eqref{Hmat} and \Eqref{Htildemat} into 
\Eqref{hessbtrs}, then \Eqref{hessbtrs} into \Eqref{flinhess}, 
we obtain
\Align{}{
F_{lin}&=\int_M\Bigg\{\frac{1}{2e^4\rho_0^4}
  (\|\d J\|^2+e^2\rho_0^2\|J\|^2)  \nonumber \\
&+\frac12(|\d\sigma|^2+(\lambda+2\eta)\sigma^2)
+\frac12\rho_0^2\left(|\d y|^2+y^T{\cal M}y\right)\Bigg\}
}
where the mass matrix is
\beq
{\cal M}=\frac{\lambda\rho_0^2}{6}
\left(\begin{array}{cccc}
1&0&1&0\\
0&1&0&-1\\
1&0&1&0\\
0&-1&0&1
\end{array}\right)+3\eta\left(\begin{array}{cccc}
0&0&0&0\\
0&0&0&0\\
0&0&1&0\\
0&0&0&1
\end{array}\right).
\eeq
The squared masses of the bosons tangent to $\CP^2$ 
are the eigenvalues of this matrix, namely
\beq
m_{\pm}^2=\frac{\lambda\rho_0^2}{6}\left\{1
+\frac{9\eta}{\lambda\rho_0^2}\pm
\sqrt{1+\left(\frac{9\eta}{\lambda\rho_0^2}\right)^2}\right\},
\eeq
each of multiplicity two. These should be compared with 
the mass of the $J$ vector boson and $\rho$ scalar boson
\beq
m_J^2=e^2\rho_0^2,\qquad
m_\rho^2=\lambda+2\eta.
\eeq

To extract information about intersoliton forces, note 
that the embedded vortex \Eqref{embvor} excites only 
the (repulsive) $J$ mode and the (attractive) $\rho$ mode, 
so one predicts the usual behaviour (\ie for the example 
considered here where there is degeneracy in couplings 
between components, at long range vortices repel if 
$m_\rho>m_J$, and attract if $m_\rho<m_J$). Note that 
in the case  when the components have different prefactors 
in $V$, there are also type-1.5 regimes with non-monotonic 
intervortex (long-range attractive, short-range repulsive) 
intervortex forces \cite{Carlstrom.Garaud.ea:11a}. Skyrmions, 
on the other hand, should in all cases excite all 6 modes, 
with a monopole source for $\rho$ and dipole (or higher) 
sources for the 4 (mixed) $Y$ modes. So an interesting 
regime would be $m_-<m_J<m_\rho$ since then intervortex
forces should be long-range repulsive, while inter-skyrmion 
forces should have an attractive channel for a certain relative 
orientations of skyrmions.

%%%%%%%%%%%%%%%%%%%%%%%%%%%%%%%%%%%%%%%%%%%%%%%%%%%%%%%%%%%%%%%%%%%%%%
%%%%%%%%%%%%%%%%%%%%%%%%%%%%%%%%%%%%%%%%%%%%%%%%%%%%%%%%%%%%%%%%%%%%%%
\section{Conclusions}\label{Conclusion}

We discussed a new kind of topological soliton which we term 
chiral $GL^{(3)}$ skyrmions. These solitons occur  in 
three-component superconductors when time reversal symmetry 
is spontaneously broken. In contrast to vortices, these skyrmions 
are characterized by a \CPtwo topological charge. These skyrmions 
have a definite {\it chirality} associated with them: \ie the order 
of the constituent fractional vortices matters, different orders 
giving inequivalent solutions. We described two situations
\begin{itemize}
\item A type-II BTRS superconductor can form a vortex lattice 
as a ground state in applied magnetic field. However in contrast 
to usual vortex states, all the regimes investigated by us 
possessed other flux-carrying topological defects of a higher 
energy: metastable $GL^{(3)}$ skyrmions characterized 
by a \CPtwo topological charge. The system thus can form 
infinitely many complex metastable states in external fields 
where vortices coexist with the $GL^{(3)}$ skyrmions solitons.
Thermal, or  magnetic field quench can force the system to fall 
into one of these states.
\item BTRS three-band superconductors in principle can have 
also a different regime where in external field \CPtwo solitons 
are energetically cheaper than vortices. In that case the system 
cannot form vortices since they are unstable against decay into 
skyrmions.  Such regimes occur for example when the free 
energy has bi-quadratic interaction terms of the form 
$\gamma_{ab}|\psi_a|^2|\psi_b|^2$.
\end{itemize}

In the regimes where  chiral $GL^{(3)}$ skyrmions are metastable
they can  spontaneously form from `collisions' of vortices, where 
intervortex interaction energy can be larger than energy of 
potential barrier of forming a skyrmion. We investigated several 
hundred regimes and found that skyrmions typically easily form 
in the energy minimization process where a system is relaxed 
from various higher energy states (such as dense groups of 
ordinary vortices). Our study indicates that the ``capture basin" 
of these solutions can in certain cases be very large. We find that 
these defects  very easily form during a rapid expansion of a vortex 
lattice (which should occur  when magnetic field is rapidly lowered, 
or if a system is quenched through $H_{c2}$). Formation of 
solitons in this process can signal  a state with Broken Time Reversal 
Symmetry. Also the potential barriers between Skyrmions and 
vortices or between different skyrmionic states can be overcome 
due to thermal fluctuations.

As shown in \Figref{Signature}, these skyrmions have very 
particular magnetic signature and thus, under certain 
conditions, may be observed in 
high-resolution scanning SQUID, Hall, or magnetic force 
microscopy measurements. A tendency for vortex pair formation, 
yielding magnetic profile similar to that shown on \Figref{Fig:Pair1} 
was observed in $\rm Ba(Fe_{1-x}Co_x)_2As_2$, \cite{Kaliski} 
as well as vortex clustering in $\rm BaFe_{2-x}Ni_xAs_2$ \cite{Li}. 
These materials have strong pinning which can naturally 
produce disordered vortex states \cite{Li}, although the 
possibility of ``type-1.5" scenario for these vortex 
inhomogeneities was also voiced in Ref.~\onlinecite{Li} .
(Note that in three band (or higher number of bands) superconductors 
with frustrated Josephson coupling, type-1.5 regimes are easily 
obtainable even if Josephson coupling is very strong 
\cite{Carlstrom.Garaud.ea:11a}.)
The vortex pairs observed in Ref.~\onlinecite{Kaliski} can be discriminated 
from $\Q=2$ solitons by quenching the system in a stronger 
magnetic field and observing whether or not it forms vortex 
triangles, squares, pentagons, such as shown on \eg 
\Figref{Signature} which correspond to flux profile of 
higher-$\Q$ solitons. Besides multiband superconductors,
another class of systems which can  support chiral $GL^{(3)}$ 
skyrmions is a Josephson coupled sandwich of an $s_\pm$ 
and $s$-wave superconductor.

The work is supported by the Swedish Research Council, 
by the Knut and Alice Wallenberg Foundation through the 
Royal Swedish Academy of Sciences fellowship and by
NSF CAREER Award No. DMR-0955902, and by the 
UK Engineering and Physical Sciences Research Council.
The computations were performed on resources 
provided by the Swedish National Infrastructure for Computing 
(SNIC) at National Supercomputer Center at Linkoping, Sweden.

%%%%%%%%%%%%%%%%%%%%%%%%%%%%%%%%%%%%%%%%%%%%%%%%%%%%%%%%%%%%%%%%%%%%%%
%%%%%%%%%%%%%%%%%%%%%%%%%%%%%%%%%%%%%%%%%%%%%%%%%%%%%%%%%%%%%%%%%%%%%%
%%%% Bibliography
%Merlin.mbs v4.21 2009-07-09.
%

%%%%%%%%%%%%%%%%%%%%%%%%%%%%%%%%%%%%%%%%%%%%%%%%%%%%%%%%%%%%%%%%%%%%%%
%%%%%%%%%%%%%%%%%%%%%%%%%%%%%%%%%%%%%%%%%%%%%%%%%%%%%%%%%%%%%%%%%%%%%%
\clearpage
\appendix
%%%%%%%%%%%%%%%%%%%%%%%%%%%%%%%%%%%%%%%%%%%%%%%%%%%%%%%%%%%%%%%%%%%%%%
\renewcommand{\theequation}{\Alph{section}.\arabic{equation}}

\section{Fractional vortices have linearly divergent energy in the presence 
of Josephson coupling}
\label{Kink}
Here we discuss 
 fractional flux vortices in three band systems. Consider the case of one 
fractional vortex in which $\psi_1$ winds through $2\pi$ and neither 
$\psi_2$ nor $\psi_3$ winds. We assume that the configuration is spatially 
localized around $r=0$, so that on any annulus $\Omega=\{r_0<r<r_1\}$, 
with $r_0$ sufficiently large, the densities $|\psi_a|$ are close to their 
ground state values (\ie we assume the London limit). It follows from 
expression \Eqref{freeEnergy} that the total free energy of any 
configuration satisfies the lower bound
\Align{}{
&F-\FGS\geq \FSG\,,  \\
\text{with}~~~&\FSG:=\sum_{a<b}\nu_{ab}\int_\Omega
|\nabla\varphi_{ab}|^2+\frac12m_{ab}^2(1-\cos\varphi_{ab})\nonumber
}
where $\nu_{ab}=|\psi_a|^2|\psi_b|^2/\varrho^2$, 
$m_{ab}^2=2\eta_{ab}\varrho^2/|\psi_a||\psi_b|$, and 
$\varrho^2=\sum_a|\psi_a|^2$. $\FGS$ denotes the energy of the 
vortex-less ground state. In the London limit, the field densities 
assume their ground state values, so $\nu_{ab}$ and $m_{ab}$ 
are constants. In this limit, $\FSG$ simplifies to a sum of sine-Gordon 
energies (hence the subscript $SG$). Note that $|\nabla\varphi_{ab}|^2\geq 
r^{-2}(\cd\varphi_{ab}/\cd\theta)^2$, with $r$ and $\theta$, the polar 
coordinates around the vortex center. Hence, 
\Align{}{
\FSG&\geq\sum_{a<b}\nu_{ab}\int_\Omega\left\{\frac{1}{r^2}
\left(\frac{\cd\varphi_{ab}}{\cd\theta}\right)^2
+m_{ab}^2\sin^2\frac{\phi_{ab}}{2}\right\}\\
&=\sum_{a<b}\nu_{ab}\int_\Omega\left\{\left(
\frac{1}{r}\frac{\cd\varphi_{ab}}{\cd\theta}
-m_{ab}\sin\frac{\varphi_{ab}}{2} \right)^2 \right.\nonumber \\
&~~~~~~~~~~~~~~~~~~~~~~~~~~~~~~\left.+\frac{2m_{ab}}{r}
\frac{\cd\varphi_{ab}}{\cd\theta}\sin\frac{\varphi_{ab}}{2}
\right\}\\
&\geq\sum_{a<b}2m_{ab}\nu_{ab}\int_{r_0}^{r_1}dr\, r\int_{0}^{2\pi}
\frac{1}{r}\frac{\cd\varphi_{ab}}{\cd\theta}\sin\frac{\varphi_{ab}}{2}\\
&=8(m_{12}\nu_{12}+m_{13}\nu_{13})(r_1-r_0)
}
where we have used the boundary conditions that $\varphi_{12}$ 
and $\varphi_{13}$ wind once, while $\varphi_{23}$ does not wind. 
So $\FSG$, and hence the total free energy $F-\FGS$, grows (at least) 
linearly with the system size, $r_1$.

Note that our lower bound on $\FSG$ cannot be attained, because for this
to happen, one would need $\varphi_{ab}$ to satisfy
\beq
\frac{1}{r}\frac{\cd\varphi_{ab}}{\cd\theta}
=m_{ab}\sin\frac{\varphi_{ab}}{2}
\eeq
and no solutions to this PDE with the correct boundary behaviour
($\varphi_{12}(r,2\pi)-\varphi_{12}(r,0)=2\pi$ for all $r$) exist.

%%%%%%%%%%%%%%%%%%%%%%%%%%%%%%%%%%%%%%%%%%%%%%%%%%%%%%%%
\section{Finite element energy minimization}\label{Numerics}

\begin{figure*}[!htb]
 \hbox to \linewidth{ \hss
  \includegraphics[width=0.75\linewidth]{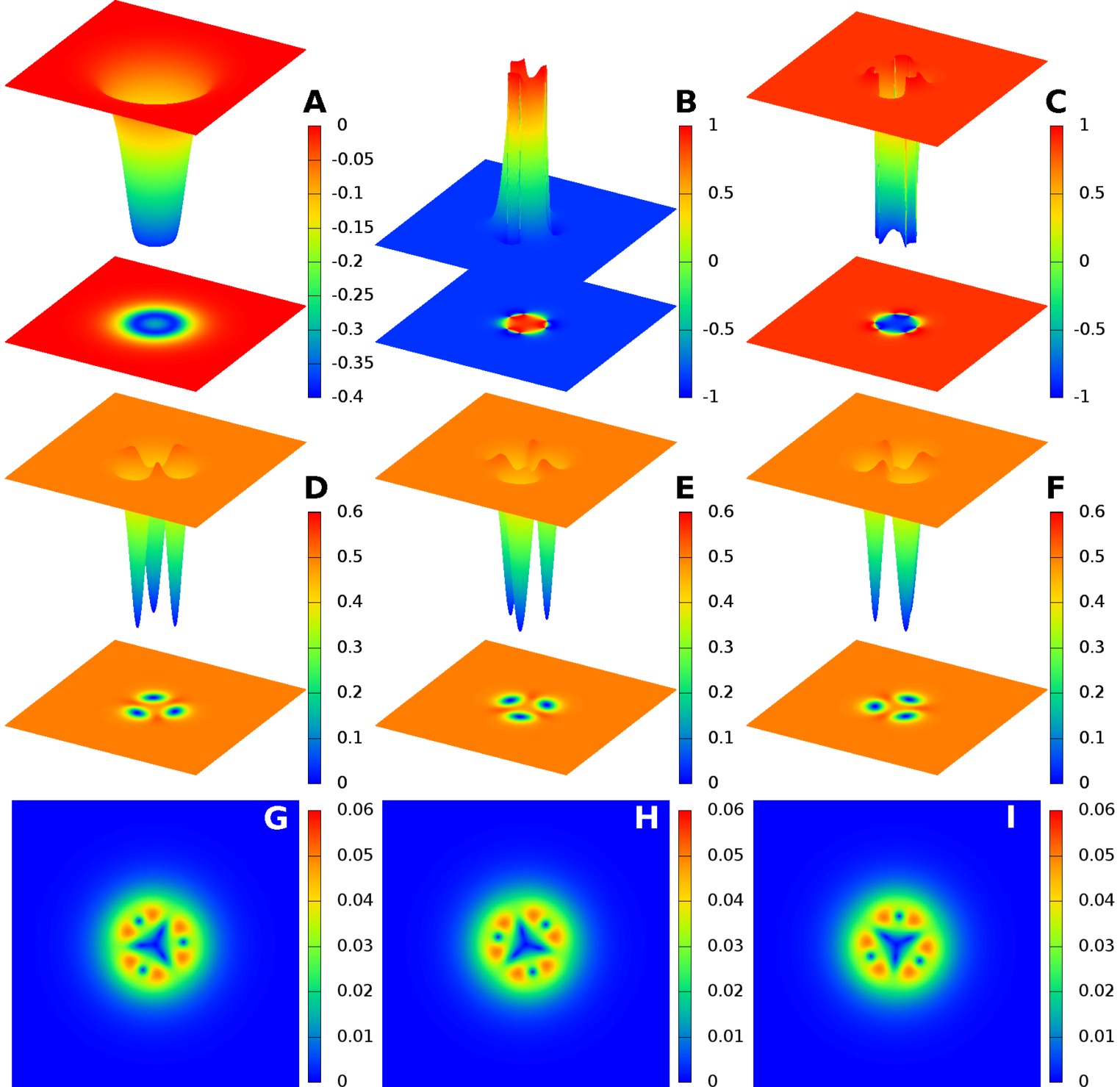}
  \hss}
\caption{
(Color online) -- A $\Q=3$ quanta soliton in a system with three identical 
passive bands as in \Figref{Fig:Single1}, except that there are no 
density-density interactions $\gamma_{ab}=0$ and  $e=0.3$. Since the 
three bands are identical, the soliton makes a homogeneous ringlike 
configuration. Displayed quantities are the same as in rest of the paper.
}
\label{Fig:Q=3}
\end{figure*}

The chiral skyrmions are either {\it global} or {\it local}  minima of the 
Ginzburg-Landau energy \Eqref{freeEnergy}. In the later case, this means that 
a good enough initial guess is necessary. In both cases, the functional minimization 
of \Eqref{freeEnergy}, from an appropriate initial guess carrying several flux 
quanta, should lead to a chiral skyrmion (if it exists as a stable solution). 
We consider the two-dimensional problem  \Eqref{freeEnergy} defined on 
the bounded domain $\Omega\subset\mathbbm{R}^2$ with $\partial\Omega$ 
its boundary. In practice we choose $\Omega$ to be a disk. 
Actually, the particular shape of the domain is not important. Indeed 
it is much larger than the typical size of solitons. Moreover, neither solitons 
nor initial guess coincide with some grid symmetry. For example, skyrmions 
are never placed at the center of the domain (the vizualization scheme 
re-centers the window around the soliton). This is an addtional argument 
that skyrmions are not boundary artefacts. One some occasions, we 
doubled checked on square domains that our solutions are  unaffected
by boundaries.

The problem is supplemented by the boundary condition 
$\bs n\cdot\bs D\Psi_a=0$ with $\bs n$ the normal vector to $\partial\Omega$. 
Physically this condition implies there is no current flowing through the 
boundary.
Since this boundary condition is gauge invariant, additional constraint can 
be chosen on the boundary to fix the gauge. Our choice is to impose the radial 
gauge on the boundary $\bs e_\rho\cdot\bs A=0$ (note that with our choice of 
domain, this is equivalent to $\bs n\cdot\bs A=0$). With this choice, (most of) 
the gauge degrees of freedom are eliminated and the `no current flow' condition 
separates in two parts
\Equation{BC}{
   \bs n\cdot\nabla \psi_a=0~~~~~~\text{and}~~~~~~\bs n\cdot\bs A=0\,.
}
Note that these boundary conditions allow a topological defect to escape 
from the domain, since there is no pressure of an external applied field. 
Because they are topological defects, vortices (and skyrmions) cannot 
unwind. However, they can be `absorbed' through the boundary in order 
to further minimize the energy. To prevent this, the numerical grid is 
chosen to be large enough so that the attractive interaction with the 
boundaries is negligible. The size of the domain is then much larger than 
the typical interaction length scales. Thus in this method one has to use 
large numerical grids, which is computationally demanding. The advantage 
is that it is guaranteed that obtained solutions are not boundary pressure 
artifacts.

The variational problem is defined for numerical computation using a finite 
element formulation provided by the Freefem++ library \cite{Hecht.Pironneau.ea}. 
Discretization within finite element formulation is done via a (homogeneous) 
triangulation over $\Omega$, based on Delaunay-Voronoi algorithm. Functions 
are decomposed on a continuous piecewise quadratic basis on each triangle. 
The accuracy of such method is controlled through the number of triangles, 
(we typically used $3\sim6\times10^4$), the order of expansion of the basis 
on each triangle (2nd order polynomial basis on each triangle), and also the 
order of the quadrature formula for the integral on the triangles. 

Once the problem is mathematically well defined, a numerical optimization 
algorithm is used to solve the variational nonlinear problem (\ie to find the 
minima of $F$). We used here a  nonlinear conjugate gradient method. 
The algorithm is iterated until relative variation of the norm of the gradient 
of the functional  $F$ with respect to all degrees of freedom is less than 
$10^{-6}$. 

\subsection*{Initial guess for obtaining metastable configurations}

\begin{figure*}[!htb]
 \hbox to \linewidth{ \hss
  \includegraphics[width=0.75\linewidth]{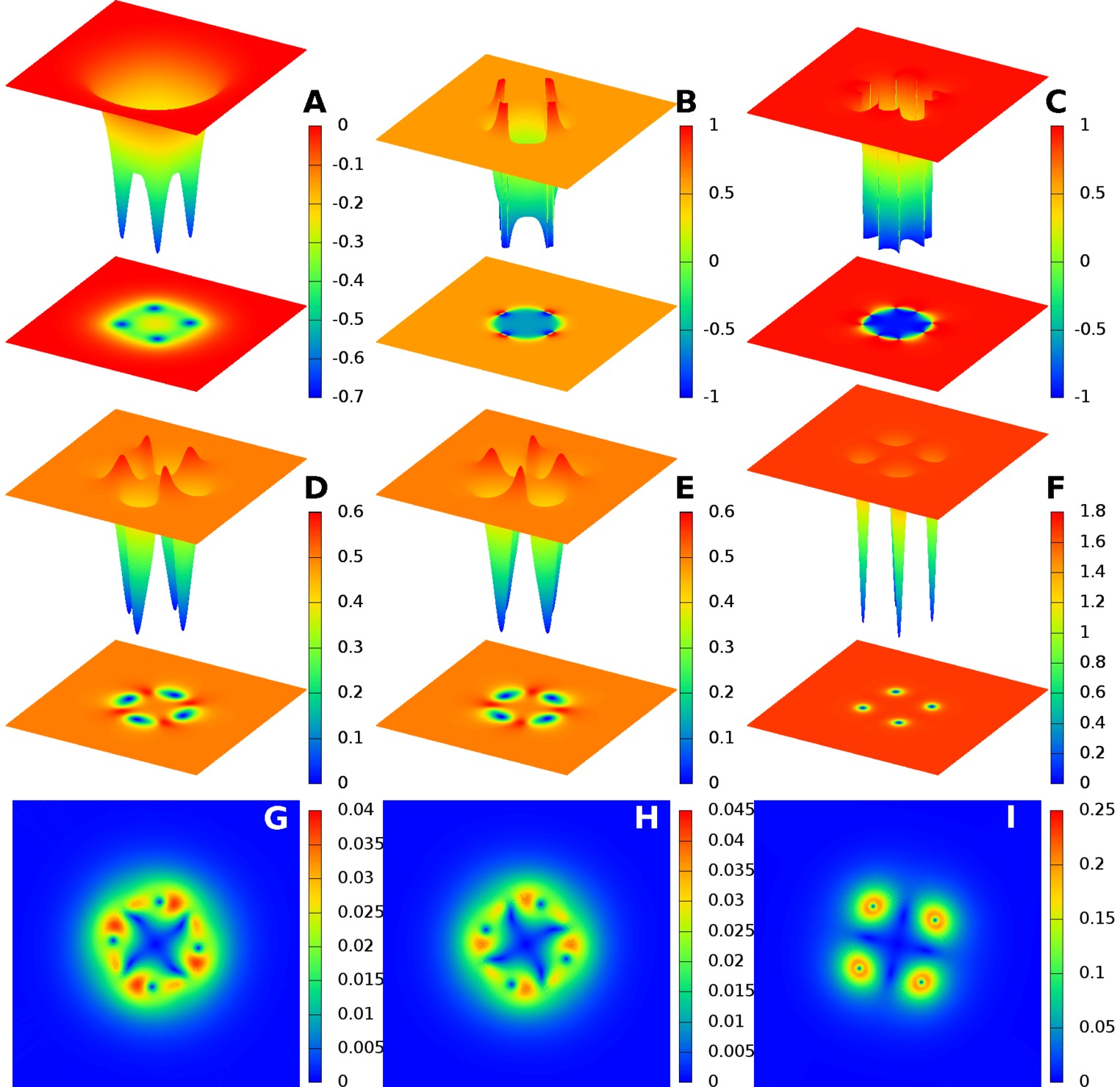}
  \hss}
\caption{
(Color online) -- A $\Q=4$ quanta soliton in a system with two identical 
passive bands as in \Figref{Fig:Pair1} coupled to a third active band 
with disparity in the ground state densities $(\alpha_3,\beta_3)=(-1.5,1)$. 
Josephson coupling constants are $\eta_{23}=-3$ and 
$\eta_{12}=\eta_{13}=1$. $e=0.2$ and $\gamma_{ab}=0$. 
}
\label{Fig:Q=4}
\end{figure*}

As discussed in the paper, $N$ quanta chiral skyrmions can be more 
energetically expensive than $N$ ordinary (type-II) vortices.  In that 
case the initial guess should be within the attractive basin of the chiral 
skyrmions. Otherwise the configuration converges to ordinary type-II 
vortices which have the same total phase winding but cost less energy. 
The initial field configuration carrying $N$ flux quanta is prepared by 
using an ansatz which imposes phase windings around spatially separated 
$N$ vortex cores in each condensates. 
\Align{Initial_Guess1}{
\psi_1&= |\psi_1|\mathrm{e}^{ i\Theta_1} \, ,
\psi_2= |\psi_2|\mathrm{e}^{ i\Theta_2+i\Delta_{2}} \, ,
\psi_3= |\psi_3|\mathrm{e}^{ i\Theta_3+i\Delta_{3}} \, ,~~  \nonumber \\
|\psi_a| &= u_a\prod_{k=1}^{N_v} 
\sqrt{\frac{1}{2} \left( 1+\tanh\left(\frac{4}{\xi_a}({\cal R}^a_k(x,y)-\xi_a) 
\right)\right)}\, ,
}
where $a=1,2,3\,$ and $u_a\,$ is the ground state value of each 
condensate density. The parameters $\xi_a$ parametrize the 
core size while 
\Align{Initial_Guess2}{
\Theta_a(x,y)&=\sum_{k=1}^{N}
      \tan^{-1}\left(\frac{y-y^a_k}{x-x^a_k}\right)  \,,\nonumber\\
{\cal R}^a_k(x,y)&=\sqrt{(x-x^a_k)^2+(y-y^a_k)^2}\,. 
}
$(x^a_k,y^a_k)$ determines the position of the core of $k$-th vortex 
of the $a$-condensate.The functions $\Delta_{a}$ are used to seed 
a domain wall. As an initial guess we generally choose 
$\Delta_{2}=-\Delta_{3}\equiv\Delta$, with $\Delta$ defined as
\Equation{PhaseDiff}{
   \Delta=\frac{\pi}{3}\left( H({\mbf r}-{\mbf r}_0)-1\right)\,,
}
where  $H({\mbf r}-{\mbf r}_0)$ is a Heaviside function. Thus in 
the initial guess  the domain wall has infinitesimal thickness. It takes 
only a few steps from this initial guess to relax to a true domain wall 
during the simulations. Consequently, it is entirely sufficient to use 
Heaviside functions for the initial guesses for domain walls. 
The starting configuration of the vector potential is determined by 
solving Amp\`ere's law equation of \Eqref{EOM} on the background 
of the superconducting condensates specified by 
\Eqref{Initial_Guess1}--\Eqref{PhaseDiff}. Being a linear equation in 
$\bs A$, this is an easy operation.

Once the initial configuration defined, all degrees of freedom are relaxed 
simultaneously, within the `no current flow' boundary conditions discussed 
previously, to obtain highly accurate solutions of the Ginzburg-Landau 
equations. In a strongly type-II system when the initial guess was either 
(a) vortices placed on a closed domain wall or (b) closed domain wall 
surrounding a densely packed group of vortices, the system almost always 
formed chiral skyrmions. We used also initial guesses (c) without any 
domain walls ($\Delta=0$). In that case we observed chiral skyrmion 
formation, if in the initial states vortices were densely packed. This again 
indicates that the chiral skyrmions in the three component GL model 
represent (local) minima with wide capture basin in the free energy 
landscape.

\section{Additional Material}\label{Extra-graphics}

\begin{figure*}[!htb]
 \hbox to \linewidth{ \hss
  \includegraphics[width=0.75\linewidth]{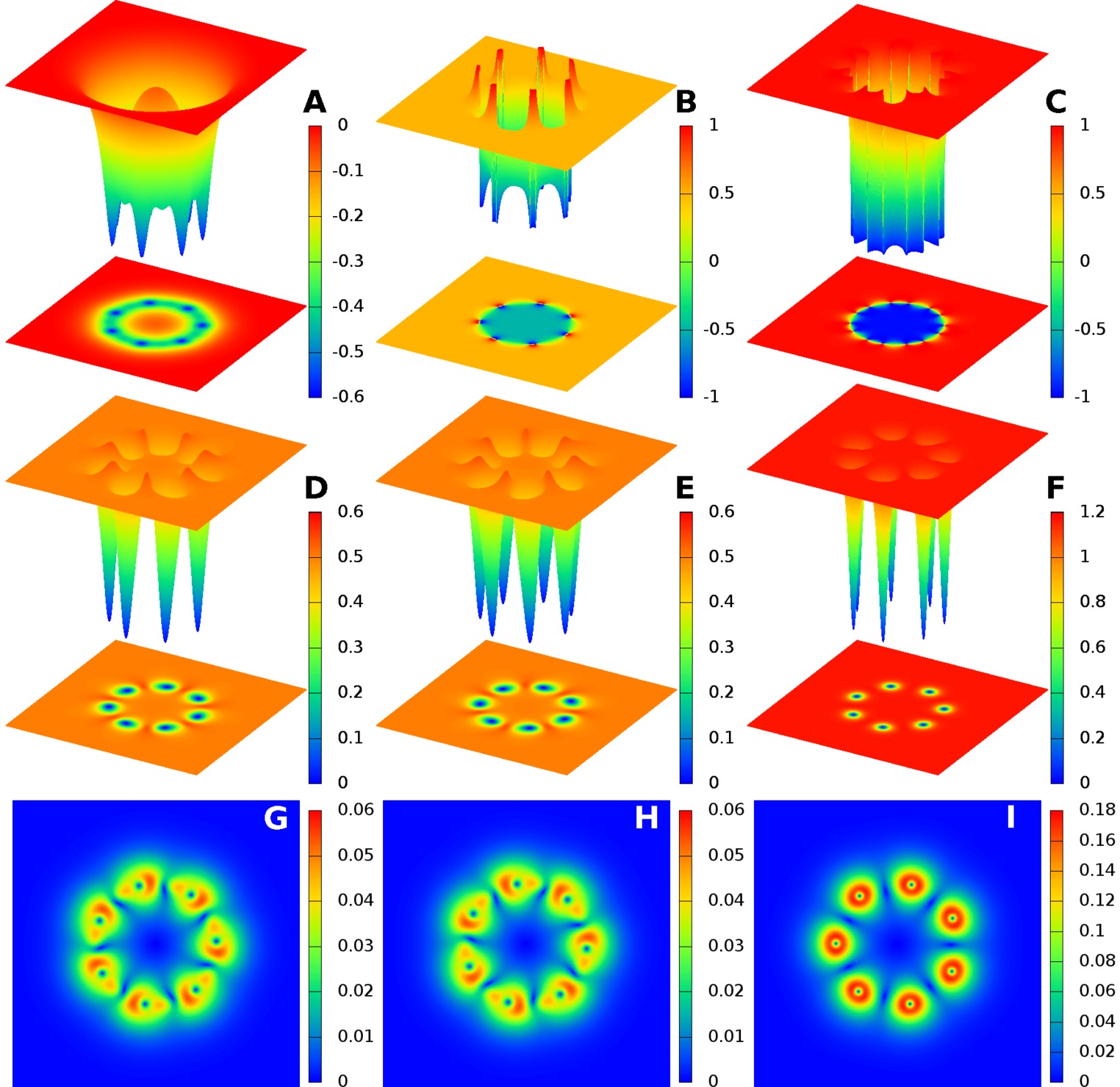}
  \hss}
\caption{
(Color online) -- A $\Q=7$ quanta soliton in a system with two identical 
passive bands as in \Figref{Fig:Pair1} coupled to a third active band with 
disparity in the ground state densities $(\alpha_3,\beta_3)=(-1,1)$.
Josephson coupling constants are $\eta_{23}=-3$ and 
$\eta_{12}=\eta_{13}=1$. $e=0.3$ and $\gamma_{ab}=0$.  
}
\label{Fig:Q=7}
\end{figure*}

In this appendix we show few additional solutions \Figref{Fig:Q=3}, 
\Figref{Fig:Q=4} and \Figref{Fig:Q=7} for chiral skyrmions. Parameters 
sets, or number of flux quanta used here are different from the 
ones considered in the main body of the paper.

%%%%%%%%%%%%%%%%%%%%%%%%%%%%%%%%%%%%%%%%%%%%%%%%%%%%%%%%%%%%%%%%%%%%%%
\end{document}